\documentclass[12pt, reqno]{article}

\usepackage{rotating} %sidewaystable
\usepackage{amssymb} %\renewcommand{\bold}{\textbf}
\usepackage{amsmath}
\usepackage{amsxtra} %\numberwithin{equation}{section}
\usepackage{geometry} %to change margins for a few pages
\usepackage{appendix}

\makeatletter
\@addtoreset{chapter}{part}
\@addtoreset{@ppsaveapp}{part}
\makeatother

\usepackage[longnamesfirst]{natbib} % Bibliography formatting
\usepackage{lscape} %\begin{landscape}  % \end{landscape}

\usepackage{color}  %\textcolor{blue}{ \beta_{3} }  

\usepackage{float} 
\usepackage{standalone} %include{} a chunk of another tex file, ignore preamble etc
\definecolor{ChadDarkBlue}{rgb}{.1,0,.2}
\definecolor{ChadBlue}{rgb}{.1,.1,.5}
\definecolor{ChadRoyal}{rgb}{.2,.2,.8}
\definecolor{ChadGreen}{rgb}{0,.4,0}    % Dark Green
\definecolor{ChadRed}{rgb}{.5,0,.5}  % purple 

\usepackage[format=plain, 
font=sc,  
labelfont=bf, 
skip=10pt, 
justification=centering
]{caption}
\usepackage{hyperref}
\hypersetup{
	colorlinks=true,        % kills boxes
	linkcolor=ChadRed,
	urlcolor=ChadGreen,
	citecolor=ChadRed,
	pdftitle={CollateralValue},
	pdfauthor={ },
	pdfstartview={FitV},
	pdfpagemode={UseNone},
	pdfnewwindow=true,      % links in new window
}

\usepackage{setspace}
\newcommand{\Lag}{\ensuremath{\mathcal{L}}}

\renewcommand{\epsilon}{\varepsilon}

\newcommand{\E}{\ensuremath{\mathbb{E}}}	% To get blackboard E

\DeclareMathOperator*{\argmin}{argmin}

\providecommand{\norm}[1]{\lVert#1\rVert}

\usepackage{booktabs,caption}
\usepackage[flushleft]{threeparttable}

\title{
\textbf{Does Collateral Value Affect Asset Prices?
\\
Evidence from a Natural Experiment in Texas}\thanks{
	I thank my discussants 
	Marc Francke,
	Sanket Korgaonkar, Adam Nowak, and Johannes Stroebel  for detailed comments.
	I thank
	Sumit Agarwal,
	Brent Ambrose,
	Zahi Ben-David,
	Sean Chu,
	Anthony DeFusco,
	Gilles Duranton,
	Vadim Elenev,
	Alex Gelber,
	Todd Gormley, 
	Ben Keys,
	Olivia Mitchell,
		Jonathan Parker,
	Tomasz Piskorski,
	Nikolai Roussanov,
	Todd Sinai,
	Boris Vabson,
	Stijn Van Nieuwerburgh,
	Jessica Wachter,
		Susan Wachter,
	Vincent Yao, 
	and Yildiray Yildirim, 
	as well as	seminar participants at
	AMES, % 2017,
	AREUEA, % 2018,
	ASSA, % 2018,
	Baruch, %Baruch Finance, Baruch Real Estate, 
	Brandeis,  
	Dallas Fed, %
	EFA, %2018
	EMES, % 2017,
	Federal Reserve Board, 
	Houston Finance,
	IAAE, %2018
	Johns Hopkins, 
	Maryland, 
	MMM, %2018 Wisconsin
	NASMES, %2018
	NYCREC, % 2017,
	Penn State, 
	PFMC,  % 2016, 
	Philadelphia Fed, %2019
	Rice,
	ReCapNet, %2020
	SEM,  %2017 
	UT Dallas,
	Wharton, 
	and 
	XXVI Finance Forum %2018
	for helpful comments.
	I thank Albert Saiz for sharing his data. 
	I thank Guido Imbens for sharing his synthetic control code. 	 
	I thank Cheng Chen and Sung Son for excellent research assistance. 
	I gratefully acknowledge funding from the PSC
	CUNY Research Foundation under grant 60202-00-48.
	All errors are my own. 
	\newline
	 Send correspondence to 
	 \newline
	 Albert Alex Zevelev, 1 Bernard Baruch Way, New York, NY 10010. 
	 \newline
	 Email: Albert.Zevelev@baruch.cuny.edu. 
}
}
\author{
\textbf{Albert Alex Zevelev}
\\
Baruch College, City University of New York	
}

\date{} %November 18, 2017

\usepackage{graphicx} %\includegraphics

\begin{document}
%\onehalfspacing
\maketitle
%\section*{}  %to locate abs in TOC
\begin{abstract}
Does the ability to pledge an asset as collateral, after purchase, affect its price?
This paper identifies the impact of collateral service flows %collateral value 
on house prices, exploiting a plausibly exogenous  
constitutional amendment in Texas that legalized home equity loans in 1998. 
The law change increased Texas house prices 4\%; 
this is price-based evidence that households are credit-constrained 
and value home equity loans to facilitate consumption smoothing.
Prices rose more in locations with inelastic supply,   %population, 
higher prelaw house prices, higher income, and lower unemployment. 
These estimates reveal that %wealthier 
richer households value the option to pledge their home as collateral more strongly. 
(JEL R0, R3, G0, E21, E44, G2)
%%% %%
\end{abstract}
\onehalfspacing
\newpage
%\tableofcontents
%
\section{Introduction} %and Literature Review	
\label{Introduction}
Real estate is the largest source of collateral used by American households \citep{FRBNY}.
A home can be pledged as collateral for a loan at the time of purchase or after purchase via
a home equity loan 
(HEL).\footnote{Other products that allow 
	future home equity extraction, 
	including cash-out refinance loans and reverse mortgages,
	will be referred to collectively as ``HELs" for ease of exposition.} 
Theory predicts that an asset's price should reflect all of its benefits 
including  options to pledge it as collateral.
This paper seeks to quantify if the ability to pledge 
an asset as collateral, after purchase, affects the price.

Prior to 1998, 
a home in Texas could only be pledged as collateral for 
a purchase mortgage or home improvement loan. 
Beginning in 1998, Proposition 8 (a Texas constitutional amendment 
inspired by federal tax reform and a circuit court ruling) greatly expanded 
the set of mortgages available to Texans. 
Texas homeowners gained access to  
HELs, cash-out refinance loans, and reverse mortgages; however, 
the total value of all new liens on the home after purchase
could not exceed 80\% of its appraised price.\footnote{See \hyperref[timeline]{Appendix \ref*{timeline}} for a timeline of relevant laws.} 
Because Texas was the only state with such restrictions, 
this law change provides a unique source of exogenous variation;
it expanded future debt capacity (allowing homeowners to extract home equity via HELs) 
without affecting purchase debt capacity (the amount homebuyers could borrow at purchase was unaffected).

While many papers study the impact of 
home equity borrowing on consumption and investment (Section \ref{Literature}),
the current paper examines the impact on house prices. 
There are three related questions. 
First, what was the total impact of the law change on house prices? 
Second, which locations were more affected?  
Third, what was the mechanism through which the law change affected house prices?  
The law change could have raised house prices directly 
through demand for collateral service flows
or indirectly by affecting other variables related to house prices. 
The identifying assumption behind the first question, the causal effect, 
is that the law change was exogenous conditional on fixed effects and controls. 
The identifying assumption behind the direct mechanism
is that the law change did not affect other variables related to house prices.
%While this paper's main focus is to study the causal effect %total impact 
%of the HEL legalization on house prices, it also provides evidence on the mechanism. 
This paper does not estimate willingness to pay for the embedded option to 
pledge a home as collateral.\footnote{\cite{Kuminoff} show that capitalization 
	effects are not equal to willingness to pay.}

This paper exploits this plausibly exogenous law change 
to identify the impact of collateral service flows on house prices. 
The estimation requires detailed Texas house price data, 
which are notoriously hard to access %find %get %since %as 
because Texas is a nondisclosure state; 
recently available house price indexes from 
the FHFA (17,936 five-digit zip codes), however, make this analysis possible.

Difference-in-differences estimates show that the impact of the HEL legalization on 
Texas house prices was positive.
%(1) positive, (2) heterogeneous and (3) direct.%
The law change raised Texas house prices 4.13\%. %3.6\%.%2.5\%-3.7\%. 
%This is equal to almost 2 years of Texas house price growth. 
%over%3 %an extra 3.4 %2.5 %-3.5 %real (average real house price growth in Texas is 1.05\%).  
This result is robust across specifications and sample restrictions.
%The rise in house prices was gradual,%and 
There is no evidence of an effect before implementation.
%No evidence anticipation of the law affected prices before ennactment
In addition, synthetic control estimates find similar effects in 
Texas zip codes and no evidence of a placebo effect in untreated zip codes in border states.

The treatment effect was heterogeneous along several dimensions. 
Consistent with theory, the effect was smaller in %relatively 
elastic locations (where it is relatively easy to build); 
each unit of the \cite{Saiz2010} measure of supply  elasticity 
corresponds to a 0.9\%  %1\% %0.66-3.38\% %1.4\% 
lower rise in prices.\footnote{This paper uses supply elasticity as a source of heterogeneity,
	not for identification.}
Furthermore, zip codes with higher prelaw house prices, %population, 
higher income, and lower unemployment  rates
saw a greater rise in prices,\footnote{The values of these variables are averaged for each zip code before the law change.} indicating that households in richer locations valued this option more.
% This is evidence that households in richer locations  valued this option more. 
While it has been shown that an expansion in purchase mortgage 
debt capacity has  a greater impact on ex ante lower-priced 
properties \citep{Landvoigt}, 
this paper shows that an expansion in future HEL debt capacity has a greater impact 
on ex ante higher-priced properties. 
This can be explained by 
(1) the greater tax shield available to rich households,
(2) the tendency of rich households to be more financially literate and aware of these financial products \citep{Lusardi}, and 
(3) the tendency of rich households to be more likely to qualify for HELs 
as they tend to have better credit and more stable income \citep{DefuscoMondragon}.

The heterogeneity analysis refines the interpretation of the main result. 
The impact of HEL legalization on house prices is not just a measure of the extent to which households expect to be credit-constrained, 
but also a measure of the extent to which households 
%value the ability to use
expect to qualify for HELs to smooth consumption. 
For example, a recently  unemployed homeowner may want to borrow but not 
%be able to 
qualify for a HEL.
On the other hand, a homeowner 
with high stable income (hence a high marginal tax rate) 
and good credit is more likely to be approved for a bigger loan at a 
lower effective interest rate.\footnote{\cite{Stolper2015}
	found that richer homeowners are more likely to use HELs for their 
children's tuition.}

The mechanism through which HEL legalization increased house prices
could have occurred through two  broad types of channels: 
\begin{enumerate}
	\item 
	\underline{Direct Channel}: 
	the	law caused a rise in demand for 
	owner-occupied housing due to the new option 
	allowing homes to be pledged as collateral after purchase.
	%Extensive or intensive margin?
	\item 
	\underline{Indirect Channel}:	
	the law affected other variables that affect house prices. 
	For example, if the law 
	increased  investment enough to stimulate the local economy, 
	this increase could have raised demand for housing, consequently raising the price. 
	%	what percent of a house price is due to the collateral option value?	
\end{enumerate}
The mechanism is investigated in two ways. 
First, falsification tests show that the law change did not affect 
observed variables related to prices including rent, population, and income. 
There was a small (0.3\%) but statistically significant rise in unemployment, which
works against the indirect channel. %driving the effect/rise in house prices
Second, estimates in border cities (such as Texarkana) are 
similar to the bigger samples, providing further support %in favor of
for the direct channel. 
Border cities are often viewed as one economy. 
Hence, if the law change stimulated the economy,
the indirect channel would affect both the Texas and control sides of the border city,  
while the direct channel would only affect the Texas side.
Together, these results provide evidence that most 
of the effect was through the direct channel.
%\textcolor{red}{
%%As 
%Because
%it is impossible to know for certain 
%whether the law affected house prices through 
%other unobserved variables, these results provide indicative 
%rather than definitive proof of the mechanism through which 
%collateral service flows affect house prices.	}

Finally, %theory predicts that 
in partial equilibrium the law change should have increased 
demand for owner-occupied housing relative to renting (in Texas)
because owners can pledge their home as collateral but renters cannot.  
On the other hand, in general equilibrium, the rise in house prices from this 
law change should have caused a reduction in demand. 
The net effect of this law change on homeownership is theoretically ambiguous. 
Estimates show that Texas homeownership rates, single-family building permits, 
and population %growth
were unaffected by the law. % change. 
%This is evidence 
These results indicate that house prices rose enough to keep 
the marginal homebuyer indifferent between owning and renting.

Inference about external validity to the collateral value of other assets, in other locations, 
at other times should be made with caution.
This paper finds that after-purchase  collateral service flows had 
a positive impact on the price of owner-occupied housing in Texas after 1998.
The conceptual framework predicts that there should be a positive effect for other assets, 
in other locations, at other times. 
In particular, the effect is likely higher for housing in other states,
because Texas  is the only state 
with an 80\% limit on future home equity extraction.
Unfortunately, it is hard to quantify the collateral value of other assets because
it is rare to find large discontinuous changes in lending laws.

This paper makes three contributions: 
\begin{enumerate}
	\item 
%It is the first paper to 
It empirically  identifies if, and to what extent, %collateral value 
the option to pledge a home %an asset %-- any asset -- 
as collateral, after purchase,  affects its price.
	\item 
It provides %the first 
\textit{price-based} evidence that households are borrowing-constrained.
The treatment effect is a measure of the extent to which households 
value HELs to facilitate future consumption smoothing.
% the option to borrow in the future against their home.  
%
\item 
It provides new evidence on the distribution  of credit supply.
Richer households that get a greater mortgage interest tax shield, 
are more financially literate, and are more likely to qualify 
value HELs more strongly.
\end{enumerate}

%Table of Contents (optional)
%The rest of the paper proceeds as follows: 
%Section 1 presents background on the determinants of
%credit limits and describes our credit card data. 
%Section 2 discusses our regression discontinuity research
%design. 
%Section 3 verifies the validity of this research design. 
%Section 4 presents our estimates of the
%marginal propensity to borrow. 
%Section 5 provides a model of credit limits. 
%Section 6 presents our estimates
%of the marginal propensity to lend. 
%Section 7 concludes.

\subsection{Literature review}
\label{Literature}
Many strands of literature study borrowing constraints and collateral.
Theoretical literature has shown that
frictions such as adverse selection, moral hazard, 
and the inability to fully pledge future labor income
restrict borrowing.\footnote{ \cite{Barro1976,
		 HartMoore1994, 
		Lustig2005}. } 
Borrowers often pledge collateral to mitigate these frictions. 
%Creditors demand collateral
%collateral is used to alleviate financial frictions stemming from moral hazard and adverse selection effects 

A second strand of literature studies the relationship between 
housing, credit constraints, and consumption.% 
\footnote{ \cite{ AHY, Agarwal2016, Bhutta2015, 
		Carroll2011,
		Chen2013, DeFusco2016, 
		Hurst2004,
		 Leth2010, 
		Mian2011, Sodini}. }
%Theory predicts that 
Households that are, or fear they will be, 
credit-constrained %in the future
should have a stronger demand for 
assets %(such as real estate)  which 
that facilitate their future ability to borrow.
If prices reveal information, %\citep{Hayek1945},
the magnitude of the treatment effect estimated 
in this paper can be interpreted as a 
measure of the extent to which households 
value HELs as a tool to relax credit constraints.

%Another 
A third strand of literature studies %how 
the relationship between property prices, 
firm investment, and entrepreneurship.\footnote{\cite{Chaney,
		Kerr, % CSFij, Gan, 
		Schmalz, Wu}.}
%and firm investment through the collateral channel. 
%Real estate has been shown to be a crucial source of collateral for 
%firms. %collateral turns out to be crucial. 
There is evidence that rises and declines in property values, 
which affect debt capacity, 
amplify firm investment in the United States and Japan but not in China.
While theory predicts that firms  prefer to own assets 
that serve as better collateral, 
none of these papers empirically identifies 
if and to what extent asset prices reflect this.

A fourth strand of literature compares the  %collateralizability %pledgeability 
collateral service flows of different %financial 
assets.
The interest rate borrowers pay in the repo market depends on 
the type of collateral they use
\citep{Bartolini}.
For example, borrowers who %use 
pledge treasuries as collateral 
often borrow at ``special" repo rates.
\cite{Duffie} showed that ``specialness"  should raise the
price of the underlying security by the 
present value of interest rate savings. 
The estimates below can be interpreted through the lens
of \cite{Duffie}: the collateral value of housing should reflect 
the interest rate savings on a HEL relative to an unsecured loan.

In addition to the interest rate savings, 
the collateral value should also reflect the greater 
debt capacity of housing
via a lower margin requirement (higher loan-to-value [LTV] ratio) 
relative to unsecured credit and debt secured by other forms of collateral. 
In related work, \cite{Garleanu} and \cite{Jylha}
showed that a security's margin also affects its return.

The collateral value of other %non-financial
assets -- such as airplanes, cars, fine art, gold, and patents -- has also been 
studied.\footnote{\cite{Argyle, Benmelech2009, Benmelech2011,
		Huang, Mann, McAndrew, chen2019pledgeability}.} 
In particular, \cite{Huang} argued that 
gold is a better source of collateral than platinum for historical and institutional reasons. 
For example, gold is formally recognized as collateral by the Basel Accords 
and is accepted as collateral by broker dealers, while platinum is not.
They contended that, in times when the probability of a consumption disaster is high, 
agents prefer gold for its collateral benefits, which is reflected in 
the price.  
The results in this paper provide support for the ``Flight to Collateral" phenomenon
\citep{FostelGeanakoplos2008} 
as distinct from ``Flight to Liquidity", as homeowners value 
the collateral privileges of real estate despite its lack of liquidity.

A fifth strand of literature studies the impact of 
total mortgage leverage (including purchase mortgages) on 
house prices.\footnote{\cite{Adelino, 
		An, Anenberg, 
		arslan2020credit,  
		DiMaggio, 
		Favara, 
		FLVN,
		Greenwald2019,
		Orlando,
	Rebucci}.} %, Labonne
In particular, \cite{FLVN} showed that (total) mortgage leverage played a quantitatively important 
role in explaining the housing boom-bust cycle.

It is worth emphasizing that this paper focuses 
on nonpurchase mortgage leverage
as opposed to total or purchase leverage.
A purchase mortgage can be used to 
buy a house and to otherwise smooth consumption (since money is fungible).
The Texas law change studied here did not affect a household's ability to finance 
the original purchase, but rather its ability to pledge its home as collateral %,
in the future. 
%While 
There is evidence that HEL debt capacity increased 
purchase mortgage debt capacity 
for some households that used second liens (``piggy-back mortgages")
to avoid mortgage insurance and obtain bigger 
loans \citep{Lee}.\footnote{Conforming loans with LTV $>80\%$ require private mortgage insurance (PMI).
	For example, a borrower who wants a 90\% LTV loan 
	can avoid PMI by getting a first lien mortgage with 80\% LTV
	and a second lien mortgage with 10\% LTV.} 
This was not a relevant factor in Texas because the law included an 80\% LTV 
limit for all liens after purchase.\footnote{The 80\%  %LTV 
	limit is based on the appraised value of the property 
	at the time of future equity extraction.}

Note that there is different terminology in the literature for how 
the ability to pledge an asset as collateral affects its price. 
\cite{FostelGeanakoplos2008} call this the collateral value, 
\cite{Brumm} call this the collateral premium, 
and \cite{HeWrightZhu2015} call this the liquidity premium.

%%%%%%%%%%%%%%%%%%%%%%%%%%%%%%%%%%%%%%%%%%%%%%%%%%%%%%
%%%%%%%%%%%%%%%%%%%%%%%%%%%%%%%%%%%%%%%%%%%%%%%%%%%%%%
%\newpage
%%%%%%%%%%%%%%%%%%%%%%%%%%%%%%%%%%%%%%%%%%%%%%%%%%%%%%
\section{Institutional Setting}
\label{Institution}
Texas lending laws have been studied widely in the history, legal, and recently economic 
literature.\footnote{\cite{AEJ2012, Forrester, Kumar, McKnight, TLC}.}
Restrictions on mortgage lending existed in Texas  before it became 
a U.S. state. % in 1845. 
These laws, protecting homes from creditors, %were in
carried over to the state's original constitution in 1845
and to  subsequent versions (\hyperref[timeline]{Internet Appendix \ref*{timeline}}).  
%Article XVI, Section 50 of the 
Before 1998, the Texas constitution protected homes from foreclosure except 
for nonpayment of property taxes, 
the purchase mortgage, 
mechanics' liens for home improvement,\footnote{Home improvements require building permits 
	and raise property taxes based on the reassessed value of the home; hence borrowers were unlikely to use mechanics' liens 
	to extract equity for consumption.} and
refinance loans. 
Under Texas law, refinance loans were only allowed up to the balance, permitting homeowners to get a lower payment (if interest rates fell) 
but not to borrow an additional amount.\footnote{Hence cash-out refinance loans were illegal before 1998.}
%Texas laws severely restricted refinance loans 
%allowing homeowners to borrow up to their existing loan balance,
%%However, refinance mortgages were restricted up to the loan balance, 
%hence cash-out refinance loans were not allowed. 
Suppose a Texan borrowed \$200,000 to buy a home, and 
five years later %her
the remaining balance was \$150,000. 
If interest rates fell, 
the Texan could refinance into a new mortgage for at most \$150,000. 
This allowed Texas homeowners to lower their mortgage payments, 
but not to extract equity  (i.e., borrow an additional amount). 
This was a binding restriction.  \cite{Chen2013}
find that 61\% of all refinance mortgages in the United States are cash-out loans. 
The fraction of %refinance mortgages involving 
cash-out loans  was  over 80\% in years 
when the interest rate incentive was high.
Before 1998, the only way for a Texas homeowner to extract home equity was to sell the home
at substantial transaction costs.\footnote{The transaction cost 
	of selling a home is $\approx10\%$ (\href{https://www.zillow.com/blog/cost-of-transaction-fees-143806/}{www.zillow.com/blog/cost-of-transaction-fees-143806/}).}

The literature (\hyperref[timeline]{Internet Appendix \ref*{timeline}})
%(notably \cite{AEJ2012} and \cite{Forrester})
has linked the movement to amend Texas lending laws
%these law changes 
to two main factors: federal tax reform and a circuit court ruling.
%three major %exogenous 
%factors. 
%These factors include the 
First, 
the Tax Reform Act of 1986
%which 
made mortgage interest the only form of  interest on consumer credit that 
is tax deductible. 
This tax shield made HELs more attractive than 
other forms of debt such as auto loans, personal loans, and credit card debt. 
While HELs expanded throughout the United States after this tax reform, 
Texans did not have access to this additional tax shield.  
%
%For obvious reasons, home equity lending
%expanded throughout the US after the 1986 law, as Texans looked on in envy
%
%Second,
% the US Court of Appeals for the 
Second, the Fifth Circuit (with jurisdiction over Louisiana, Mississippi, and Texas)
%\footnote{which has jurisdiction over Louisiana, Mississippi and Texas} 
ruled in  1994
that federal regulations superseded the Texas constitution. 
This temporarily overturned 
Texas restrictions on HELs. 
Even though subsequent actions by %congress 
lawmakers quickly reestablished the restrictions, 
this ruling brought attention 
%and publicity 
to amending Texas mortgage laws. 
%Third, %the rise in 
%\textbf{\textcolor{red}{Details about Circuit law suit}}

Since these lending restrictions were in the Texas constitution, they could only be overturned 
by a constitutional amendment. 
This requires 
%Constitutional amendments in Texas require 
a joint resolution from both the state Senate and House. % (with two-thirds vote in each),
If the resolution is approved with a two-thirds vote in each,
it becomes a proposition, 
which must then be passed by a majority of the state's voters in a referendum.

Joint Resolution 31, %proposed 
to amend the state constitution and legalize home equity lending, 
was passed by the Texas House and Senate in May 1997.\footnote{At the time of the vote, Democrats had a majority in the Texas 
	%the Texas legislature was split between 1996-2000
	House, and Republicans had a majority in the Senate
	(\href{http://www.ncsl.org/documents/statevote/legiscontrol\_1990\_2000.pdf}
	{http://www.ncsl.org/documents/statevote/legiscontrol\_1990\_2000.pdf}).} 
This resulted in Proposition 8, which was approved by Texas voters 
%in a referendum 
in November 1997. 
The amendment became effective January 1, 1998, legalizing after-purchase mortgages 
with a combined balance of up to 80\% of the appraised market value of the home. 
Suppose that after 1998, a Texan's home was appraised to be worth \$200,000, and the original purchase
mortgage had an outstanding balance of \$100,000. The homeowner could borrow up to 
80\% $\times$ \$200,000 =\$160,000 and get a second lien HEL (or reverse mortgage) for up to \$60,000 or 
a cash-out refinance loan for up to \$160,000. 

%\textcolor{red}{Other TEXAS ballots, economy, Taxes etc?}

There is evidence in the literature that Texans took advantage
of these newly available products after Proposition 8 was passed. 
\cite{AEJ2012}  %,  Table 4, 
showed that Texans took out home equity loans,
and \cite{Kumar} %, Figures 7 and A6, 
showed that Texans  used cash-out refinance loans.
%,
%albeit less than in border states which didn't have the 80\% restriction.
%Note that although the 80\% LTV limit did not apply to purchase mortgages, these types of highly 
%leveraged loans were uncommon in Texas 
%where the average purchase LTV was $77.6\%$.
%\citep{Kumar}.
%\footnote{In a sample of 
%	non-prime mortgages, which are the riskiest and highest leveraged, 
%	in 1998 the initial LTV on purchase mortgages 
%	was just under 80\% (\hyperref[Kumar]{Kumar, 2018}).}. 

%%%%%%%%%%%%%%%%%%%%%%%%%%%%%%%%%%%%%%%%%%%%%%%%%%%%%%
%%%%%%%%%%%%%%%%%%%%%%%%%%%%%%%%%%%%%%%%%%%%%%%%%%%%%%
%\newpage
%%%%%%%%%%%%%%%%%%%%%%%%%%%%%%%%%%%%%%%%%%%%%%%%%%%%%%
\section{Conceptual Framework}
\label{Model}
This section constructs a model to help interpret the results below.
The model has three main goals: 
(1) to illustrate how %an asset's 
pledgeability affects the standard asset pricing equation, 
(2) to 
%help explain
study the heterogeneity of collateral service flows, 
(3) to clarify the different mechanisms through which collateral service flows affect house prices.  
%for houses at different price points.
Consider a household 
%who
that values 
nondurable consumption $c_{t}$
and durable housing $h_{t}$,
which depreciates at rate $\delta_{t}$. %\footnote{Following Jeske, Krueger, Mitman}
The household can borrow 
up to 
%a fraction 
$\kappa_{t} 
\equiv 
%= 
\min\left\{  \overline{\kappa}_{t} , LTV_{t}  \right\}$
%$\kappa_{t}$ 
of its 
house value,
%home equity,
where
$\overline{\kappa}_{t}$ is the legal limit
and 
$LTV_{t}$ is the most lenders will lend at time $t$.
%\footnote{The loan to value 
%	(LTV) ratio is
%	%. It is 
%	determined endogenously 
%	in credit markets (\hyperref[Geanakoplos2009]{Geanakoplos 2009}).}.
%$\kappa_{t} = \min\left\{  \overline{\kappa}_{t} , LTV_{t}  \right\}$
%of its house value.\footnote{ $\kappa_{t}$ is the highest loan to value (LTV)
%	ratio 
%	%lenders allow 
%	at time $t$.} 
%The variable  $\kappa_{t}$ is the highest loan to value (LTV)
%ratio lenders allow at time $t$. 
%The household's problem is below. 
%The household 
It solves %the  following problem.
\begin{align*}
\max \text{ } \E_{0} \sum_{t=0}^{\infty} \beta^{t} u(c_{t} , h_{t})
&  
\text{     } \text{     } \text{     } \text{     } \text{         s.t.}
%\\
\\
c_{t}+p_{t}h_{t+1} + a_{t+1}
&
\leq 
y_{t} + p_{t} h_{t} (1-\delta_{t}) + (1+r_{t}) a_{t}
\tag{DBC $\lambda_{t}$}
\\
-a_{t+1} 
&\leq 
\kappa_{t} p_{t} h_{t}
\tag{CC $\mu_{t}$}
%
%\\
%\kappa_{t}
%&=
%\min\left\{  \overline{\kappa}_{t} , LTV_{t}  \right\}
%%\min\left\{  \overline{\kappa}_{s,t} , LTV_{t}  \right\}
\end{align*}
where $u(c_{t} , h_{t})$ is the household's flow utility from consumption and housing, 
assumed 
%to be 
twice continuously differentiable and strictly concave in each argument.
$\lambda_{t}$ is the 
%Lagrange 
multiplier on the 
dynamic budget constraint (DBC)
and 
$\mu_{t}$ is the 
%Lagrange 
multiplier on the 
collateral constraint (CC).\footnote{The collateral constraint can be written in different ways.
See \hyperref[Comparison]{Appendix \ref*{Comparison}} for a comparison.}
$c_{t}$ is the num\'eraire, and  $p_{t}$ is the price per unit of housing.
%Housing depreciates at rate $\delta_{t}$. %\footnote{Following Jeske, Krueger, Mitman}
$a_{t}$ is the amount saved or borrowed at  interest rate $r_{t}$, and $y_{t}$ is income.

%Two frictions:%\\
%1. time-varying depreciation $\delta_{t}$ [Jeske, Krueger, Mitman, JME]
%2. time-varying leverage $\kappa_{t}$

% % % % % % % % % % % % % % %
The solution to the household's problem 
 (derived in \hyperref[deriv1]{Appendix \ref*{deriv1}})   
 reveals that house prices reflect service flows from shelter and collateral: 
 %housing delivers service flows from shelter and collateral: 
 %the    equilibrium price of housing is:
% % % % % % % % % % % % % % %
\begin{align} 
\label{eq:FOC}
\underbrace{p_{t}}_{\text{price}}
&=
\E_{t}
\left[ 
\underbrace{ 
	\beta   
	\frac{  u_{c}( t+1 ) }{ u_{c}( t )   }
}_{\text{ discount factor}} %_{sdf}
\times 
\left( 
\underbrace{ 
	\frac{  u_{h}(  t+1  ) }{ u_{c}( t+1 )   }
}_{ 
\substack{\text{housing} \\ \text{service flow (rent)}}
}
%_{\text{service flow--rent}}
+
\underbrace{ 
	\frac{ \mu_{t+1} \kappa_{t+1}  p_{t+1} }{ u_{c}( t+1 ) }    
	%\kappa_{t+1}  p_{t+1}
}_{
%\text{collateral value}
\substack{\text{collateral} \\ \text{service flow}}
}
+   
\underbrace{ (1-\delta_{t+1} ) p_{t+1} }_{\text{resale price}}
\right) 
\right] 
\tag{1a}
\\
\label{eq:CC}
%\underbrace{ 
\mu_{t+1} 
%}_{\text{price}}
&=
u_{c}( t+1  ) -  \E_{t+1}[ \beta \left(1+r_{t+2}\right) u_{c}(t+2) ]
\tag{1b}
%u_{c}( t  ) -  \E_{t}[ \beta \left(1+r_{t+1}\right) u_{c}(t+1) ]
%
%\\ & =
%PDV_{t}\left( \text{s} \right) + PDV_{t}\left( \text{CSF} \right)
\end{align} 
Equations (\ref{eq:FOC}) and (\ref{eq:CC}) come from the first-order conditions (FOCs). % of the household.
Following \cite{FLVN}, the periodic service flow from housing can be interpreted 
as a cash flow equal to rent. 
%For a collateral constrained household 
If a household is credit-constrained,
then
$\mu_{t+1}>0$,
%hence $CSF$
and its FOCs reflect the collateral service flow. 
Any model with a collateral constraint
has this type of collateral term (denoted $CSF_{t}$).\footnote{\cite{BianchiBozMendoza2012, Greenwald2016, HeWrightZhu2015, Iacoviello2005}.}
		%, Piazzesi2016 
Even though it is ubiquitous, it is hard to quantify,
%this term has not been estimated in the literature
as it is rare to find a setting with a large exogenous shock to the pledgeability of 
housing $\kappa_{t}$.
%separately identify the housing service flow 
%(or cash flow if the property is rented) from the collateral service flow.

In equilibrium, the only way the collateral service flow can be positive 
is if there is at least one unconstrained household that is lending with $\mu_{t+1}=0$.
%If there 
Hence a general equilibrium model requires 
%some kind of 
heterogeneity 
for $CSF_{t+1}>0$.
A convenient
%and tractable
 way to model this heterogeneity 
is with an impatient borrower and a patient lender
\citep{Iacoviello2005, KiyotakiMoore1997}.
%Many papers capture this with an impatient borrower and patient lender.
An alternative way to model positive collateral service flow is to 
allow interest rates to be exogenous 
and to assume the representative agent borrows 
from a deep-pocket, risk-neutral, international lender
\citep{BianchiBozMendoza2012}.

%Mention seminal work by SVN and write two summations 

The results above can help illustrate heterogeneity in the collateral service flow. 
First, index each household with superscript $j$. 
Second, rewrite the effective interest rate for household $j$ as 
$r_{t}^{j}=(1-\tau_{t}^{j}) i_{t}^{j}$, where $\tau_{t}^{j}$ is household $j$'s marginal income tax rate
and $i_{t}^{j}$ is the contract rate charged to household $j$. 
Finally, index household $j$'s credit constraint
$\kappa_{t}^{j} \equiv \min\left\{  \overline{\kappa}_{t} , LTV_{t}^{j}  \right\}$.
Note that %the effective interest rate 
(1) households with higher marginal income tax rates $\tau_{t}^{j}$ enjoy a bigger 
mortgage interest deduction\footnote{A progressive income tax implies a regressive mortgage interest deduction.} 
and (2) different households are charged different interest rates and credit limits depending on their risk 
(measured by their credit score and stability of income).
The collateral service flow for household $j$ can be decomposed 
%into %three %parts
from Equations (\ref{eq:FOC}) and (\ref{eq:CC}): 
%this equation: %has three components
% % % % % % % % % % % % % % %
\begin{align} \label{eq:CSFdecomp}
\underbrace{
	CSF_{t}^{j}
}_{
%\text{collateral value}
\substack{\text{collateral} \\ \text{service flow}}
}
&=
\frac{ 1 }{ u_{c}( t ) }
\times 
\underbrace{ 
	%\mu( t+1 )
	\mu_{t}^{j}
}_{\text{desire to borrow}}
\times 
\underbrace{     
	\kappa_{t}^{j}
}_{\text{fraction you can borrow}}  
\times 
\underbrace{     
	p_{t}
}_{\text{value of collateral}}  
\tag{2a} 	
%}_{\text{collateral value}}
\\
\label{eq:CSFdecomp2}
&=
\frac{ 1 }{ u_{c}( t ) }
\times 
%\underbrace{ 
\left(	u_{c}( t  ) -  \E_{t}[ \beta \left(1+
(1-\tau_{t}^{j}) i_{t}^{j}
\right) u_{c}(t+1) ] \right) 
%}_{\text{desire to borrow}}
\times 
\kappa_{t}^{j}
%\min\left\{  \overline{\kappa}_{t} , LTV_{t}^{j}  \right\}
\times 
	p_{t}
\tag{2b} 	
\end{align} %CAVEAT HOME EQUITY
% % % % % % % % % % % % % % % %
%% \kappa_{t}^{hel} = Legal_{t}^{hel} \times LTV_{t}^{hel} %
% % % % % % % % % % % % % % % %
This decomposition shows that the value of being able to pledge housing 
as collateral depends on the desire to borrow $\mu_{t}^{j}$ and the 
debt capacity $ \kappa_{t}^{j} p_{t} $ (the amount a homeowner can borrow),
which depends on $j's$ access to credit.

%\\
%If you have no desire for a 
%The component $\mu$ depends on...
%\mu = f( market incompleteness, desire to use HELs)
The component $\mu_{t}^{j}$ (the multiplier on the collateral constraint) 
%depends on 
is a measure of the desire to borrow.
If there is no 
%desire 
demand
for HELs (or if markets are complete), then
%,
%$\left( \mu_{t+1}=0 \right)$  
$\mu_{t}^{j}=0 $, 
%$\mu_{t+1}=0$ 
making $CSF_{t}^{j}=0$.
%$CSF$ 
The collateral service flow can still affect property prices 
today if there is a desire for HELs in the future.
Observe from equation \ref{eq:CSFdecomp2} that the desire to borrow depends negatively on the effective interest rate that household $j$ faces.
Households that face lower effective interest rates 
(because they have better credit scores or more stable documented income) should have a greater demand for HELs. 
The estimates in Section \ref{ResultsHTE} are consistent with this prediction.

The component 
$\kappa_{t}^{j} \equiv \min\left\{  \overline{\kappa}_{t} , LTV_{t}^{j}  \right\}$
depends on 
(a) the amount households can legally borrow
$(\overline{\kappa}_{t})$
%whether HELs
%%, HELOCs 
%and reverse mortgages are legal
and 
(b) the amount lenders are willing to lend $(LTV_{t}^{j})$.
%how much lenders are willing to lend you. 
%Hence $\kappa_{t+1}$ can be a function of 
%%your 
%the borrower's current LTV, debt to income etc. 
%
If HELs are illegal 
$\left( \overline{\kappa}_{t}=0 \right)$, 
%making 
equation \ref{eq:CSFdecomp}
implies
$CSF_{t}=0$,
as 
%was the case 
in Texas before 
1998.\footnote{The Texas constitution set $\overline{\kappa}_{t}=0$ for after-purchase mortgages
%	for all mortgages 
%	except purchase mortgages 
	prior to 1998, and $\overline{\kappa}_{t}=0.80$ after.} 
\cite{AEJ2012} and \cite{Kumar}
showed that HELs and cash-out refinance loans were indeed utilized by Texans after 1998. 
It is worth noting that the amount lenders are willing to lend, $LTV_{t}^{j}$, depends on the 
borrower's characteristics: borrowers with better credit scores and stable documented income 
are more likely to qualify for bigger loans \citep{Brueggeman}.
These underwriting requirements help explain the heterogeneity results in Section \ref{ResultsHTE}.

The model above does not distinguish between 
the ability to pledge 
%the
an asset as collateral at the time of purchase 
and in the future. 
\hyperref[Model_simple]{Appendix \ref*{Model_simple}}
considers a three-period setting 
where purchase mortgage collateral service flows are disentangled 
from HEL collateral service flows.
In this setting, the borrower's HEL debt capacity is equal to 
$
\kappa_{t} p_{t} h_{t}   - B_{t}^{pm} 
%\times  \left( p_{t} h_{t}   - B_{t} \right) 
$, 
where $B_{t}^{pm}$ 
is the remaining balance on the purchase  mortgage
used to buy the home. 
For example,
consider a household that owns a home appraised to be worth 
$p_{t} h_{t}= \$ 100k$ 
with a remaining balance on its purchase mortgage  of $B_{t}^{pm}= \$ 50k$.
If this household can borrow up to 
%such that all liens are less than or equal to
$\kappa_{t}=80\%$ of the price, 
then its HEL debt capacity 
%is 
would be
$80\% \times  \$ 100k- \$ 50k= \$30k$.

\section{Empirical Strategy, Identification, and Data}
\label{Empirical}

\subsection{Empirical strategy}
\label{EmpiricalStrategy}
This paper estimates the impact 
of the HEL legalization on various outcome variables 
with a generalized difference-in-differences (DID) methodology. 
The main analysis 
%compares estimates in
uses
four geographically nested samples (United States, Border-State, Border, and CBCP), explained
below. 
Four border cities (notably Texarkana) are also studied.
Further analysis constructs a synthetic control group for each treated zip code
and 
conducts placebo tests for untreated zip codes. 
In all samples, the treated group consists of all locations in Texas 
and the control group includes all locations outside Texas.
The U.S. sample includes all zip codes 
with data six years before and after the treatment year, 1998.
The Border State sample %includes 
restricts the U.S. sample to all zip codes in Texas and its border states 
(Louisiana, Arkansas, Oklahoma, and New Mexico). 
The Border sample %is constructed from 
restricts the Border State sample to
all zip codes within 
a 50-mile radius of each Texas zip code, using distance data from the NBER and Census.\footnote{\href{http://www.nber.org/data/zip-code-distance-database.html}{http://www.nber.org/data/zip-code-distance-database.html}.} Only zip codes that 
include both Texas and non-Texas locations within a 50-mile radius are kept. 
The fourth, and most local, sample  
is the Contiguous Border County Pair (CBCP) sample.\footnote{\cite{Dube, heider2015certain, severino2017personal}.}
This sample consists of all zip codes that belong to a 
contiguous county on either side of the Texas border. 
Since a given Texas county may border more than one non-Texas county, 
data in this sample are stacked as in \cite{Dube}, 
making it possible to identify county pair by year 
fixed effects.\footnote{This stacking procedure creates more observations in the CBCP sample	
	than in the Border sample, despite the fact that the CBCP sample is nested 
	in the Border sample.}

The local samples can help control for unobserved heterogeneity 
to the extent that houses near each other are more likely to be affected by the same local 
shocks such as hurricanes and factory shutdowns.
%zip codes near each other are more likely to be similar. 
%For example, 
In particular, the border city Texarkana, the only Metropolitan Statistical Area (MSA) in Texas 
that includes another state,\footnote{\href{https://www.census.gov/population/cen2000/phc-t29/tab03a.pdf}{https://www.census.gov/population/cen2000/phc-t29/tab03a.pdf}.}
%which is at the intersection of Texas and Arkansas, 
can be viewed as one economy. 
So if the law change 
%increased consumption and investment 
affected the economy
on the Texas side of Texarkana, 
it should have also affected the economy on the Arkansas side.
%\textbf{\textcolor{red}{If the treatment effect is positive in the bigger samples, but 
%%much smaller or 
%zero in Texarkana, 
%that would provide evidence in favor of the  indirect mechanism. 
%If the treatment effect remains positive in Texarkana, that would suggest 
%the indirect mechanism played a smaller role. }
%}

This paper estimates: 
\begin{align*}
 y_{i,s,t} 
&
=
%\beta_{0}
%+
%\gamma_{s}
\alpha_{i}
+
\theta_{t}
%
%+ \nu_{pt}
%
%+
%\beta_{T} Texas_{s} 
%+
%\beta_{P} Post_{t}
+
\textcolor{ChadRed}{\beta_{DID}}  Texas_{s} \times Post_{t}
%\textcolor{ChadRed}{\beta_{3}} Texas_{s} \times post_{t}
%+
%\alpha_{i}
%+                
%\zeta_{s}
%+
%\delta_{t}
+
\Gamma X_{i,s,t}
+
\epsilon_{i,s,t}
\tag{static DID} %\tag{CC $\mu_{t}$}
%\end{equation}
\\
%\begin{equation}
 y_{i,s,t} 
&
=
%\beta_{0}
%+
%\beta_{1} post_{t}
%+
%\beta_{2} Texas_{s}
%\beta_{0}
%+
%\beta_{1} Texas_{s} 
%+
\alpha_{i}
%\gamma_{s}
+
\theta_{t}
%\delta_{t}
+
\sum_{k\neq 1997}
\textcolor{ChadRed}{\eta_{k}}  Texas_{s} \times  1_{k}
%\textcolor{ChadRed}{\eta_{k}}  Texas_{s} \times 1_{k}
%\textcolor{blue}{\eta_{t}} 
%+
%\alpha_{i}
%+
%\zeta_{s}
%+
%\delta_{t}
+
\Gamma X_{i,s,t}
+
\epsilon_{i,s,t},
\tag{dynamic DID}
%\tag{dynamic DID} %\tag{CC $\mu_{t}$}
%
%
%\\
\end{align*}
where $y_{i,s,t}$ is the outcome variable. 
In the main regressions, it is the log real house price index. 
In further regressions, real house price growth, log real rent, log population, log real income per capita, 
%the 
unemployment rate, homeownership rate, and 
log single-family building permits 
are also considered. 
The index $i$ corresponds to the most local level  of the outcome variable. 
%and $\alpha_{i}$ is a location fixed effect. 
For house price regressions, $i$ is the  zip code. 
For population,  income, employment, and permit regressions, 
$i$ is the county 
%FIPS
 Federal Information Processing Standard (FIPS)  
code. 
For rent and homeownership rate regressions, $i$ is the MSA.
%Metropolitan Statistical Area (MSA).
The index $s$ corresponds to the level of treatment, which is the state 
in all regressions. 
$Texas_{s}$ is the treatment group indicator variable, 
which is equal to $1$ if $s = Texas$. 
%t is the time index, year 
$Post_{t}$ is the treatment period indicator variable, 
which is equal to $1$ for $t\geq 1998$.
$1_k$ is a year indicator variable, 
which is equal to $1$ for $t= k$.
$\alpha_{i}$ and $\theta_{t}$ are the location and time fixed effects.
$\epsilon_{i,s,t}$ is an error term  
%unobserved determinants of $y_{i,s,t}$
assumed to be conditionally uncorrelated with the treatment.

Specifications in the main analysis include 
location fixed effects, time fixed effects, time trends, and
national oil prices interacted with MSA dummies. 
The CBCP sample also includes county pair by year fixed effects.
%%%%%%%%%%%%%%%%%%%%%%%%%%%%%%%%\textcolor{red}{AZ REVISION 3}  
Robustness tests also consider 
national interest rates interacted with state dummies as well 
as real income per capita and population.

This paper also investigates heterogeneity in the treatment 
%effect,
effect -- that is, 
whether the law change had a different impact in different locations. 
To study 
the
%heterogeneity of the treatment effect
 sensitivity 
 of the 
effect
%price impact 
to  various observable measures of heterogeneity $H_{i}$, 
%(such as supply elasticity), 
this paper estimates:
\begin{align*}
y_{i,s,t} 
&
=
%\beta_{0}
%+
%\beta_{1} Texas_{s} 
%+
%\beta_{2} Post_{t}
%+
%\beta_{3} H_{i}
%\\
%&
\alpha_{i}
+
\theta_{t}
+
\textcolor{ChadRed}{\beta_{H,0}}   Texas_{s} \times Post_{t}
%\beta_{4}  Post_{t} Texas_{s}
+
\beta_{5} Texas_{s} \times  H_{i}
+
\beta_{6} Post_{t} \times  H_{i}
\\
&
+
%\textcolor{ChadRed}{\beta_{H,DDD}} Post_{t} Texas_{s} H_{i}
\textcolor{ChadRed}{\beta_{H}} Texas_{s} \times Post_{t} \times H_{i}
%
%\textcolor{ChadRed}{\beta_{3}} Texas_{s}  post_{t}
%\textcolor{ChadRed}{\beta_{3}} Texas_{s} \times post_{t}
%+
%\alpha_{i}
%+
%\zeta_{s}
%+
%\delta_{t}
%+
+
\Gamma X_{i,s,t}
+
\epsilon_{i,s,t}
\tag{DDD}
%\tag{static DDD}
%\tag{static DDD} %\tag{CC $\mu_{t}$}
\end{align*} 
%The ATE from this model is 
In this specification, the average treatment effect (ATE)
%This specification implies the average treatment effect (ATE)
%is no longer constant, but
 is an affine function of $H_{i}$:  
\begin{align*}
\text{ATE}\left(H_{i}\right) 
&= 
\textcolor{ChadRed}{\beta_{H,0}} + \textcolor{ChadRed}{\beta_{H}}  H_{i}  
\end{align*}
The coefficient 
$\textcolor{ChadRed}{\beta_{H,0}}$ 
is the estimated average treatment effect if $H_{i} =0$,
and  
$\textcolor{ChadRed}{\beta_{H}} 
= \frac{\partial \text{ATE}\left(H_{i}\right)  }{ \partial H_{i}}  $ is 
the sensitivity of the average treatment effect 
to a rise in $H_{i}$.
%Specifications with quadratic ATEs are also considered. 

For example, 
theory predicts that a rise in demand should have 
a smaller impact on house prices in elastically supplied locations 
where it is easier to build real estate (\hyperref[ElasticSupply]{Figure \ref*{ElasticSupply}}). 
This corresponds to the hypothesis 
$\textcolor{ChadRed}{\beta_{elasticity}} <0$.
The coefficient $\textcolor{ChadRed}{\beta_{elasticity,0}}$
is the estimated impact of 
%collateral value 
the law change
on prices in a hypothetical location where 
the asset (housing) is in perfectly inelastic supply. 
%$\textcolor{ChadRed}{\beta_{H,DDD}} <0$.

This paper investigates treatment effect heterogeneity 
in elasticity, 
\textit{prelaw} median house prices, %population, 
income, and unemployment.
Prelaw variables are set equal to their average value 
%in 1992 (the first year in the sample) 
before 1998 to ensure 
they are unaffected by the treatment.

\subsection{Identification}
\label{Identification}

The main goal of this paper is to identify 
the impact of the HEL legalization on house prices.
The  identifying assumption is 
that the error term is conditionally uncorrelated with 
the treatment. 
The identifying assumption can be defended because 
%these three factors 
the %law change
constitutional amendment, approved by the state House, 
Senate, and ultimately by Texas voters in a referendum, 
%has been attributed to a 
was motivated by
federal tax reform and a circuit court ruling (Section \ref{Institution}). 
%(\cite{AEJ2012} and \cite{Forrester}, 
These factors are not   %likely to affect %clearly 
linked to   %factors that are unrelated to 
Texas house prices and other outcome variables studied 
in this paper. 
%
%%from the Office of Thrift Supervision 
%, thereby %overturning the state’s restrictions on home equity laws
% above 
 \cite{Forrester} reviews the extant literature detailing the adoption 
of Proposition 8, and none of the legislative histories 
suggest that the law was passed with any intent to stimulate the economy.

Several further steps are taken to defend the identification of the main effect: 
the impact of the law change on house prices. 
\begin{enumerate}
	\item 
	Various levels of fixed effects control for different types of unobserved heterogeneity.  
	%Omitted variables
	\\
	Zip code fixed effects remove time-invariant zip code-specific differences.
	\\
	Time fixed effects remove location-invariant time-specific differences.
	\\
	County pair by year fixed effects (in the CBCP sample) remove border county pair by year differences. 
	\\
	Estimates remain positive and statistically significant across specifications. 
%	\\
%	Note: none of these fixed effects solve the problem if there was another even in Texas at the same 
%	time as the law change that affected HP!!!
	\item Four geographically nested samples and four border city samples
	control for local unobserved heterogeneity such as hurricanes and factory shutdowns.
	%variables. % that may have affected house prices.%
	\\
	Estimates remain positive and statistically significant across sample restrictions.
	\item 
	Various covariates at national and local levels are included to 
	control for observed time-varying sources of heterogeneity. 
	%Additional estimates investigate if the effect is robust to the inclusion of
	%national and local control variables.
	%	A battery of regressions 
	%	Robustness tests investigate if 
	%	A battery of  
	%	which include national and local control variables are conducted.}	
	\\	
	The estimates remain positive and statistically significant.
	\item
	This paper re-estimates the main effect with the ZHVI, 
	which is constructed differently from the FHFA index,\footnote{\url{https://www.zillow.com/research/zhvi-methodology/}.\\The main analysis does not use ZHVI data because it begins too late in April 1996.}  
	to test if the results are robust to the method used for measuring house prices. 
	\\	
	The estimates remain positive, statistically significant, and similar to the main estimates 
	using the FHFA data. 
	\item 
	%pretrends 
	Dynamic estimates are analyzed to rule out the risk of upward sloping pretrends. 
	\\
	Pretrends are parallel across samples conditional on fixed effects and controls (national oil prices interacted with MSA dummies). 
	For two of the samples, Border State and CBCP, 
	the oil price interactions in the baseline specification are required for parallel pretrends.
	\item 
	%Third, 
	Synthetic control estimates are analyzed to further investigate the pretrends,
	and 
	placebo tests are conducted in untreated zip codes.
%	 are investigated to determine if there was a placebo effect.
%	. 
%	are conducted to %further 
%	test for robustness. 
	\\
	The synthetic control estimates are	similar to the main results and there is 
	no placebo effect in untreated zip codes.
	\item 
	%
	%Fifth, 
	A heterogeneity analysis investigates whether zip codes located in relatively inelastic MSAs
	(where it is harder to build) experienced a bigger treatment effect. 
	This is a sanity check to see if the results are
	%which is %standard
	consistent with %other estimates in 
	standard theory
	(\hyperref[ElasticSupply]{Figure \ref*{ElasticSupply}}).\footnote{While some authors criticize using this measure as an instrument for house prices \citep{Davidoff}, 
		the estimates here do not use supply elasticity for identification 
		but as a source of heterogeneity.} 
	\\ Zip codes in relatively inelastic MSAs had a bigger treatment effect. 	
%	and empirical estimates	in the literature \citep{Mian2011}.
	\item Falsification tests 
	%are conducted on
	investigate if other outcome variables 
	(log real rents, real income per capita, and unemployment rates) were affected.
	If the law change was correlated 
	with policies designed to stimulate the economy,
	%to an expected rise in house prices, 
	these other variables 
	would be positively affected as well. 
	\\ There was a small $(0.3\%)$ but statistically significant rise in the unemployment rate.
	The other variables were not affected.
	Moreover, \cite{Kumar2019} found that the law change had no statistically 
	significant effect on GDP growth in Texas.
	\item 
	Population  %migration 
	is investigated to see if %there was 
	households moved to Texas as a result of this law change.
	\\ Texas population was not affected.
	Furthermore, \cite{Kumar}  finds no evidence of migration using  
	IRS tax return data.
\end{enumerate}

The identifying assumption for a causal effect interpretation is that the 
law change was exogenous conditional on fixed effects and controls,
$\E[ \varepsilon_{i,s,t} | \text{Texas}_{s}\times\text{Post}_{t}, \alpha_i,\theta_t, X_{i,s,t} ] = 0$.
While the fixed effects and border samples help control for several levels of local unobserved heterogeneity, 
they alone cannot eliminate state specific %within-state % 
time-varying differences. 
Within-state time-varying fixed effects, such as county by year fixed effects, 
cannot be included because they would 
wash out the treatment variable $\text{Texas}_{s}\times\text{Post}_{t}$.
For example, suppose there was a tax cut passed in Texas 
but not in other states around the same time as the HEL legalization. 
This tax cut could affect Texas house prices and would not be
absorbed by the fixed effects. 
In this case, it is not possible to separately identify 
the impact of the HEL legalization from the tax cut. %;
If this type of Texas-specific shock existed at the same time as the treatment, we could not interpret the coefficients 
$\textcolor{ChadRed}{\beta_{DID}}$  and $\textcolor{ChadRed}{\eta_{k}}$
as the causal effect of the HEL legalization on house prices.

%For example, suppose there was a homebuyer tax credit passed in Texas 
%but not in other states around the same time as the HEL legalization. 
%This tax credit would likely affect Texas house prices and would not be
%absorbed by the fixed effects. 
%In this case it is not possible to separately identify 
%the impact of the HEL legalization from the tax credit. %;
%If this type of Texas-specific shock existed at the same time as the treatment, we could not interpret the coefficients 
%$\textcolor{ChadRed}{\beta_{DID}}$  and $\textcolor{ChadRed}{\eta_{k}}$
%as the causal effect of the HEL legalization on house prices.

This paper argues that this type of contemporaneous 
Texas-specific shock is not likely 
based on analysis of the law change and falsification tests. 
The paper explores the institutional setting in the legal, history, and economic literature 
in Section \ref{Institution} and \hyperref[timeline]{Internet Appendix \ref*{timeline}}. 
Federal tax reform and a circuit court ruling motivated the HEL legalization. 
None of the legislative histories suggest that the law was passed with any intent to stimulate the economy.
Moreover, if such a contemporaneous Texas-specific shock existed, 
it would have likely affected variables related to house prices. 
The falsification tests discussed above, 
both in this paper and in other papers (\cite{Kumar, Kumar2019}), 
find that these variables were not affected in a way that would stimulate house prices. 

A secondary goal is to investigate the mechanism behind the main effect. 
The mechanism is investigated in two ways. 
First, falsification tests are conducted on observed local economic outcome variables. 
If the law change had a positive effect on these variables, that would provide evidence 
that the indirect mechanism played an important role.
Estimates presented in \hyperref[LHS_ECON]{Table \ref*{LHS_ECON}}
show that the HEL legalization did not lead to  economically or statistically significant
changes in real rents, population, or real income per capita.
There was a small (0.3\%) but statistically significant rise in unemployment, which
works against the indirect channel. %driving the effect/rise in house prices
Second, the indirect effect should at least partially cancel out in the border city samples,  
to the extent that they can be viewed as one economy. 
Estimates in border cities remain positive, providing additional evidence 
that the effect was mostly direct.
%
%However, since it is impossible to know for certain whether other unobserved variables were affected, the 
%evidence on the mechanism is not definitive. 

%
%\textcolor{red}{\textbf{ 
%$\eta_{k} = \frac{ PDV(CSF) }{ p _{t} } $  
Under additional (stronger) assumptions,
the regression coefficients can be linked %tied 
to the model %above 
(\hyperref[deriv1]{Appendix \ref*{deriv1}}).
If the law had no
indirect impact on prices,
housing supply was perfectly inelastic and
households were fully aware of and understood the new home equity extraction products,
and 
if the functional forms for the representative Texas homebuyer 
are as specified in the model, 
then 
the first coefficient in the dynamic difference-in-differences regression
%coefficients 
can be tied to the model:
\begin{align*}
%\eta_{1998}%AZ: use hat? 
\textcolor{ChadRed}{\eta_{1998}}
%&=
%\left( \frac{ \Delta p_{98} }{ p_{97}  } \right)_{\text{Texas}}
%- 
%\left( \frac{ \Delta p_{98} }{ p_{97}  } \right)_{\text{control}}
%%\frac{ \Delta p_{98}^{\text{control}} }{ p_{97}^{\text{control}} }
%\\
&=
\left( \frac{  PDV_{98}(CSF)  }{ p_{97} } \right)_{\text{Texas}}
%\frac{  PDV_{98}^{\text{TX}}(CSF)  }{ p_{97}^{\text{TX}} }
%\\
%&
=
\frac{ 
	\E_{98}
	\left[
	\sum_{j=1}^{\infty} 
	(1-\delta )^{j-1} 
	M_{99,98+j} \times 
	CSF_{98+j} 
	\right]
}{ p_{97} }
%
%\textcolor{ChadRed}{\beta_{3}}
%
\end{align*}
Under these assumptions, the regression coefficient
$\textcolor{ChadRed}{\eta_{1998}}$ 
is equal to the present value of HEL collateral service flows 
(which themselves depend on future house prices) 
divided by the prelaw price.

The model teaches us that house prices are the present value of 
housing service flows (rent) and collateral service flows. 
The HEL legalization had an indirect effect if it affected housing service flows. 
Housing service flows can be directly measured by rents, or indirectly measured
by variables that determine housing demand and housing supply. 
%Estimates in \hyperref[LHS_ECON]{Table \ref*{LHS_ECON}} 
The falsification tests discussed above 
show that the law change did not affect rents directly, 
nor did it affect economic outcome variables that affect 
housing demand.\footnote{Except for a small rise in the unemployment rate, which if anything would reduce housing demand and thus the indirect effect.} 
Further estimates show that the law change did not affect housing supply, as single-family 
building permits were not affected.  

In summary, there are several nested levels of interpretation 
of the treatment effect coefficients
$\textcolor{ChadRed}{\beta_{DID}}$ and $\textcolor{ChadRed}{\eta_{k}}$.
If we assume that the law change was exogenous conditional on fixed effects and controls, 
$\textcolor{ChadRed}{\beta_{DID}}$ is the causal effect of the HEL legalization on Texas house prices. 
If we also assume that the law change did not have a significant effect 
on other variables that affect house prices,
$\textcolor{ChadRed}{\beta_{DID}}$ is the causal effect of the HEL legalization on Texas house prices
through the direct, collateral service flow, channel. 
If we also assume the structure of the model described above 
and in \hyperref[deriv1]{Appendix \ref*{deriv1}},
$\textcolor{ChadRed}{\eta_{1998}}$ 
is equal to the present value of HEL collateral service flows 
(which themselves depend on future house prices) 
divided by the prelaw price.

\subsection{Data}
\label{Data}
The data used in this paper are summarized in \hyperref[data_table]{Table \ref*{data_table}}.
The main outcome variable used in this study is the log real house price index. 
It is notoriously hard to access detailed Texas house price data
because Texas is a nondisclosure state. 
Household-level data sets such as CoreLogic (and DataQuick) do not have good 
Texas data for the relevant time periods.  
The Zillow Home Value Index (ZHVI) is not used in the main analysis because 
the data begin too late in April 1996. 
Zillow data are used to test whether the main results are robust to the method used for 
constructing the house price index. % and
Zillow data are also used in the heterogeneity analysis to investigate whether 
zip codes with higher prelaw median house price levels were affected differently by the law change.  

This paper measures 
house prices using the 
recently available 
Federal Housing Finance Agency
\href{http://www.fhfa.gov/DataTools/Downloads/pages/house-price-index-datasets.aspx}
{(FHFA)} 
five-digit zip code data, 
which contain 17,936 zip codes in the United States, %and DC
 including 
%and includes
%1,640 
918 zip codes in Texas.
% and border states.  
%The trade-off is that more local data is lower frequency. 
The trade-off %of
for
this lower level of geographic aggregation (zip code) 
is a higher level of time aggregation (annual).
Like the 
S\&P/CoreLogic/Case-Shiller home price indices,
%S\&P/Case–Shiller price indices, 
the FHFA series 
%attempts to correct 
corrects
for the changing quality of houses being
sold at any point in time by estimating price changes with repeat sales.
The data set only includes zip codes and years with enough repeat sales 
to construct the index.\footnote{For details about 
	%the construction of 
	this new data set, see \cite{Bogin}.}
%
%\textbf{\textcolor{red}{
Zip codes in sparsely populated locations tend to be larger than zip codes in densely populated areas. 
%\footnote{https://www.fhfa.gov/DataTools/Tools/Pages/HPI-ZIP5-Map.aspx}
While parts of the Texas border are sparsely populated, 
the %restricted FHFA 
Border sample has 110 zip codes %(1,430 zip code year observations)
(within 50 miles of the Texas border)
and the CBCP sample has 73 zip codes (in contiguous border counties). 
%including 67 zip codes in Texas and 43 in border states.
%The FHFA data has almost half of the zip codes in Texas
%

Data used for controls, heterogeneity analysis, and falsification tests %other outcome variables
come from several different sources.
Supply elasticity data are available at the MSA level 
from \cite{Saiz2010}.
Employment data at the county level are  from the BLS. %county 
Median house price data at the zip code level and rent data at the MSA 
level are from Zillow. 
Income data at the county level are from the BEA. %county
Population data at the county level, 
single-family building permit data at the county level, 
and homeownership rate data at the MSA level 
are from the Census.
Homeowner survey data at the household level are from the American Housing Survey (AHS).
%The population and permit data are at the county level, %while
%whereas the homeownership rate data %isare at the MSA level. 
One must be careful in merging the data sets since 
the same zip code can be located in more than one county. 
Each zip code is assigned to the county with the maximum 
allocation factor 
(e.g., if $75\%$ of zip code $z$ is in county $A$ and $25\%$ is in county $B$, 
then zip code $z$ is assigned to county $A$).
U.S. oil price data are from the EIA. %US annual, monthly, quarterly etc   %PPI from BLS
U.S. interest rates are constructed as in 
\cite{HMS}
by correcting the 
%ten 
10-year Treasury bond rate for inflation 
with the Livingston Survey of inflation expectations.
%%using data from the Livingston survey and Treasury. %US annual
%
%Real interest rates are constructed following the strategy outlined in HMS
%(2005). That is, we start with the ten- year Treasury bond rate and then correct
%for inflation with the Livingston Survey of inflation expectations. A long
%rate is used to approximate the duration of most mortgages. The Treasury
%rate rather than the actual mortgage rate is employed to reduce the feedback
%between events in the housing market and market rates. However, we have
%used alternative interest rates measures and found quite similar results.
Nominal variables
%house prices and income
 are deflated using the CPI for all urban consumers 
from the BLS as 
in \cite{GGG}.

%The control  data is taken from the Census and BEA. 
%The control variables are
%the log of per capita real disposable income, 
%%the state sales tax rate, 
%%the log of real house prices, 
%mortgage interest rates, and real oil prices following Abdallah \& Lastrapes 2012.

  %ddddddddd
%\text{ }   
%\newpage
\section{Estimates}
\label{Estimates}
%%%%%%%%%%%%%%%%%%%%%%%%%%%%%%%%%%%%%%%%%%%%%%%%%%%%%%
%%%%%%%%%%%%%%%%%%%%%%%%%%%%%%%%%%%%%%%%%%%%%%%%%%%%%%
%\newpage
%%%%%%%%%%%%%%%%%%%%%%%%%%%%%%%%%%%%%%%%%%%%%%%%%%%%%%
This section presents and discusses the empirical analysis. %estimates. 
First, Section \ref{Main_summ} %analyzes
examines summary statistics and raw house price growth
in the treatment and control groups.  
Next, Section \ref{Main} studies the impact of the HEL legalization 
on Texas house prices, the main question of the paper. 
Section \ref{ResultsHTE} investigates heterogeneity of the treatment effect. 
Section \ref{Main_channels} conducts falsification tests and investigates
the mechanism by analyzing the law's impact on other outcome 
variables. 
Section \ref{Main_MarginalBuyer} studies whether the HEL legalization affected 
the marginal homebuyer, examining the impact on 
homeownership, population, and single family building permits.  
Finally, Section \ref{External} discusses external validity and 
compares the main results to the quantitative macroeconomic literature. 

\subsection{Summary statistics and raw data}%Control Group: 
\label{Main_summ}
\hyperref[Summary]{Table \ref*{Summary}} presents 
summary statistics, comparing variables in Texas and border states
before and after the law change.
%
%Diff se w/ robust se as described in the paper
%
%the treatment group (Texas) 
%the %main 
%control group %(the US excluding Texas) using the main sample restriction.
%in the four geographically nested samples.
%
%Texas variables tend to be more similar to the control group in more local samples (border-state)
%than in the US sample.
%In the sample period, 
%
Before the law change, 
the average LTV for primary purchase mortgages was 83.17\% in Texas
and 79.29\% in border states. 
After the law change, Texas LTV fell 5.6\% to 77.57\%
while 
LTV in the border states rose slightly from 2.23\% to 81.52\%. 
The drop in LTV makes intuitive sense. 
Before 1998, Texas homeowners had one shot to pledge their home as 
collateral,\footnote{The only way to extract home equity after purchase was to sell the home 
at significant transaction costs.}
so they naturally borrowed more for their home purchase. 
After 1998, Texas homeowners gained future opportunities to 
extract equity, so they did not need to borrow as much at the time of purchase. 
%
%It is important to bear in mind that second purchase liens 
%were illegal in Texas before 1998, and were not advantageous after
%due to the 80\% CLTV limit. 
%Hence in Texas LTV=CLTV. 

Mortgage interest rates were similar in Texas and border states both before
and after the law change.
%
%Texas %had lower %mean real house price growth than
%house price growth
%was lower before the law change,  
%but higher afterwards. 
%the rest of the US ($1.05\%$ vs $2.36\%$), 
%but also lower house price volatility ($3.55\%$ vs $5.10\%$).
%
%Texas house price growth is more similar to the control groups 
%in the Border-State sample (and in more local samples). 
% and more local samples. 
Texas had higher median real house price levels than border states
both before and after the law change.
Real rents in Texas were similar to rents in border states, both before and 
after the law change. 
Texas had slightly higher population growth than border states before and after the law change.
%\textcolor{red}{population is included as a control in Table}
Both Texas and border states had a small drop in population growth after the law change. 
Real income per capita in Texas was slightly higher than in 
border states, both before and after the law change. 
Texas and border states had similar unemployment rates before and after the law change. 
The unemployment rate fell slightly less in Texas
than in border states after the law change. 
Texas had a lower homeownership rate than border states, 
both before and after the law change. 
%but experienced a similar rise after the law change. 
%
Texas had more single-family building permits than border states 
before the law change and experienced a larger rise in building permits
after. 
In this raw analysis, comparing differences in means pre- and post-treatment, 
it appears the HEL legalization raised building permits in Texas.
\hyperref[LHS_ECON]{Table \ref*{LHS_ECON}}, column 7, 
presents regression estimates from a more refined analysis, controlling 
for location and time fixed effects, revealing that the HEL 
legalization did not have a statistically significant impact on log single family building permits. 
%This reveals that the fixed effects wash away this effect.
%  
%Consistent with its reputation of being a large state with a lot of space, 
%Texas MSAs have a higher average 
%supply elasticity than MSAs in the US control group, but slightly lower than MSAs in the 
%Border-State control group.
Finally, Texas MSAs have a slightly lower supply elasticity than MSAs in  
border states. 
%

\begin{figure}[!ht]%[center]
	\centering
	\small \caption{Annual Percent Change in Real House Prices (demeaned) 
		in the United States, Texas, and Border States} 
	\label{AT}
	\includegraphics
	[scale=1]
	{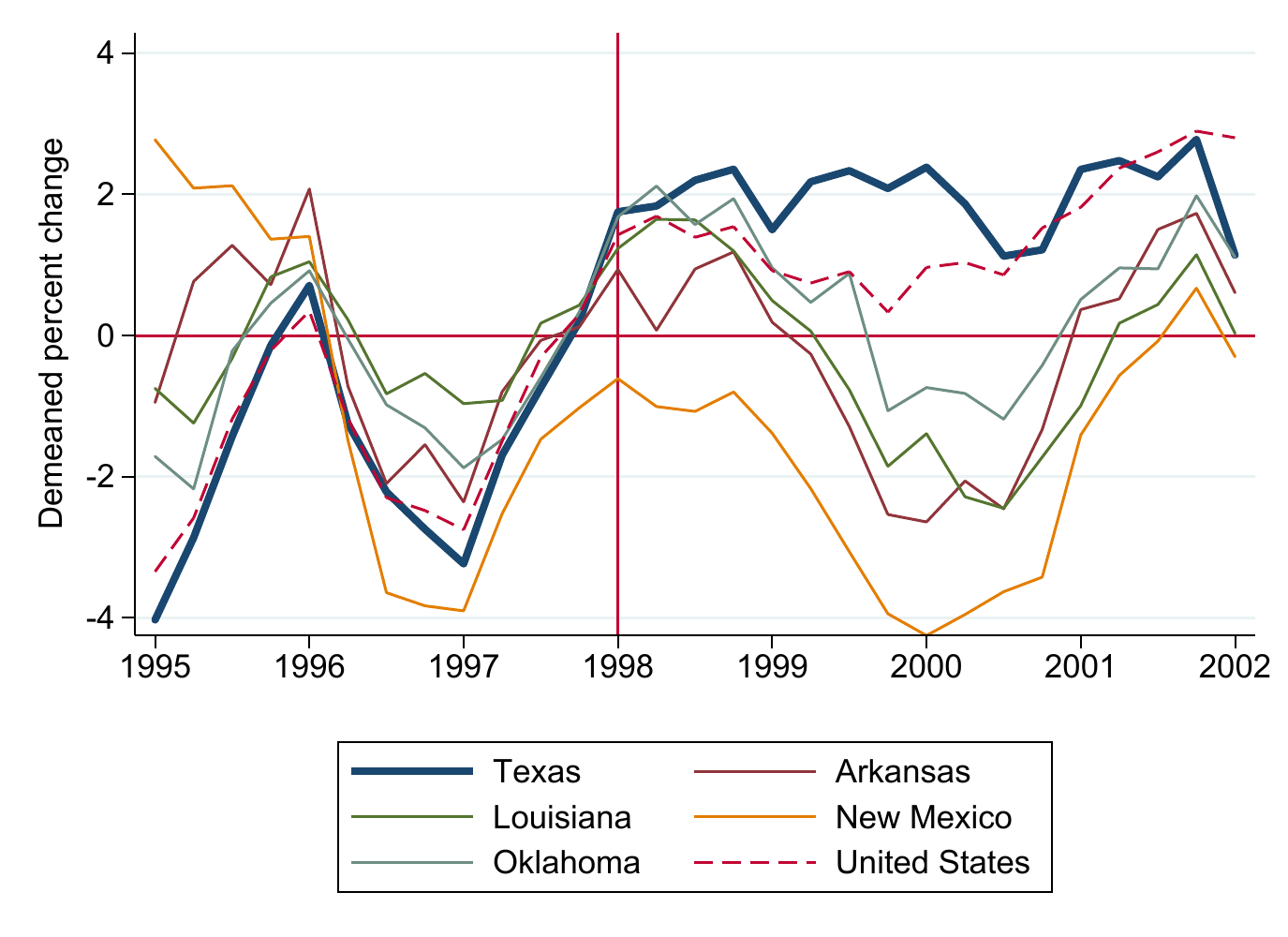} %ATdemean  %AT_combined
%\vspace{3pt plus 1pt minus 1pt} %\vspace{10pt}
\medskip

\begin{minipage}{0.8\textwidth}
	{\footnotesize 
		This figure plots the annual demeaned percent change in
		real house 
		prices in  the United States, Texas, and its four border states.
		There is a vertical red line in 1998, the year home equity loans were legalized in Texas.	
		The house price data are from the FHFA AT Index. 
		House prices are deflated by the CPI-U, as explained in the paper. 
		Data sources can be found in \hyperref[data_table]{Table \ref*{data_table}}.
		\par}
\end{minipage}
\end{figure}

%\label{Percent}	
%Before the estimates are studied, 
%For a preview of the results
\hyperref[AT]{Figure \ref*{AT}}
plots the
demeaned annual percent change in real house prices 
using %coarse 
raw data in Texas, the United States, and four 
states bordering Texas.
%border states. 
%The series are plotted between 1992 and 2004, 
%%six
%6 years before and after the 
%treatment.  
The only time in this sample 
%(in this sample period) 
when Texas had greater real house price 
growth than either the full United States or  its border states was for a few years after 
the law change in 1998.

\subsection{Impact of HEL legalization on Texas house prices}%Control Group: 
\label{Main}
This section presents the main results of this paper, 
the impact of HEL legalization on Texas house prices. 
Estimates are presented for four geographically nested 
samples and the data are weighted by 
the inverse of the number of zip codes in each state.
The baseline specification contains 
year fixed effects, 
%location
zip code fixed effects, 
a time trend interacted with 
%a location 
an MSA dummy, 
and national log real oil prices interacted with 
%a location 
an MSA dummy.
The specification in the CBCP sample also includes 
contiguous border county pair by year fixed effects as in \cite{Dube}. 
In the main analysis,
robust standard errors are computed three ways: 
(1) clustered by state, 
(2) clustered by zip code, 
and 
(3) spatially correlated (\cite{Conley}).\footnote{In 
the CBCP sample, standard errors are 
(1) double clustered by state and border county pair, 
(2) double clustered by zip code and border county pair, 
and 
(3) spatially correlated (\cite{Conley}).} 
This strategy is conservative %, except for the US sample, 
in the border samples
because there are relatively few clusters to precisely estimate standard errors 
clustered at the state level.\footnote{This follows 
\cite{Cameron}: ``The consensus is to be conservative and avoid bias and to use bigger and more aggregate clusters when possible, up to and
including the point at which there is concern about having too few 
clusters.''} 
%}%
%
All samples have enough zip codes to precisely estimate standard errors 
clustered at the zip code level.
The largest standard error of the three is reported. % presented. 
Estimates from the baseline specification will be presented first, 
followed by an exploration of robustness.

Estimates from static regressions %Equation blah \ref{static DID}
across the four samples 
(\hyperref[US_T1]{Table \ref*{US_T1}}, columns 1--4)
show that the HEL legalization raised Texas house prices %rose an extra
$3.5\%-6.16\%$.  %$2.5\%- 5.5\%$.  %$3.5- 4.56\%$ %$3.8\% - 4.2\%$ 
All estimates are statistically significant %at least 
at the $5\%$ level. 
The nested samples %used in columns 1-4 
are progressively refined; 
column 1 uses the most aggregate sample (U.S.) 
whereas column 4 uses the most local sample (CBCP). 
Border samples help eliminate local time-varying differences 
to the extent that zip codes near each other are more likely affected 
by the same local shocks such as hurricanes and factory shutdowns. 
The effect was $3.5\%$ in the U.S. sample, $6.16\%$ in the Border State sample, 
$4.76\%$ in the Border sample, and $4.13\%$ in the CBCP sample. 
There is no systematic pattern in the size of the effect as the sample becomes more refined. 
%
%\textcolor{red}{AZ: fill in details, strategy nested samples, no systematic pattern}
%
The preferred estimate, 
$\textcolor{ChadRed}{\hat{\beta}_{DID}} = 4.13\%$, is chosen from the CBCP sample  
because it is the most local sample with 
the most comprehensive battery of fixed effects.
%a large number of 
%observations. 
%
%This is equal to an extra 3.4 years of real house price growth.\footnote{Average annual real house price growth 
%	in Texas was 1.05\%.}
%

%
Next, estimates 
from %the  
dynamic regressions %are in 
(\hyperref[US_T1]{Table \ref*{US_T1}}, columns 5--8)
help rule out the biggest concern: upward pretrends. 
If Texas house prices were on an upward pretrend (relative to control zip codes)
before the law change, 
that would raise concerns that some other factor was causing them to rise 
and that they would have continued to rise without the treatment.

\begin{figure}[!ht]
	\centering

	\caption{Impact of HEL Legalization on Texas House Prices in Four 
	Geographically Nested Samples}
\label{US_F1}
%	\subfigure%[Second caption]
	\includegraphics
[scale=1.25]
{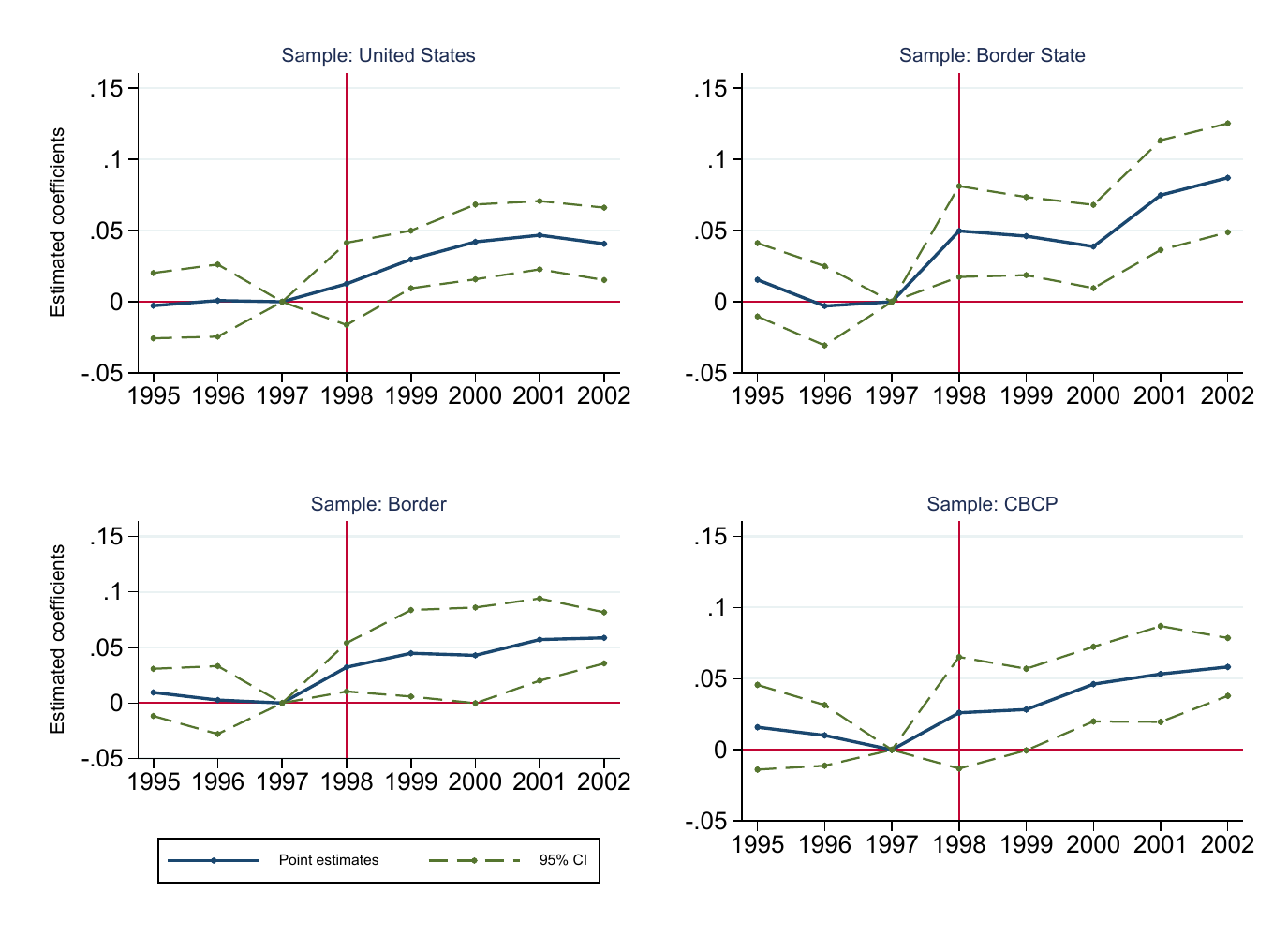} %ATdemean  %AT_combined
%	{
%		\includegraphics [scale=.63]
%		{../../4_output_analysis/main/US-Main-Cluster-StateZip5OLS}
%		%		\label{fig:first_sub}
%	}
%	%\\[-1.5em]
%	%\hspace{-2.65em}%
%%	\subfigure%[Second caption]
%	{
%		\includegraphics
%		[scale=.63]
%		{../../4_output_analysis/main/Border-Main-Cluster-StateZip5OLS}
%		%		\label{fig:second_sub}
%	}
%	%\hspace*{\fill}%
%	\\ %[-1.5em]
%%	\subfigure%[Third caption]
%	{
%		%		\includegraphics[width=1.0in]{imagefile2}
%		\includegraphics
%		[scale=.65]
%		{../../4_output_analysis/main/Texarkana-Main-Cluster-StateZip5OLS}
%		%		\label{fig:third_sub}
%	}
	%	\label{fig:sample_subfigures}
		\medskip
	
	\begin{minipage}{0.8\textwidth}
		{\footnotesize 
			This figure plots point estimates 
			$\textcolor{ChadRed}{\hat{\eta}_{k}}$ and 95\% confidence 
			intervals from 
			the dynamic regression 
			in \hyperref[US_T1]{Table \ref*{US_T1}}. %: columns 4-6.
			There is a vertical red line in 1998, the year of the law change.
			Data sources can be found in \hyperref[data_table]{Table \ref*{data_table}}.
			\par}
	\end{minipage}

\end{figure}

The coefficients and 95\% confidence intervals from the dynamic regressions 
for all four samples are plotted in
\hyperref[US_F1]{Figure \ref*{US_F1}}
and show a positive $\textcolor{ChadRed}{\hat{\eta}_{k}} > 0$ and statistically significant 
effect after the law change. % in all samples. 
In the U.S. and CBCP samples, the effect was positive in 1998 
but did not become statistically significant until 1999 in the U.S. sample and 2000 in the CBCP sample. 
The pretrends are parallel 
%at the 95\% confidence level
in all samples and show no evidence of anticipation
before the law was implemented.  
Estimates in two of the samples, Border State and CBCP, 
have parallel pretrends only after including the oil price MSA interactions. 
The oil price MSA interaction is a valuable control because 
oil prices play an important economic role in Texas and its border 
states (\cite{murphy2015plunging}).

\subsubsection{Robustness}%Control Group: 
\label{Main_Robustness}
This section investigates whether the main results in 
\hyperref[US_T1]{Table \ref*{US_T1}}
are robust to
(1) the method used for estimating standard errors, 
(2) the inclusion of covariates  and an alternative specification of the outcome variable, 
(3) the restriction to border city samples, and
(4) the method used to construct the house price index.
Estimates in \hyperref[SE_robust]{Table \ref*{SE_robust}}
explore whether  the treatment effect remains significant 
when standard errors are conventional (OLS), robust (EHW), 
clustered 
by five-digit zip code, three-digit zip code, county FIPS code, 
MSA, state, or spatially correlated (SHAC).
For the CBCP sample, standard errors are also clustered 
by county pair, 
double clustered by county pair and zip code, 
and double clustered by county pair and state.\footnote{I thank an anonymous
referee for this suggestion.} All estimates are statistically significant at least at the $5\%$ significance level.
The only clear pattern is that the main treatment effect is significant at the $1\%$ level
in all samples when standard errors are spatially correlated as in \cite{Conley} or
clustered at the three-digit zip code or smaller level. 
There does not seem to be an obvious pattern as standard errors are clustered at larger levels. 
In the U.S. and CBCP samples, the standard error falls as we move from MSA to state clusters,
whereas the standard error estimates rise slightly in the Border State and Border samples.

Estimates in \hyperref[SE_Specification]{Table \ref*{SE_Specification}}
%presents estimates to 
investigate whether the %main
treatment effect in \hyperref[US_T1]{Table \ref*{US_T1}}
is robust to the inclusion of covariates including 
national interest rates interacted with state dummies, 
real income per capita, 
and population. 
The table also explores an alternative
specification of the outcome variable
using annual real house price growth
(as opposed to log real house prices).  
The treatment effect remains positive and statistically significant 
across samples and specifications.
%and %very %similar 
%to the effect estimated in \hyperref[US_T1]{Table \ref*{US_T1}} %all 
In specifications using log real house prices as the outcome variable, 
including covariates slightly reduces the estimates in the U.S. and Border State samples
and slightly raises the estimates in the Border and CBCP samples. 
In specifications using real house price growth as the outcome variable, 
the estimates are larger in the U.S. sample and smaller in the other three samples.

Estimates in \hyperref[AlternativeBorder]{Table \ref*{AlternativeBorder}}
study the impact of HEL legalization on house prices
in four border city samples. 
The advantage of these smaller samples is they can potentially provide 
better control for local unobserved heterogeneity. 
In addition, these estimates can help %rule out
alleviate concerns about the indirect mechanism 
to the extent that border cities can be viewed as one economy. 
The samples are
(1) Texarkana, the only %official 
MSA in Texas with zip codes in another state, 
(2) the Dallas-Fort Worth (DFW) %, TX-OK 
Combined Statistical Area (CSA), 
which includes zip codes in Bryan County, Oklahoma,\footnote{Not to be confused with the DFW MSA, which is entirely in Texas.}
% restricted to zip codes within 50 miles of the Texas border, 
(3) the El Paso-Las Cruces CSA, 
which includes zip codes in Do\~{n}a Ana County, New Mexico,\footnote{El Paso-Las Cruces was delineated as a CSA by the Office of Management and Budget in 2013.}
and 
(4) the Texoma area, 
which includes counties in Oklahoma near Lake Texoma. 
%Unfortunately the FHFA index does not include observations on both sides of Texhoma and Texico. 
%
%
The estimates in these  border samples are between $3.75\%$ and $5.97\%$ and statistically significant.\footnote{Standard errors are clustered by zip code 
	in these smaller border city samples as they only contain two states.}

% one informal area.\footnote{Two other informal areas on the Texas border are 
%	Texhoma (not to be confused with Texoma) and Texico. 
%	Unfortunately the FHFA index does not include observations in these areas.}

Estimates in \hyperref[ZHVI_robust]{Table \ref*{ZHVI_robust}}
investigate whether the treatment effect in \hyperref[US_T1]{Table \ref*{US_T1}}
is robust to the method used for constructing the house price index. 
The outcome variables in this table are from the ZHVI, which is 
constructed differently\footnote{\url{https://www.zillow.com/research/zhvi-methodology/}.\\The main analysis does not use ZHVI data because it begin too late in April 1996.} from 
the FHFA index used in the main analysis. 
These estimates remain positive, statistically significant, and similar to the main estimates. 
Another advantage of the ZHVI index is that it can be 
directly interpreted as the median dollar house value in a zip code year. 
\hyperref[ZHVI_robust]{Table \ref*{ZHVI_robust}} also presents estimates 
of the HEL legalization on real ZHVI levels, 
finding an inflation-adjusted effect between $\$2,714.41$ and $\$5,363.34$.

\subsubsection{Synthetic controls and placebo tests}%Control Group: 
\label{Syn}
%\textcolor{red}{AZ: Only cite Doudchenko Imbens}
This section investigates the impact of HEL legalization on Texas house 
prices using the synthetic control method %developed by 
\citep{Abadie}.
\cite{Athey} called this ``the most important innovation in the policy evaluation 
literature in the last 15 years. This method builds on difference-in-differences estimation, 
but uses systematically more attractive comparisons."
\cite{Cavallo} extend the method to allow for %more than one
multiple units to experience treatment. %and at possibly different times.
This is useful because each Texas zip code received the treatment.
\cite{Doudchenko} generalize the synthetic control method 
to allow nonconvex weights and a permanent additive difference between the treated and 
control units. They show that this powerful generalization nests many existing approaches as special 
cases including classical  difference-in-differences and matching methods.

Let $Y_{z,t}$ denote the outcome variable for treated zip code $z$ in year $t$  
and $Y_{j,t}$ 
%is 
the outcome %in year $t$ in 
for untreated zip code $j$.
Let   $Y_{z,pre}$ and  $Y_{j,pre}$ be vectors of the outcome variables in the pretreatment years. 
Let $Y_{C,pre}$ be a matrix of predictors  whose columns 
consist of $Y_{j,pre}$
(outcome variables for all control zip codes in the pretreatment years)
as well as other control variables (national oil prices).
The  \cite{Doudchenko} estimator 
minimizes the distance between the treated outcome 
and an affine combination of the untreated outcome for the pretreatment period, regularized by the 
elastic-net (en) penalty \citep{Zou}: 
\begin{align*}
\left(\hat{\mu}^{en}, \hat{\omega}^{en} \right)
&= 
\underset{ \mu, \omega }{\argmin} \text{ }
\norm{ Y_{z,pre} - \mu -Y_{C,pre} \cdot  \omega   }_{2}^{2}
+\lambda \cdot \left( \alpha \norm{\omega}_{1} + (1-\alpha) \norm{\omega}_{2}  \right)
\end{align*}
The parameter $\lambda$ determines the amount of regularization, 
and   $\alpha$ determines the type. 
The case $\alpha =1$ corresponds to a LASSO
%Least Absolute Shrinkage and Selection Operator (LASSO)
penalty function, which captures a preference for parsimony
via a small number of nonzero weights. 
The case $\alpha =0$ corresponds to a Ridge penalty function, which captures a preference for 
smaller weights. 
%, then relative importance of 
\cite{Doudchenko} propose a cross-validation procedure 
to select the regularization parameters
$\lambda$ and $\alpha$ that minimize the average mean squared prediction error for all 
%control
untreated units.

These estimates give the counterfactual outcome for  %the 
treated zip code $z$ if it did not receive
the treatment as a function of the control zip codes: 
\begin{align*}
\hat{Y}_{z,t}(0) &= \hat{\mu}^{en} + Y_{C,t} \hat{\omega}^{en},
\end{align*}
where $Y_{C,t}$ is a row vector consisting of outcomes for the control zip codes
and national oil prices in year $t$.  
The identifying assumption is that the relationship between 
the treated and control outcome variables, %(house prices),
given by $\hat{\mu}^{en}$ and $\hat{\omega}^{en}$,
would have remained the same in absence of the treatment.
While this is defended in the same way as in Section \ref{Identification}, 
the advantage of the synthetic control estimator is a more attractive control group with
tighter pretrends.

The estimated treatment effect for zip code $z$ is 
the gap (i.e., difference) between the observed and counterfactual outcome  
$
\textcolor{ChadRed}{ \hat{\eta}_{z , t}} 
= 
Y_{z,t} - \hat{Y}_{z,t}(0)
$.
%\begin{align*}
%\textcolor{ChadRed}{ \hat{\eta}_{z , t}} 
%&= 
%Y_{z,t} - \hat{Y}_{z,t}(0)
%\end{align*}
The extension by \cite{Cavallo}  allows for more than one 
%unit
zip code to experience treatment.
%and at possibly different times. 
Let  $z \in \left\{ 1, \dots, Z \right\}$ be the index for all treated zip codes;
%that never undergo treatment.
then the average treatment effect in Texas (across all Texas zip codes) in year $t$ is given by
$
\textcolor{ChadRed}{ \hat{\eta}_{TX , t}} 
= 
\frac{1}{Z} \sum_{ z=1 }^{Z}   \textcolor{ChadRed}{ \hat{\eta}_{z , t}} 
$.
%
%\begin{align*}
%\textcolor{ChadRed}{ \hat{\eta}_{TX , t}} 
%&= 
%\frac{1}{Z} \sum_{ z=1 }^{Z}   \textcolor{ChadRed}{ \hat{\eta}_{z , t}} 
%\end{align*}
The average post-treatment effect is defined similarly
$
\textcolor{ChadRed}{ \hat{\eta}_{TX}} 
= 
\frac{1}{\#\left\{t\geq 1998 \right\}} \sum_{ t\geq 1998 }   \textcolor{ChadRed}{ \hat{\eta}_{TX , t}} 
$.
%\begin{align*}
%\textcolor{ChadRed}{ \hat{\eta}_{TX}} 
%&= 
%\frac{1}{\#\left\{t\geq 1998 \right\}} \sum_{ t\geq 1998 }   \textcolor{ChadRed}{ \hat{\eta}_{TX , t}} 
%\end{align*}

\begin{figure}[!ht]%[center]
	\centering
	\small \caption{Impact of HEL Legalization on Texas House Prices Using Synthetic Controls} 
	\label{Synth} 
		\textbf{Panel A: Zip Codes in Texas}
	{
		\includegraphics[trim=0 27 0 47.5, clip,scale=.625]
		%{../../4_output_analysis/main/SCM_TX_NoP}		
		{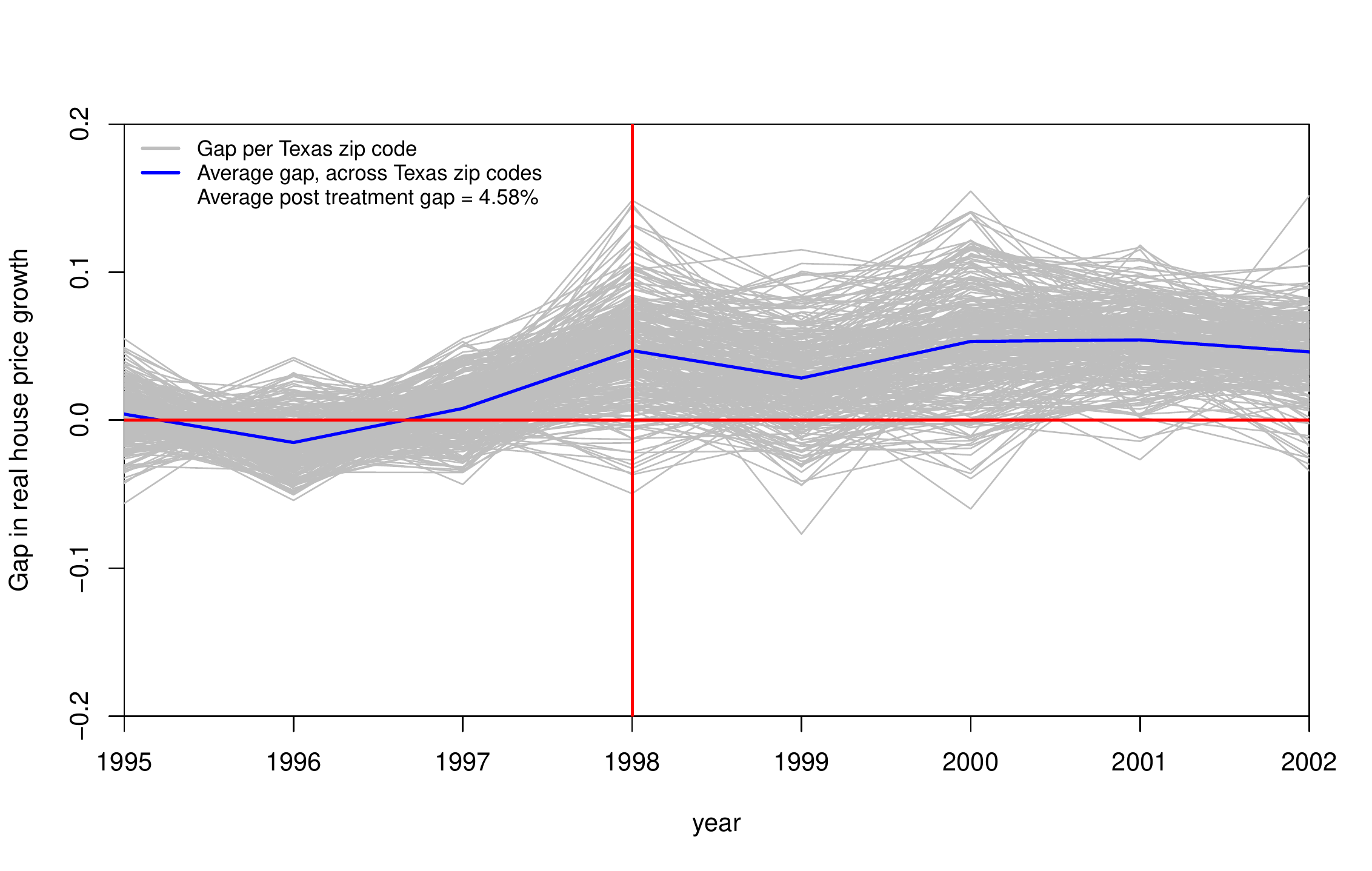}		
%		{../../4_output_analysis/main/SynthTexas}
%
%		\includegraphics[trim=0 27 0 47.5, clip,scale=.625]{../../4_output_analysis/main/SynthAllZips2}
	}
%	 \\ [-1.1em]
	%	\subfigure%[Third caption]
			\textbf{Panel B: Zip Codes in Border States}
	{
\includegraphics[trim=0 17.5 0 47, clip,scale=.625]
{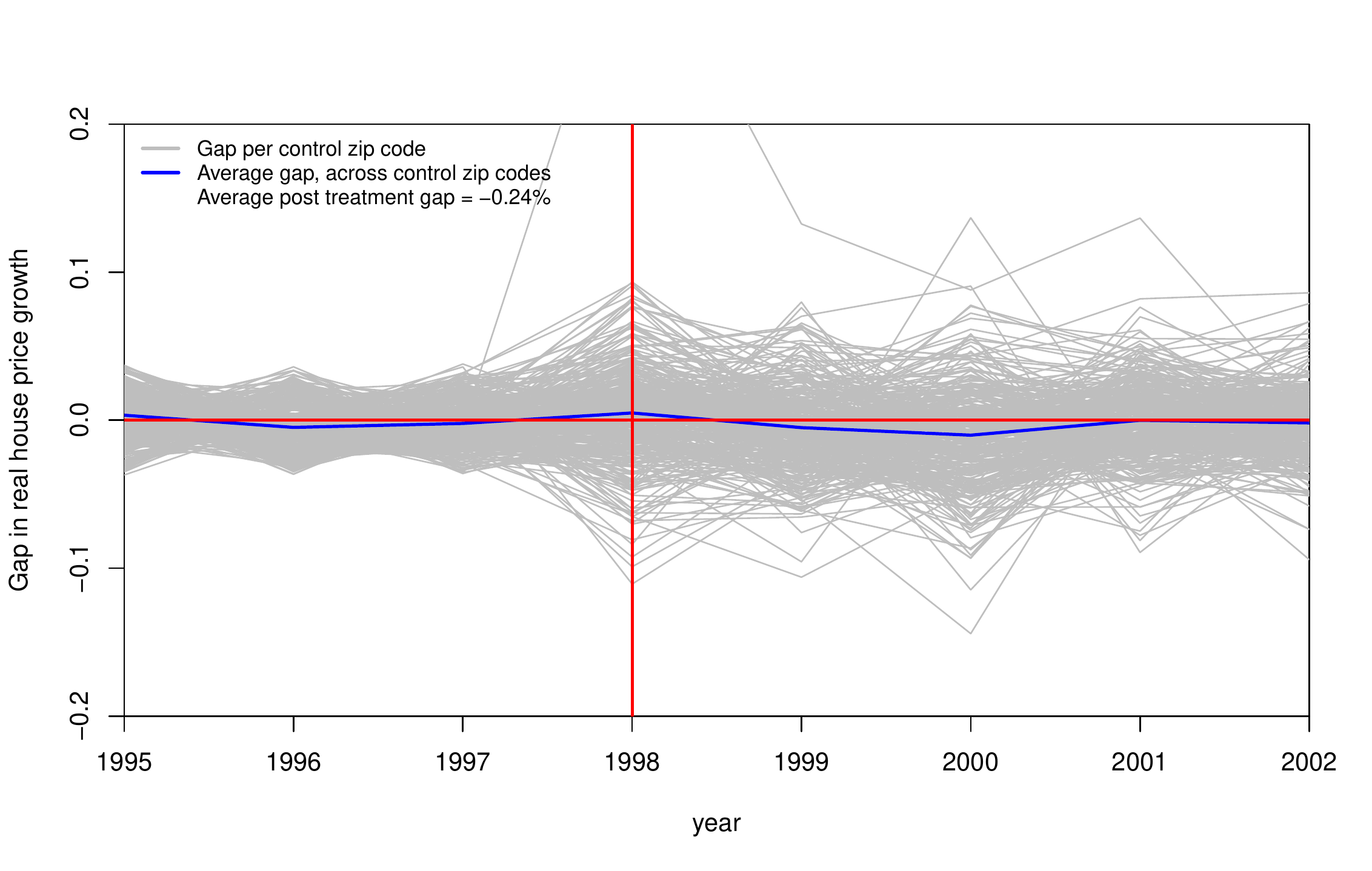}		
%{../../4_output_analysis/main/SynthControl}		
%\includegraphics[trim=0 17.5 0 47, clip,scale=.625]{../../4_output_analysis/main/SynthATEControl2}		
	}
	%	{../../4_output_analysis/0_clean_12_25/dynamic-1-dynPAPER-allstates-Clusterstate-lnrhpi-MSATrend-from1992-to-2004-Min1992-allstates-1992-2004}
	%	\includegraphics[scale=.75]{1992_2015_6_29_Texarkana_DynamicPanel}
	%\floatfoot{A note}
	\begin{minipage}{0.75\textwidth}
		{\footnotesize 
			Panel A presents 
			the gap (treatment effect) in real house price growth for each 
			treated zip code $\textcolor{ChadRed}{ \hat{\eta}_{z , t}} $ (gray) 
			estimated using a synthetic control sample consisting of zip codes in border states, as 
			explained in Section \ref{Syn}.
			The blue curve is the average gap across all treated zip codes each year 
			$\textcolor{ChadRed}{ \hat{\eta}_{TX, t}} $. % (blue).
			The average post-treatment effect in Texas is
			$\textcolor{ChadRed}{ \hat{\eta}_{TX}}=4.58\% $.
%			and each control zip code (gray) located in a border state. 
			%
			Panel B presents the corresponding placebo gaps for 
			%control 
			zip codes
			in border states that did not receive the treatment.  
			The average post-treatment effect in the control group is
			$\textcolor{ChadRed}{\hat{\eta}_{\text{C}}}=-0.24\%$.
			There is a vertical red line in 1998, the year of the law change.
			Data sources can be found in \hyperref[data_table]{Table \ref*{data_table}}.			
%
%			Panel B presents the average gap each year across all treated zip codes 
%			$\textcolor{ChadRed}{ \hat{\eta}_{TX, t}} $ (blue) and 
%			control zip codes (dark green). Since the control zip codes did not receive 
%			a treatment in 1998, the dark green curve gives the placebo effect.
%			The legend in Panel B also includes the average post-treatment effect 
%			$\textcolor{ChadRed}{ \hat{\eta}_{TX}}=2.86\% $. 
%			housing demand on house prices in 
%			two cities with different supply elasticities.
%			Price (P) is on the y-axis and quantity (Q) in on the x-axis. 
%			Initially, the price of housing is the same in both cities. 
%			A rise in demand 
%			%(due to the law change) 
%			causes prices
%			to rise more in the inelastic 
%			city relative to the elastic city.
			\par}
	\end{minipage}
	%\fnote{\textit{Note}. This figure compares the impact of a rise in demand on prices in 
	%two cities with different supply elasticities.
	%Initially, the price of housing is the same in both cities. 
	%After a rise in demand (due to the law change), prices rise more in the inelastic 
	%city relative to the elastic city.}
	%
	%\caption*{A note}
	%\small \caption*{The impact of a rise in demand on locations with different supply elasticity} 
\end{figure}

\begin{figure}[!ht]%[center]
	\centering
	\small \caption{The impact of a rise in demand on house prices
		in cities with different supply elasticities} 
	\label{ElasticSupply}
	%\ref{Texarkana_dynamic_fig}
	\includegraphics
	[scale=1]
	{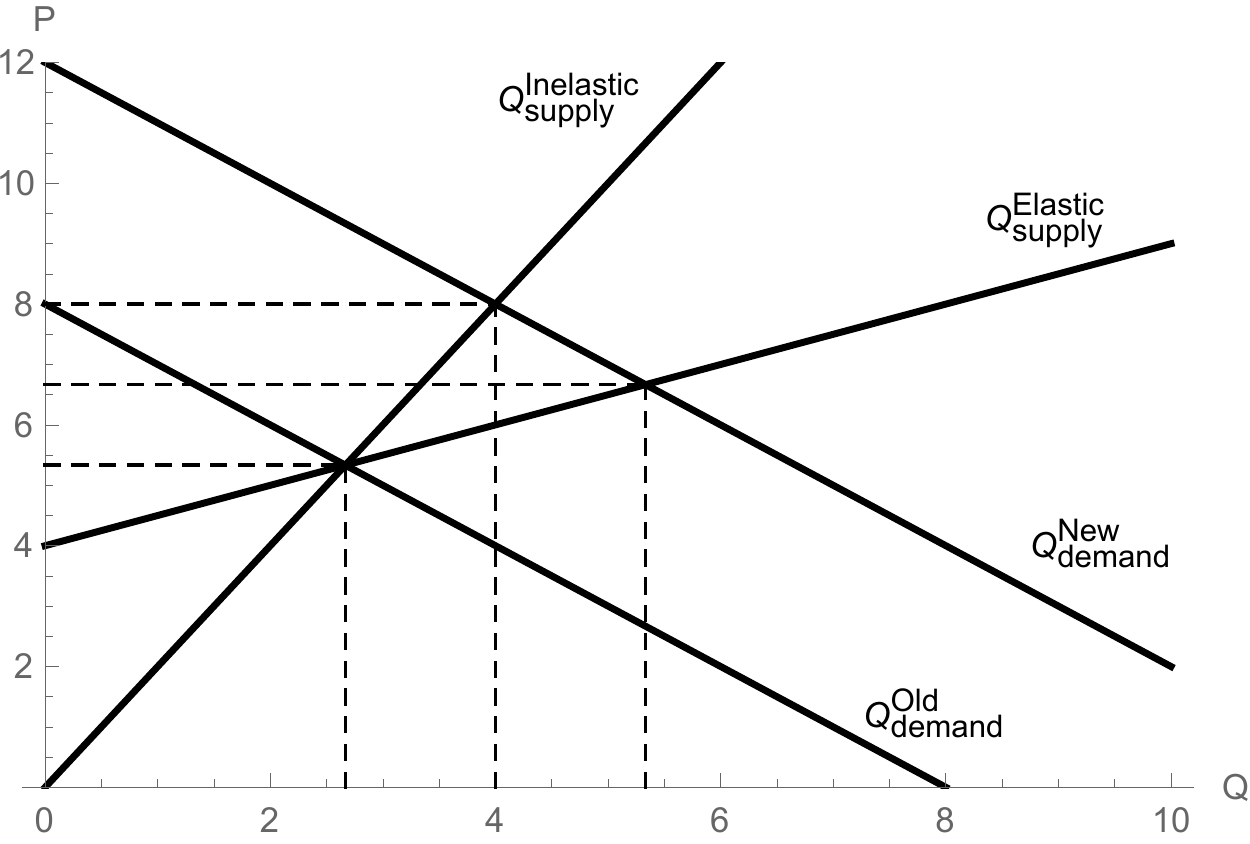}
	%	{../../4_output_analysis/0_clean_12_25/dynamic-1-dynPAPER-allstates-Clusterstate-lnrhpi-MSATrend-from1992-to-2004-Min1992-allstates-1992-2004}
	%	\includegraphics[scale=1]{../../DATA/dynamic-texarkana-clusterzip5-1992-2004-quadstrend}
	%	\includegraphics[scale=1]{texarkana_95_90}
	%	\includegraphics[scale=.75]{6_29_Texarkana_DynamicPanel}
	%	\includegraphics[scale=.75]{1992_2015_6_29_Texarkana_DynamicPanel}
	%\floatfoot{A note}
	\begin{minipage}{0.7\textwidth}
		{\footnotesize 
			This figure compares the impact of a rise in
			housing demand on house prices in 
			two cities with different supply elasticities.
			Price (P) is on the vertical axis, and quantity (Q) is on the horizontal axis. 
			Initially, the price of housing is the same in both cities. 
			A rise in demand 
			%(due to the law change) 
			causes prices
			to rise more in the relatively inelastic city. %relative to the elastic city.
			This figure is from a discussion by Loretta Mester at the 
			Federal Reserve Bank of Philadelphia.
			\par}
	\end{minipage}
	%\fnote{\textit{Note}. This figure compares the impact of a rise in demand on prices in 
	%two cities with different supply elasticities.
	%Initially, the price of housing is the same in both cities. 
	%After a rise in demand (due to the law change), prices rise more in the inelastic 
	%city relative to the elastic city.}
	%
	%\caption*{A note}
	%\small \caption*{The impact of a rise in demand on locations with different supply elasticity} 
\end{figure}

The synthetic control estimates using the Border State sample 
are presented in \hyperref[Synth]{Figure \ref*{Synth}}.
%
%%%
%
Panel A plots the treatment effect  for each Texas
zip code $\textcolor{ChadRed}{ \hat{\eta}_{z , t}} $ in gray and 
%placebo treatment effects for each control zip code in gray.
the  treatment effect 
averaged across all Texas zip codes $\textcolor{ChadRed}{ \hat{\eta}_{TX , t}}$ in blue.
Panel B plots the corresponding placebo estimates for each untreated zip code in gray 
as well as the average across untreated zip codes in blue.

Observe that before the treatment year, 1998, 
the treatment effect in both the Texas and control zip codes 
is approximately zero as intended. 
%The weights are chosen to achieve this. 
%This is by construction: the weights are chosen to minimize the distance 
%between the outcome and 
After 1998, the treatment effect in Texas zip codes is mostly positive, %for most zip codes
whereas the placebo treatment effects in control zip codes are equally positive and negative.

The treatment effect averaged over all post-treatment years
in Texas is $\textcolor{ChadRed}{ \hat{\eta}_{TX}} = 4.58\%$. 
It is heartening to see that the average treatment effect across 
Texas zip codes is positive and of a similar magnitude as the estimates 
in the main analysis. 
The average post-treatment effect in the control group 
is $\textcolor{ChadRed}{ \hat{\eta}_{C}} =-0.24\%$, 
indicating no evidence of a placebo effect.

\subsection{Treatment effect heterogeneity}%Control Group: 
\label{ResultsHTE}
This section explores if and in what ways 
the effect differed across treated zip codes. 
\hyperref[HTE_HIST]{Figure \ref*{HTE_HIST}}
presents histograms and summary statistics of the treatment effect 
for each treated zip code in the four geographically nested samples. 
These estimates are from regressions in the baseline specification
(\hyperref[US_T1]{Table \ref*{US_T1}}, columns 1--4), except the term 
$\text{Texas} \times \text{Post}$ is interacted with an indicator for 
each zip code.
%First, note that 
The histograms reveal that there is heterogeneity in the effect across zip codes. 

\begin{figure}[!ht]%[center]
	\begin{center}
\small \caption{Impact of HEL Legalization on Texas House Prices by MSA Elasticity}
\label{HTE_Elasticity} 
\includegraphics[trim=0 10 0 5, clip,scale=.65]
{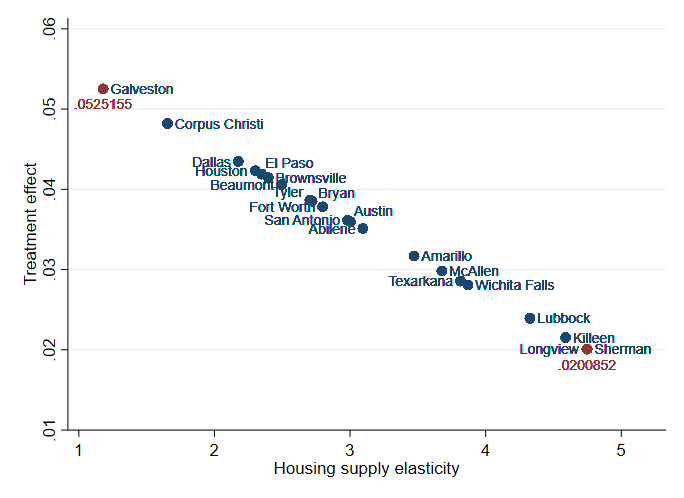}
\medskip
\begin{minipage}{0.9\textwidth}
	{\footnotesize 
		This figure plots the fitted average treatment effect 
		 $\left(
		\hat{\text{ATE}}\left(H_{i}\right) 
		= 
		\textcolor{ChadRed}{\hat{\beta}_{H,0}} + \textcolor{ChadRed}{\hat{\beta}_{H}}  H_{i}  
		\right)$		
		%point estimates 
		%fitted values for the treatment effect 
		for each MSA
		against housing supply elasticity.
		The estimates are 
		from triple-difference regressions 
		presented in \hyperref[US_DDD]{Table \ref*{US_DDD}}. %column 7
		This figure 
		%uses data from 
		%HP data from FHFA
		includes
		all 21 Texas MSAs. 
		Galveston, the most inelastic MSA, had a 5.25\% effect, 
		whereas Sherman, 
		the most elastic MSA, had a 2\% effect.
		The measure of housing supply elasticity is from \cite{Saiz2010}.
		Data sources can be found in \hyperref[data_table]{Table \ref*{data_table}}.
		\par}
\end{minipage}
	\end{center}
\end{figure}

\begin{figure}[!ht]
	\centering
	\caption{Impact of HEL Legalization on Texas House Prices, Heterogeneity Across Zip Codes}
	\label{HTE_HIST}
	\includegraphics
	[scale=1.25]
	{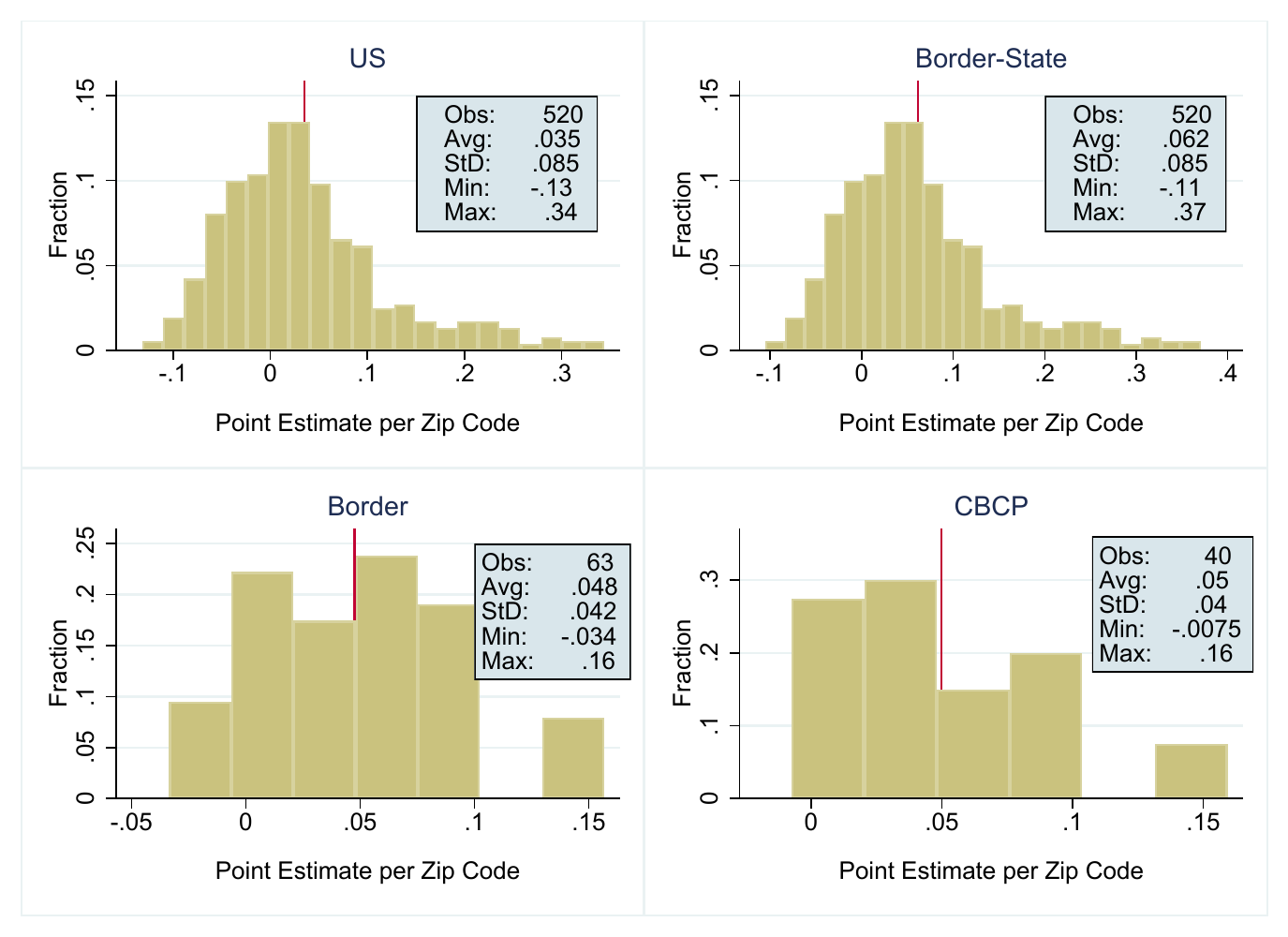} %ATdemean  %AT_combined
	
	%%	\subfigure%[Second caption]
	%	{
	%		\includegraphics [scale=.63]
	%		{../../4_output_analysis/main/HTE-Histogram-US}
	%		%		\label{fig:first_sub}
	%	}
	%	%\\[-1.5em]
	%	%\hspace{-2.65em}%
	%%	\subfigure%[Second caption]
	%	{
	%		\includegraphics
	%		[scale=.63]
	%		{../../4_output_analysis/main/HTE-Histogram-Border}
	%		%		\label{fig:second_sub}
	%	}
	%	%\hspace*{\fill}%
	%	\\ %[-1.5em]
	%%	\subfigure%[Third caption]
	%	{
	%		%		\includegraphics[width=1.0in]{imagefile2}
	%		\includegraphics
	%		[scale=.65]
	%		{../../4_output_analysis/main/HTE-Histogram-Texarkana}
	%		%		\label{fig:third_sub}
	%	}
	%	\label{fig:sample_subfigures}
	\medskip
	
	\begin{minipage}{0.8\textwidth}
		{\footnotesize 
			\textit{Note}. This figure presents histograms and summary statistics 
			of the treatment effect for each zip code in %three
			the geographically nested samples.
%			The mean treatment effect in each sample is exactly equal to the coefficient
%			on $\text{Texas} \times \text{Post}$ 
%			in \hyperref[US_T1]{Table \ref*{US_T1}}. %: columns 1-3.
			There is a vertical red line at the mean. % and at zero.
			Data sources can be found in \hyperref[data_table]{Table \ref*{data_table}}.
			%The house price data is from the FHFA, deflated by the CPI-U.
			%Data sources can be found in \hyperref[US_T1]{Table \ref*{US_T1}}.
			%Source: FHFA AT Index 
			%Use prices from sales transactions of mortgage data obtained from the Enterprises    
			%adding prices from appraisal data obtained from the Enterprises	 
			\par}
	\end{minipage}

\end{figure}

%The following regressions 
%Triple difference regressions in \hyperref[US_DDD]{Table \ref*{US_DDD}}
\hyperref[US_DDD]{Table \ref*{US_DDD}} 
presents estimates from triple-difference regressions to
investigate 
treatment effect 
heterogeneity along 
four dimensions: housing supply elasticity, income per capita, the unemployment rate, and median house price level. 
Estimates 
%from triple difference regressions %Equation blah \ref{static DID}
using the 
\cite{Saiz2010} measure of supply 
elasticity (\hyperref[US_DDD]{Table \ref*{US_DDD}}, column 1)
show that zip codes in more elastic 
MSAs saw a smaller rise in prices.\footnote{While this measure of elasticity is widely used as 
	an instrumental variable for
	%a source of exogenous variation in 
	house prices
	\citep{Mian2011}, 
	not all authors agree it is ideal
	\citep{Davidoff}.
	%The estimates in \hyperref[US_DDD]{Table \ref*{US_DDD}} do 
	This paper does not use elasticity as an instrument, 
	but as a source of heterogeneity.
}
This is consistent 
with predictions from a partial equilibrium 
model; a rise in housing demand should 
have a bigger impact on prices in locations 
where it is relatively hard to build (\hyperref[ElasticSupply]{Figure \ref*{ElasticSupply}}).
The rise in house prices was 
$0.9\%$ lower per unit of elasticity.
%If supply elasticity was zero
%the range of elasticity in Texas is 
%
If housing supply was perfectly inelastic, the average treatment effect 
would be the intercept $6.3\%$. 
%The intercepts suggest ...
%Tom Davidoff ... 
The 
%estimated
%average
fitted
 treatment effect 
 is plotted against elasticity
 $\left(
 \hat{\text{ATE}}\left(H_{i}\right) 
 = 
 \textcolor{ChadRed}{\hat{\beta}_{H,0}} + \textcolor{ChadRed}{\hat{\beta}_{H}}  H_{i}  
   \right)$
  for each Texas MSA 
in \hyperref[HTE_Elasticity]{Figure \ref*{HTE_Elasticity}}.
The treatment effect varies 
considerably
from $5.25\%$ 
in the most inelastic MSA (Galveston) %, an island city)
to $2\%$ in the most elastic MSA (Sherman).

\hyperref[US_DDD]{Table \ref*{US_DDD}}, column 2,
%\hyperref[lnrincpc1990]{Table \ref*{lnrincpc1990}}
investigates heterogeneity by  prelaw log  real income per capita.  % in 1992. 
Zip codes in higher-income counties  saw larger treatment effects.
%Counties with 
A 1\% higher prelaw real  income per capita %in 1992 
corresponds to a $0.109\%$ larger treatment effect. 
\hyperref[US_DDD]{Table \ref*{US_DDD}},  column 3,
investigates heterogeneity by prelaw unemployment.  % in 1992. 
Zip codes in counties with higher unemployment rates saw  smaller treatment effects.
%Counties with 
A 1\% higher prelaw unemployment rate %in 1992 
corresponds to a $0.7\%$ smaller treatment effect. 

\hyperref[US_DDD]{Table \ref*{US_DDD}}, column 4,
investigates heterogeneity by each zip code's prelaw
log real median house price level. 
Zillow estimated median house prices are used 
%as the measure of heterogeneity 
because 
the level of the FHFA index is not %very 
informative about median price levels.\footnote{Zillow data are not used in the main analysis because the 
sample begins too late in April 1996.}
%%%%%%%
Ex ante pricier zip codes saw a bigger treatment effect.
A 1\% higher prelaw   real median house price level
%saw 
corresponds to 
a $0.036\%$ larger treatment effect. 
The coefficient on $\text{Texas} \times \text{Post}$
is not interesting in this specification because there were no observations with 
zero house price levels.

These results complement \cite{Landvoigt}. % (LPS).
While \cite{Landvoigt} find that the credit expansion 
%(2000-2005)
during the housing boom 
had a bigger impact on ex ante lower-priced homes in San Diego,  
this paper finds that the (exogenous) HEL legalization had a bigger 
impact on ex ante higher-priced homes in Texas.
%These results are not inconsistent. 
The credit expansion in \cite{Landvoigt} increased access to borrowers seeking all loans 
secured by housing;
in particular purchase mortgages became available 
to many households that previously did not qualify. 
These previously purchase-constrained households tended to buy lower-priced homes.
In contrast, the Texas law change 
did not affect access to purchase mortgages,
%leverage, 
but rather the ability of existing homeowners 
to extract equity by borrowing after they were already homeowners.

%The triple-difference regressions 
Together, the estimates in \hyperref[US_DDD]{Table \ref*{US_DDD}} (columns 2--4)
provide evidence that households in more prosperous zip codes %with stronger economic conditions
(with \textit{ex ante} higher house price levels, higher income, and lower unemployment)
value the option to pledge their home as collateral more strongly. 
%Consistent with evidence that HELs are an upper-middle class product (Tracy)
There are several possible explanations.
First, richer households typically enjoy a bigger tax shield on mortgage interest 
because (a) they have higher marginal tax rates and 
(b) they are more likely to itemize deductions. 
Second, these households tend to be more financially literate and are more likely to be aware of these equity extraction products \citep{Lusardi}.
Third, rich households are more likely to qualify for these new loans as 
they tend to have better   
credit and more stable income \citep{DefuscoMondragon}. %DTI, LTV

To be clear, these results do not show that poorer households do not value HELs.
%this evidence is not inconsistent with literature 
%that shows homeowners smooth consumption better than renters
%after job loss  or disability when local house prices  are rising 
%\citep{HryshkoLuengoPradoSorensen2010}.
The treatment effect was still positive (but smaller) in poorer zip codes. 
These results reveal that households expect to borrow more 
when they are not under distress (e.g., for education or entrepreneurship).
%This is likely because households under economic distress are less likely to be approved 
%for home equity loans. 
%
These results highlight the asymmetric benefits of %collateral value
home equity borrowing due to mortgage underwriting 
requirements: households are less likely to qualify for HELs when they are experiencing 
economic distress.

%%%%%%%%%%%%%%%%%%%%%%%%%%%%%%%%%%%%%%%%%%%%%%%%%%%%%%
%%%%%%%%%%%%%%%%%%%%%%%%%%%%%%%%%%%%%%%%%%%%%%%%%%%%%%
%\newpage
%%%%%%%%%%%%%%%%%%%%%%%%%%%%%%%%%%%%%%%%%%%%%%%%%%%%%%
\subsection{Channels} %Different Outcome Variables
\label{Main_channels}
%Next
This section investigates the 
%channels
mechanism behind the treatment effect on house prices
%are investigated 
by 
%replacing house prices with different 
studying the impact of %the  
HEL legalization on various other
outcome variables.
The treatment effect could have occurred 
through two channels: 
\begin{enumerate}
	\item 
	\underline{Direct Channel}: 
	the
	%new options to borrow
	law %change
	caused a rise in demand for %homeownership
	owner-occupied housing due to the new option 
	allowing homes to be pledged as collateral.
	%Extensive or intensive margin?
	%This shift in demand caused the price of housing to rise.
	%
	%, raising the price of housing. 
	%what is the 
	%	total
	%	impact of 
	%	%collateral value 
	%	collateralizability
	%	on house prices?
	\item 
	\underline{Indirect Channel}:	
	the law 
	%otherwise 
	affected 
	other variables %which
	that affect house prices. 
	%the economy in other ways, affecting the price. 
	For example, if the law 
	%increased borrowing, 
	%which 
	increased %consumption and 
	investment enough to stimulate
	the local economy, 
	%that
	this could have raised demand for housing, thus raising the price. 
\end{enumerate}
If variables known to affect house prices such as 
%-- 
rent, population, income, and 
unemployment %-- 
were not affected by the law change,
%treatment, 
that would provide evidence
that the direct channel 
%is driving 
drove the treatment effect. 
%
%Evidence in favor of the direct channel 
%would show that variables know

%Regressions using different outcome variables are estimated to 
%investigate whether the treatment affect occurred through 
%the direct real options channel.
Estimates presented in \hyperref[LHS_ECON]{Table \ref*{LHS_ECON}}
%,  columns 1-4
show that the law did not lead to  economically or statistically significant
changes in real rents, population, or real income per capita.
%Economic theory implies that 
The rent regressions are particularly informative
as classical economic theory predicts that 
the price of housing should have been equal to the present discounted value of 
rents before the legalization of HELs: 
$p_{t} = 
\E_{t} 
%\sum_{j=1}^{\infty} \frac{  Rent_{t+j} }{(1+r)^{j} }  
\sum_{j=1}^{\infty} \frac{  Rent_{t+j} }{
(1+r_{t+1}) \times \cdots \times (1+r_{t+j})
	%(1+r)^{j} 
}  
$.
%\footnote{For simplicity the discount rate is assumed constant.
%	In reality it varies. See appendix \ref{deriv1}.}
After the HEL  legalization, the price should reflect 
both the housing service flow (rent) and the additional collateral service flow: 
$p_{t} = 
\E_{t} 
\sum_{j=1}^{\infty} 
\frac{  Rent_{t+j} + 
	\textcolor{ChadRed}{\textbf{CSF}_{\textbf{t+j}}}
	%\textcolor{red}{\textbf{CSF}_{\textbf{t+j}}}
	%\textbf{$\textcolor{red}{CSF_{t+j}}$}	
	%\textcolor{ChadRed}{CSF_{t+j}}	
	% CSF_{t+j} 
}{
%(1+r)^{j} 
(1+r_{t+1}) \times \cdots \times (1+r_{t+j})
}  
%+
%\frac{    CSF_{t+j} }{(1+r)^{j} }  
$.
The rent regression helps reassure us that the treatment effect 
is 
%due to the demand for HELs and 
not due to indirect effects on rent. 
%The price should reflect both the 
%
There was a small but statistically significant rise in the unemployment rate 
of $0.3\%$
(\hyperref[LHS_ECON]{Table \ref*{LHS_ECON}}, column 4), 
but this 
works against the indirect channel as higher unemployment is associated with lower house price growth. 
In addition, \cite{Kumar2019} find that Texas GDP was unaffected by this law 
change.
%

%\textbf{\textcolor{red}{
		While the falsification tests address concerns about observed economic impact 
		of the law change, estimates in border cities help control for unobserved economic impact 
		to the extent that a border city (such as Texarkana) is one economy.
Estimates in the border city samples 
(\hyperref[AlternativeBorder]{Table \ref*{AlternativeBorder}})
are %shown in the main analysis to be 
positive and statistically significant, providing additional support in favor of the direct channel.
%}}

%The set of all 
Various other
concerns regarding the 
%channels
mechanism behind 
the
treatment effect are considered and addressed below: 
\begin{enumerate}
	\item 
	\underline{Home Improvement Loans}: if Texans used HELs to improve their homes,
	the rise in house prices might be due to the higher quality of the properties 
	and not due to the demand for HELs. 
	\\
	A: Home improvement loans were available 
	before 1998 (Section \ref{Institution}). 
%	In fact, home improvement loans were the only option 
%	for equity extraction before Proposition 8.
	\item 
	\underline{Piggy-Back Loans}: HELs could have increased 
	purchase mortgage debt capacity for households that used 
	second liens (``piggy-back mortgages") to avoid mortgage insurance and obtain bigger 
	loans \citep{Lee}.     %, Zevelev2018	 
	\\
	A: This was not a relevant factor in Texas because of the 80\% LTV 
	limit for all liens obtained after purchase.
\end{enumerate}

Together, %these 
the falsification tests and border city 
estimates suggest that the law change did not 
have a significant impact on variables related to house prices,
providing evidence that the treatment effect occurred mainly 
through the direct channel.
%\textcolor{red}{Since it is impossible to know for certain 
%	whether the law affected house prices through 
%	other unobserved variables, these results provide indicative 
%	rather than definitive proof of the mechanism. 
%}

%%%%%%%%%%%%%%%%%%%%%%%%%%%%%%%%%%%%%%%%%%%%%%%%%%%%%%
%%%%%%%%%%%%%%%%%%%%%%%%%%%%%%%%%%%%%%%%%%%%%%%%%%%%%%
%\newpage
%%%%%%%%%%%%%%%%%%%%%%%%%%%%%%%%%%%%%%%%%%%%%%%%%%%%%%
\subsection{The marginal buyer} %Different Outcome Variables
\label{Main_MarginalBuyer}
This section investigates whether the law change affected homeownership in Texas. 
The law change created a new benefit for homeownership: 
the option to extract equity without selling the home.
%Since an owner can pledge the property as collateral, but a renter cannot, 
Hence, if households value this option, and 
if house prices were held constant, demand for owner-occupied housing would be 
expected to rise. 
In equilibrium, the rise in demand 
should have raised house prices until the marginal buyer was indifferent
between owning and renting.
The same logic applies to potential owners deciding whether 
to live in Texas (or a nearby state).

Estimates presented in 
\hyperref[LHS_ECON]{Table \ref*{LHS_ECON}},  columns 2 and 5--7,
show the law did not lead to   economically or statistically significant
changes in  population, %the 
MSA-level homeownership rates,  household-level homeownership (using AHS survey data),
%\textcolor{red}{Using Census data at the MSA level and AHS survey data at the household level}, 
or single-family building permits.
In addition, \cite{Kumar}  finds no evidence of migration using  
IRS tax return data.
These estimates provide evidence that the 
rise in house prices was  
sufficient to offset the rise in demand for ownership, 
keeping the marginal buyer indifferent between renting and owning.

%%%%%%%%%%%%%%%%%%%%%%%%%%%%%%%%%%%%%%%%%%%%%%%%%%%%%%
%%%%%%%%%%%%%%%%%%%%%%%%%%%%%%%%%%%%%%%%%%%%%%%%%%%%%%
%\newpage
%%%%%%%%%%%%%%%%%%%%%%%%%%%%%%%%%%%%%%%%%%%%%%%%%%%%%%
\subsection{External validity and discussion} 
\label{External}
This paper finds that after-purchase collateral service flows 
had a positive impact on the price of owner-occupied housing 
in Texas (over 500 treated zip codes) after 1998.
The conceptual framework predicts that collateral service flows 
should have a positive impact on the prices
of other assets, in other locations, at other times. 
In particular, the effect is likely higher for housing in other states, 
because Texas is the only state with an 80\% limit on home equity extraction.
More generally, collateral service flows for purchase loans 
are expected to have a positive effect on asset prices as well. 
Unfortunately, it is notoriously difficult to empirically identify 
the impact of purchase mortgage leverage on house prices 
due to simultaneity: 
on the one hand, a rise in purchase mortgage credit supply 
can raise housing demand and thus house prices; 
on the other hand, higher house prices require potential homebuyers 
to get bigger purchase mortgages.

There is little empirical work to compare the estimate to, because it is unusual 
to find a setting where it was illegal to pledge an asset as collateral 
after purchase. 
However, there is a growing literature in quantitative macroeconomics 
that seeks to understand whether households are liquidity-constrained 
and the role of illiquid housing wealth.
%
%Within that literature, the analysis in \cite{Gorea} 
%is the most closely related to this paper.  
%
In particular, \cite{Gorea} observe that housing is an important component 
of wealth for American households\footnote{In the 
Survey of Consumer Finances, about 70\% of U.S. households own a home,
and housing equity accounts for about 80\% of the median homeowner's wealth.} and 
seek to quantify the extent to which housing wealth is illiquid.
They study this question using a %quantitative 
life-cycle model with uninsurable idiosyncratic
risks in which they explicitly model key institutional details 
including LTV constraints, payment-to-income (PTI) constraints, 
long-term amortizing mortgages, and transaction costs %(fixed and variable) 
of refinancing.
Their model predicts that three-quarters of homeowners are liquidity-constrained, 
in that they would be better off if they could convert housing equity into liquid wealth. 
These homeowners are willing to pay five cents, on average, for every additional dollar of
liquidity extracted from their homes. 
Liquidity constraints increase the average marginal propensity to consume out of a transitory income windfall by about 40\%.
Their model predicts that frictions that prevent homeowners from tapping home equity are sizable.

\cite{Gorea} proceed to simulate the Texas HEL legalization in their model. 
They find that the addition of the option to extract home equity raises equilibrium 
house prices 5.5\%. 
This is within the range of the main estimates (\hyperref[US_T1]{Table \ref*{US_T1}}, columns 1--4) 
and slightly higher than the preferred estimate in the CBCP sample of 4.13\%. 
This can be explained because in their model housing is in fixed supply, hence
perfectly inelastic. 
Interestingly, they find that the addition of the option to extract home equity causes 
a small drop in the homeownership rate from 66\% to 64\%, 
whereas this paper finds a statistically insignificant rise of 0.3\%
(\hyperref[LHS_ECON]{Table \ref*{LHS_ECON}}).
%HEL legalization MPC 19.4 to 21.5

%This paper estimates the impact of collateral value on house prices 
%in Texas zip codes. 
%The heterogeneity %results
%of collateral value may also vary in times when underwriting standards are lower \citep{Mian2011}. 
%%Unfortunately there are no more discrete exogenous changes in collateral value that would make it possible to 
%%test this theory.

%%%%%%%%%%%%%%%%%%%%%%%%%%%%%%%%%%%%%%%%%%%%%%%%%%%%%%
%%%%%%%%%%%%%%%%%%%%%%%%%%%%%%%%%%%%%%%%%%%%%%%%%%%%%%
%\newpage
%%%%%%%%%%%%%%%%%%%%%%%%%%%%%%%%%%%%%%%%%%%%%%%%%%%%%%
%%%%%%%%%%%%%%%%%%%%%%%%%%%%%%%%%%%%%%%%%%%%%%%%%%%%%%%
%%%%%%%%%%%%%%%%%%%%%%%%%%%%%%%%%%%%%%%%%%%%%%%%%%%%%%%
%%\newpage
%%%%%%%%%%%%%%%%%%%%%%%%%%%%%%%%%%%%%%%%%%%%%%%%%%%%%%%
\section{Conclusion}
\label{Conclusion}
A large body of literature studies the impact 
of credit constraints on borrowing,  consumption, and investment.
This paper finds that there is also an impact on the prices 
of assets that can be pledged as collateral. 
Estimates using zip code data show that
the HEL legalization raised Texas house prices 4.13\%.
%Texas house prices rose 
%an extra 
%3.8\% %over 5-7 years 
%after 
%the option to pledge a home as collateral was legalized.
%the laws. 
If households fear they will be credit-constrained, 
they should value 
assets that facilitate their future ability to borrow. 
Hence, 
the treatment effect estimated in this paper
can be interpreted as price-based evidence 
that households are credit-constrained
and value HELs to facilitate future consumption smoothing.
%are credit-constrained.

Prices rose more in locations with inelastic housing supply, 
higher prelaw income, lower unemployment, and higher house price levels.
This reveals that %wealthier 
richer households 
value the option to pledge their home as collateral more strongly. 
This heterogeneity can be explained by 
(1) the greater tax shield available to rich households,
(2) the tendency of rich households to be more financially literate and aware of these financial products, and 
(3) the tendency of rich households to be more likely to qualify for HELs 
as they tend to have better credit and more stable income. 
The law change did not affect 
variables known to be related to house prices such as 
rent, population, and income. %  and employment. 
There was a small but statistically significant rise in the unemployment rate, 
which works against the indirect channel.
%is driving the treatment effect. 
Moreover, the border and border city samples, %Texarkana
which help control for local unobserved heterogeneity, 
provide further evidence that the effect was direct.  
These results indicate  that the treatment effect was mainly driven by 
the direct channel. 
%
%If the HEL legalization raised house prices through another unobserved channel, 
%that would affect the identification of the mechanism not the main result.
%
Finally, the law change did not affect Texas population, 
homeownership, or building permits. 
This offers evidence that 
%the rise in demand for ownership 
%due to this new option was offset by 
the rise in house prices was sufficient to keep the marginal homebuyer 
indifferent between owning and renting.

There are several avenues for future work.
%%
%First, research has shown that the availability of housing collateral affects 
%risk sharing (\cite{Lustig2010}).
%%(\hyperref[Lustig]{Lustig and Van Nieuwerburgh, 2010}).
%This theory can be tested by studying if risk sharing between Texas and the 
%rest of the US increased after HELs were legalized.\footnote{I 
%	thank Stijn Van Nieuwerburgh for suggesting this.}
%%
It would be interesting to estimate how  collateral service flows 
affect the price of other assets  such as stocks and Treasury bonds.
Between 1934 and 1974, the Federal Reserve changed 
the initial margin requirement (regulation T)
for the U.S. stock market 22 times \citep{Jylha}.
%(\hyperref[Jylha]{Jylh\"{a}, 2018}). 
%Since 1974, regulation T has set the 
%minimum
%initial margin requirement for stocks to 50\%. 
%However, finding a good source of exogenous variation to identify 
%this 
A good experiment %to identify the collateral value of stocks 
would compare the same stock 
(possibly traded on different exchanges)
but where certain shares of the stock are not affected by the margin requirements. 
%Finding this kind of experiment might prove difficult, if not impossible.
%
%Second, it %also 
%would be interesting for a 
%macroeconomic
%%calibrated 
%life cycle  model 
%to provide an estimate of how much a household would be willing to pay for the option 
%to pledge its home as collateral in the future. 
%These structural estimates could be compared with the 
%results
%%estimates 
%in this paper.
%It would be interesting for researchers working in asset pricing 
%to empirically identify how 
%stock pledgability affects stock prices. 
%
%Researchers working in asset pricing should think more about 
%how asset pledgability affects prices in addition to the assets' cash flow.
%
%In light of this literature this paper asks a new 
%question, 
%Finally, more work can be done 
It would also be interesting 
to disentangle the components of collateral 
service flows.
Loans secured by housing have two %types of 
benefits: a lower interest rate and a higher debt capacity. 
It would be helpful to separately identify %what
the fraction of collateral service flow  %of an asset %reflects 
that is due to interest rate savings compared 
to debt capacity. 

%This future research would be a valuable next step to build on
%the conclusion here: owner-occupied housing comes with 
%a valuable 
%option to pledge 
%the home as collateral in the future, and home prices reflect this. 

In conclusion, owner-occupied housing comes with 
a valuable 
option to pledge 
the home as collateral in the future. 
%Prices reflect this. 
The legalization of HELs in Texas 
%This paper 
provides evidence that house prices reflect this.

%Asset pricing  papers 
%Consumption 
%Tenure Choice
%
%
%Future work compare the predictions to macro models.

%This paper has implications for the 
%%buy vs rent 
%tenure choice
%decision. 
%On the one hand, laws that legalize HELs, HELOCs and reverse mortgages 
%create an additional reason to become a homeowner because it comes with 
%these valuable real options. 
%On the other hand, if prices reflect this option value the marginal owner should be 
%indifferent. 

%\newpage
%\bibliographystyle{jf}
%\bibliography{master}
%\addcontentsline{toc}{section}{Bibliography}
%{\small	\bibliography{Growth.bib} }

\newpage
\section*{}
%  \bibliographystyle{apalike}
%\nocite{*}
\bibliographystyle{jf}
\bibliography{AZ_master}
%\addcontentsline{toc}{section}{Bibliography}
%{\small	\bibliography{Growth.bib} }

%  \bibliographystyle{apalike}
%  \bibliography{bibfile}
%\newpage   
%\begin{thebibliography}{} %{9}
%
%\end{thebibliography}

\newpage
\begin{appendices}

\section{Appendix A. Figures}

%\newpage
\subsection{Percent Change in Real House Prices}
%\medbreak and \bigbreak

\newpage
%\newgeometry{left=1.5cm, right=1.5cm, bottom=0.0cm} 	
%\newpage
%\restoregeometry
\subsection{Pretrends}
%\newpage
%\restoregeometry

\newgeometry{bottom=1.50cm, top=1.50cm} 	
\newpage
\subsection{Synthetic Controls}
%%%%%%%%%%%%%%%%%%%
\restoregeometry

%\newpage
%\subsection{Treatment Effect by MSA Elasticity}

\newpage
\subsection{Supply Elasticity Theory}
%%%%%%%%%%%%%%%%%%%

\newpage
\subsection{Elasticity Fitted Values}

\newpage
\section{Appendix B. Tables}
%%%%%%%%%%%%%%%%%%%%%%%%%%%%%%%%%%%%%%%%%%%%%%%%%%%%%%
%\newgeometry{left=1.5cm, right=1.5cm, bottom=1.0cm, top=0.50cm} 	
%\newpage
%\newgeometry{left=1.5cm, right=1.5cm, bottom=0.0cm} 	
%\newpage
%\newpage

%\newgeometry{ bottom=1.5cm, top=1.5cm} 	
%\newpage
%\newpage
% \begin{landscape}
\subsection{Summary Statistics}
\begin{table}[!htb]%[center]
	\centering \small  \caption{Summary Statistics} \label{Summary}
{\footnotesize  %\scriptsize
\begin{tabular}{lllllllll} %{lcccccccc}
	\hline \noalign{\smallskip}
	Sample: 
	& \multicolumn{3}{l}{Pre-1998} 
	& \multicolumn{3}{l}{Post-1998} 
	%& \multicolumn{2}{c}{} 
	%& \multicolumn{2}{c}{}
	\\
	Variables & Texas & Control & Difference & Texas & Control & Difference &  & \\
	\noalign{\smallskip}\hline \noalign{\smallskip}LTV (\%) & 83.17 & 79.29 & 3.88* & 77.57 & 81.52 & –3.95*** &  & \\
	& \begin{footnotesize}(1.46)\end{footnotesize} & \begin{footnotesize}(0.73)\end{footnotesize} & \begin{footnotesize}(1.50)\end{footnotesize} & \begin{footnotesize}(0.99)\end{footnotesize} & \begin{footnotesize}(0.49)\end{footnotesize} & \begin{footnotesize}(0.40)\end{footnotesize} & \begin{footnotesize}\end{footnotesize} & \begin{footnotesize}\end{footnotesize}\\
	\noalign{\smallskip}Interest Rate (\%) & 7.58 & 7.58 & 0.00 & 6.80 & 6.77 & 0.03 &  & \\
	& \begin{footnotesize}(0.17)\end{footnotesize} & \begin{footnotesize}(0.08)\end{footnotesize} & \begin{footnotesize}(0.16)\end{footnotesize} & \begin{footnotesize}(0.28)\end{footnotesize} & \begin{footnotesize}(0.14)\end{footnotesize} & \begin{footnotesize}(0.04)\end{footnotesize} & \begin{footnotesize}\end{footnotesize} & \begin{footnotesize}\end{footnotesize}\\
%	\noalign{\smallskip}House Price Growth (\%) & 2.68 & 4.61 & -1.93*** & 4.24 & 3.85 & 0.39 &  & \\
%	& \begin{footnotesize}(0.07)\end{footnotesize} & \begin{footnotesize}(0.09)\end{footnotesize} & \begin{footnotesize}(0.32)\end{footnotesize} & \begin{footnotesize}(0.05)\end{footnotesize} & \begin{footnotesize}(0.06)\end{footnotesize} & \begin{footnotesize}(0.24)\end{footnotesize} & \begin{footnotesize}\end{footnotesize} & \begin{footnotesize}\end{footnotesize}\\
	\noalign{\smallskip}Real House Price (\$000) & 82.01 & 74.96 & 7.05 & 92.78 & 81.08 & 11.70* &  & \\
	& \begin{footnotesize}(3.83)\end{footnotesize} & \begin{footnotesize}(1.91)\end{footnotesize} & \begin{footnotesize}(4.73)\end{footnotesize} & \begin{footnotesize}(3.17)\end{footnotesize} & \begin{footnotesize}(1.58)\end{footnotesize} & \begin{footnotesize}(4.71)\end{footnotesize} & \begin{footnotesize}\end{footnotesize} & \begin{footnotesize}\end{footnotesize}\\
	\noalign{\smallskip}Real Rent (\$) & 482.67 & 488.56 & –5.88 & 502.77 & 502.96 & –0.19 &  & \\
	& \begin{footnotesize}(5.16)\end{footnotesize} & \begin{footnotesize}(1.38)\end{footnotesize} & \begin{footnotesize}(22.78)\end{footnotesize} & \begin{footnotesize}(4.56)\end{footnotesize} & \begin{footnotesize}(1.22)\end{footnotesize} & \begin{footnotesize}(20.65)\end{footnotesize} & \begin{footnotesize}\end{footnotesize} & \begin{footnotesize}\end{footnotesize}\\
	\noalign{\smallskip}Population Growth (\%) & 1.74 & 1.08 & 0.66** & 1.62 & 0.93 & 0.68*** &  & \\
	& \begin{footnotesize}(0.07)\end{footnotesize} & \begin{footnotesize}(0.07)\end{footnotesize} & \begin{footnotesize}(0.22)\end{footnotesize} & \begin{footnotesize}(0.09)\end{footnotesize} & \begin{footnotesize}(0.08)\end{footnotesize} & \begin{footnotesize}(0.09)\end{footnotesize} & \begin{footnotesize}\end{footnotesize} & \begin{footnotesize}\end{footnotesize}\\
	\noalign{\smallskip}Real Income Per Cap (\$) & 12,585.55 & 11,885.64 & 699.91** & 14,115.58 & 13,312.92 & 802.66*** &  & \\
	& \begin{footnotesize}(93.08)\end{footnotesize} & \begin{footnotesize}(86.11)\end{footnotesize} & \begin{footnotesize}(159.86)\end{footnotesize} & \begin{footnotesize}(105.47)\end{footnotesize} & \begin{footnotesize}(97.57)\end{footnotesize} & \begin{footnotesize}(113.15)\end{footnotesize} & \begin{footnotesize}\end{footnotesize} & \begin{footnotesize}\end{footnotesize}\\
	\noalign{\smallskip}Unemployment Rate (\%) & 6.30 & 6.38 & –0.08 & 5.41 & 5.22 & 0.19 &  & \\
	& \begin{footnotesize}(0.12)\end{footnotesize} & \begin{footnotesize}(0.11)\end{footnotesize} & \begin{footnotesize}(0.60)\end{footnotesize} & \begin{footnotesize}(0.07)\end{footnotesize} & \begin{footnotesize}(0.07)\end{footnotesize} & \begin{footnotesize}(0.41)\end{footnotesize} & \begin{footnotesize}\end{footnotesize} & \begin{footnotesize}\end{footnotesize}\\
	\noalign{\smallskip}Homeownership Rate (\%) & 56.96 & 60.88 & –3.91 & 60.80 & 64.52 & –3.72* &  & \\
	& \begin{footnotesize}(0.80)\end{footnotesize} & \begin{footnotesize}(0.90)\end{footnotesize} & \begin{footnotesize}(2.92)\end{footnotesize} & \begin{footnotesize}(0.71)\end{footnotesize} & \begin{footnotesize}(0.79)\end{footnotesize} & \begin{footnotesize}(1.43)\end{footnotesize} & \begin{footnotesize}\end{footnotesize} & \begin{footnotesize}\end{footnotesize}\\
	\noalign{\smallskip}Single-Family Permits & 798.46 & 313.24 & 485.22*** & 1,304.91 & 400.52 & 904.39*** &  & \\
	& \begin{footnotesize}(55.69)\end{footnotesize} & \begin{footnotesize}(52.80)\end{footnotesize} & \begin{footnotesize}(42.80)\end{footnotesize} & \begin{footnotesize}(88.26)\end{footnotesize} & \begin{footnotesize}(83.68)\end{footnotesize} & \begin{footnotesize}(53.88)\end{footnotesize} & \begin{footnotesize}\end{footnotesize} & \begin{footnotesize}\end{footnotesize}\\
	\noalign{\smallskip}Saiz Supply Elasticity & 3.06 & 3.60 & –0.54 & 3.06 & 3.60 & –0.54 &  & \\
	& \begin{footnotesize}(0.43)\end{footnotesize} & \begin{footnotesize}(0.47)\end{footnotesize} & \begin{footnotesize}(0.36)\end{footnotesize} & \begin{footnotesize}(0.43)\end{footnotesize} & \begin{footnotesize}(0.47)\end{footnotesize} & \begin{footnotesize}(0.36)\end{footnotesize} & \begin{footnotesize}\end{footnotesize} & \begin{footnotesize}\end{footnotesize}\\
	\noalign{\smallskip}\hline\end{tabular}
}
%\small  \caption{Summary Statistics} % 
\medskip
\begin{minipage}{0.9\textwidth}
	{\footnotesize 
		%all data is from the restricted sample used in the analysis. 
		This table presents summary statistics for the treatment group (Texas)
		and control group (border states) six years before and after the law change.
		The standard deviation of each variable is in parentheses below 
		the average. 
		Tests for differences between treatment and control groups using robust standard errors
		as described in the paper are also reported. 
		Nominal variables are adjusted for inflation, as explained in Section \ref{Empirical}. 
		Supply elasticity, as estimated by 	\cite{Saiz2010}, does not vary over time in
		the sample.
		%\\  
		Data sources can be found in \hyperref[data_table]{Table \ref*{data_table}}.
		%\\
		* p$<$.1; ** p$<$.05; *** p$<$.01.
		\par}
\end{minipage}
\end{table} 
%\end{landscape}

%\restoregeometry

\newpage
\subsection{Main}
\begin{table}[!htb]%[center]
	\centering
%\small \caption{US Sample}
\small \caption{Impact of HEL Legalization on Texas House Prices}
\label{US_T1}
\scalebox{.875}
%{\scriptsize \input  {../../4_output_analysis/main/US_NoLagOilInt.tex}	 }
{\scriptsize 	%\documentclass[]{article}\pagestyle{empty}\begin{document}
%\begin{center}
\begin{tabular}{lcccccccc}
\hline \noalign{\smallskip}
 & (1) & (2) & (3) & (4) & (5) & (6) & (7) & (8) \\
Variables & United States & Border State & Border & CBCP & United States & Border State & Border & CBCP\\
\noalign{\smallskip}\hline \noalign{\smallskip}TexasPost & 0.0350*** & 0.0616** & 0.0476** & 0.0413*** &  &  &  & \\
 & \begin{footnotesize}(0.0099)\end{footnotesize} & \begin{footnotesize}(0.0221)\end{footnotesize} & \begin{footnotesize}(0.0151)\end{footnotesize} & \begin{footnotesize}(0.0126)\end{footnotesize} & \begin{footnotesize}\end{footnotesize} & \begin{footnotesize}\end{footnotesize} & \begin{footnotesize}\end{footnotesize} & \begin{footnotesize}\end{footnotesize}\\
\noalign{\smallskip}Texas95 &  &  &  &  & -0.0027 & 0.0155 & 0.0096 & 0.0159\\
 & \begin{footnotesize}\end{footnotesize} & \begin{footnotesize}\end{footnotesize} & \begin{footnotesize}\end{footnotesize} & \begin{footnotesize}\end{footnotesize} & \begin{footnotesize}(0.0117)\end{footnotesize} & \begin{footnotesize}(0.0131)\end{footnotesize} & \begin{footnotesize}(0.0108)\end{footnotesize} & \begin{footnotesize}(0.0107)\end{footnotesize}\\
\noalign{\smallskip}Texas96 &  &  &  &  & 0.0009 & -0.0029 & 0.0027 & 0.0101\\
 & \begin{footnotesize}\end{footnotesize} & \begin{footnotesize}\end{footnotesize} & \begin{footnotesize}\end{footnotesize} & \begin{footnotesize}\end{footnotesize} & \begin{footnotesize}(0.0129)\end{footnotesize} & \begin{footnotesize}(0.0142)\end{footnotesize} & \begin{footnotesize}(0.0110)\end{footnotesize} & \begin{footnotesize}(0.0085)\end{footnotesize}\\
\noalign{\smallskip}Texas98 &  &  &  &  & 0.0126 & 0.0498*** & 0.0322** & 0.0260\\
 & \begin{footnotesize}\end{footnotesize} & \begin{footnotesize}\end{footnotesize} & \begin{footnotesize}\end{footnotesize} & \begin{footnotesize}\end{footnotesize} & \begin{footnotesize}(0.0147)\end{footnotesize} & \begin{footnotesize}(0.0163)\end{footnotesize} & \begin{footnotesize}(0.0111)\end{footnotesize} & \begin{footnotesize}(0.0166)\end{footnotesize}\\
\noalign{\smallskip}Texas99 &  &  &  &  & 0.0298*** & 0.0462*** & 0.0448** & 0.0283*\\
 & \begin{footnotesize}\end{footnotesize} & \begin{footnotesize}\end{footnotesize} & \begin{footnotesize}\end{footnotesize} & \begin{footnotesize}\end{footnotesize} & \begin{footnotesize}(0.0103)\end{footnotesize} & \begin{footnotesize}(0.0110)\end{footnotesize} & \begin{footnotesize}(0.0140)\end{footnotesize} & \begin{footnotesize}(0.0146)\end{footnotesize}\\
\noalign{\smallskip}Texas00 &  &  &  &  & 0.0421*** & 0.0388** & 0.0429* & 0.0462***\\
 & \begin{footnotesize}\end{footnotesize} & \begin{footnotesize}\end{footnotesize} & \begin{footnotesize}\end{footnotesize} & \begin{footnotesize}\end{footnotesize} & \begin{footnotesize}(0.0134)\end{footnotesize} & \begin{footnotesize}(0.0136)\end{footnotesize} & \begin{footnotesize}(0.0155)\end{footnotesize} & \begin{footnotesize}(0.0128)\end{footnotesize}\\
\noalign{\smallskip}Texas01 &  &  &  &  & 0.0468*** & 0.0749*** & 0.0571** & 0.0533**\\
 & \begin{footnotesize}\end{footnotesize} & \begin{footnotesize}\end{footnotesize} & \begin{footnotesize}\end{footnotesize} & \begin{footnotesize}\end{footnotesize} & \begin{footnotesize}(0.0122)\end{footnotesize} & \begin{footnotesize}(0.0139)\end{footnotesize} & \begin{footnotesize}(0.0133)\end{footnotesize} & \begin{footnotesize}(0.0127)\end{footnotesize}\\
\noalign{\smallskip}Texas02 &  &  &  &  & 0.0407*** & 0.0871*** & 0.0587*** & 0.0583***\\
 & \begin{footnotesize}\end{footnotesize} & \begin{footnotesize}\end{footnotesize} & \begin{footnotesize}\end{footnotesize} & \begin{footnotesize}\end{footnotesize} & \begin{footnotesize}(0.0130)\end{footnotesize} & \begin{footnotesize}(0.0143)\end{footnotesize} & \begin{footnotesize}(0.0117)\end{footnotesize} & \begin{footnotesize}(0.0104)\end{footnotesize}\\
\noalign{\smallskip}$N$ & 143,637 & 11,310 & 1,365 & 2,041 & 143,637 & 11,310 & 1,365 & 2,041\\
$R^2$ & \begin{footnotesize}0.9771\end{footnotesize} & \begin{footnotesize}0.9825\end{footnotesize} & \begin{footnotesize}0.9814\end{footnotesize} & \begin{footnotesize}0.9848\end{footnotesize} & \begin{footnotesize}0.9772\end{footnotesize} & \begin{footnotesize}0.9833\end{footnotesize} & \begin{footnotesize}0.9820\end{footnotesize} & \begin{footnotesize}0.9847\end{footnotesize}\\
\noalign{\smallskip}\hline\end{tabular} %\\
%\end{center}
%\end{document}
}
\medskip
\begin{minipage}{1\textwidth}
	{\footnotesize 
This table reports estimates of the effect of a law change 
		(which legalized HELs in Texas) on house prices in geographically nested samples. 
		Each column reports
		a separate regression estimated at the zip code year level 
		where the dependent variable is the
		log of the real house price index. 
		In columns 1–4, coefficients are reported for 
		the interaction of the Texas dummy with an indicator for
		whether the year of observation falls on or after 1998.
In columns 5–8, coefficients are reported for 
%the 
interactions of the Texas dummy with year indicators.
All specifications include zip code and year fixed effects,  
state time trends, and national oil prices interacted with MSA dummies.
Specifications using the CBCP sample also include 
contiguous border county  pair by year fixed effects.
%Data are weighted by the inverse as described in the paper
	Standard errors, computed using methods described in Section \ref{Main},
	 are reported in parentheses.
%	clustered at the state level
%	
	Data sources can be found in \hyperref[data_table]{Table \ref*{data_table}}.
	* p$<$.1; ** p$<$.05; *** p$<$.01.
		\par}
\end{minipage}

\end{table}

	\newpage
\subsection{Specification Robustness}
\begin{table}[H] %[!htb]%[center]
		\centering
	%	%\small \caption{US Sample}
\small \caption{Impact of HEL Legalization on Texas House Prices, 
	Specification Robustness}
\label{SE_Specification}	
\small \text{ } %\textbf{Panel A: Robustness to covariates and outcome}
\newline %\newline   %\\ \\
	\scalebox{1}
	{ %\footnotesize %      \scriptsize 
		%\documentclass[]{article}\pagestyle{empty}\begin{document}
%\begin{center}
\begin{tabular}{lcccc}
\hline \noalign{\smallskip}
& (1) & (2) & (3) & (4)  \\
Variables & United States & Border State & Border & CBCP
%\\
%\noalign{\smallskip}\hline \noalign{\smallskip}
\\
\noalign{\smallskip}
\hline
\\
\begin{footnotesize}A.\end{footnotesize}
& 
\multicolumn{3}{l}{Outcome: Log real house price index} 
%\underline{%}
% $Y=log(HP)$
\\
\noalign{\smallskip}
\hline
%\\ 
\noalign{\smallskip}
Baseline & 0.0350*** & 0.0616** & 0.0476** & 0.0413***\\
 & \begin{footnotesize}(0.0099)\end{footnotesize} & \begin{footnotesize}(0.0221)\end{footnotesize} & \begin{footnotesize}(0.0151)\end{footnotesize} & \begin{footnotesize}(0.0126)\end{footnotesize}\\
\noalign{\smallskip}Covariates & 0.0330*** & 0.0558** & 0.0480** & 0.0499***\\
 & \begin{footnotesize}(0.0086)\end{footnotesize} & \begin{footnotesize}(0.0196)\end{footnotesize} & \begin{footnotesize}(0.0166)\end{footnotesize} & \begin{footnotesize}(0.0159)\end{footnotesize}
 \\
\noalign{\smallskip}$N$ & 143,637 & 11,310 & 1,365 & 2,041
\\
\noalign{\smallskip}
\hline
\\
\begin{footnotesize}B. \end{footnotesize}
& 
\multicolumn{3}{l}{Outcome: Real house price growth} 
%\underline{%}
% $Y=log(HP)$
\\
\noalign{\smallskip}
\hline
\noalign{\smallskip}Baseline & 0.0389*** & 0.0563*** & 0.0317** & 0.0409***\\
 & \begin{footnotesize}(0.0071)\end{footnotesize} & \begin{footnotesize}(0.0079)\end{footnotesize} & \begin{footnotesize}(0.0110)\end{footnotesize} & \begin{footnotesize}(0.0124)\end{footnotesize}\\
\noalign{\smallskip}Covariates & 0.0374*** & 0.0539*** & 0.0288** & 0.0375***\\
 & \begin{footnotesize}(0.0055)\end{footnotesize} & \begin{footnotesize}(0.0065)\end{footnotesize} & \begin{footnotesize}(0.0100)\end{footnotesize} & \begin{footnotesize}(0.0125)\end{footnotesize}
 \\
\noalign{\smallskip}$N$ & 141,986 & 11,185 & 1,347 & 1,808
\\
\noalign{\smallskip}\hline\end{tabular}%\\
%\end{center}
%\end{document}

		% {../../4_output_analysis/main/Robust1.tex}	 
%		\input  {../../4_output_analysis/main/US-Spec-Cluster-StateZip5OLS.tex}	 
	}
\newline \newline 
%\bigskip  % \\ \\
\begin{minipage}{.95\textwidth}
	{\footnotesize 
		This table reports estimates of the effect of a law change 
		(which legalized HELs in Texas) on house prices in %the 
		geographically nested samples. 
		Each entry reports
		a separate regression estimated at the zip code year level 
		where the dependent variable is the
		log of the real house price index in panel A 
		or
		real house price growth in panel B. 
		All specifications include zip code and year fixed effects,  
		state time trends, and national oil prices interacted with MSA dummies.
		Specifications using the CBCP sample also include 
		contiguous border county  pair by year fixed effects.
Specifications in rows labeled ``Covariates'' include
 U.S. interest rates interacted with state dummies, real income per capita, and
 population. % as described in the bottom row. 	
Standard errors, computed using the methods described in Section \ref{Main},
are reported in parentheses.
		Data sources can be found in \hyperref[data_table]{Table \ref*{data_table}}.
		* p$<$.1; ** p$<$.05; *** p$<$.01.
%estimator 
%The estimation
%is implemented using the xtabond2 package in 
%STATA (\hyperref[Roodman]{Roodman, 2009}).
		\par}
\end{minipage}

\end{table}

	\newpage
\subsection{Border City Samples}
\begin{table}[H] %[!htb]%[center]
	\centering
	%	%\small \caption{US Sample}
	\small \caption{Impact of HEL Legalization on Texas House Prices in Border City Samples}
	\label{AlternativeBorder}
	%		\scalebox{1}  %NO SCALEBOX
	%	\small 
	%	\caption{US Sample: The effect of Proposition 8 on Texas House Prices}
	%\text{Sample: US}
	%	\multicolumn{7}{l}{}  %\        \midrule
	{ %\scriptsize 
		
\begin{tabular}{lcccc} \hline
 & (1) & (2) & (3) & (4) \\
Variables & Texarkana & DFWBorder & ElPasoLasCruces & Texoma \\ \hline
 &  &  &  &  \\
Treated\_post1998 & 0.0588** & 0.0405** & 0.0597*** & 0.0375** \\
 & (0.0091) & (0.0131) & (0.0133) & (0.0126) \\
 &  &  &  &  \\
\textit{N} & 39 & 65 & 195 & 78 \\
$R^2$ & 0.961 & 0.994 & 0.996 & 0.992 \\ \hline
%\multicolumn{5}{c}{ Robust standard errors in parentheses} \\
%\multicolumn{5}{c}{ *** p$<$0.01, ** p$<$0.05, * p$<$0.1} \\
\end{tabular}

		% {../../4_output_analysis/main/MainLocal.tex}
	}
	\medskip
	%  \\ \\
	\begin{minipage}{.95\textwidth}
		{\footnotesize 
			This table reports estimates of the effect of a law change 
			(which legalized HELs in Texas) on house prices in 
			various border city samples.
			Column 1 uses data from the Texarkana MSA,
			column 2 uses data from the Dallas-Fort Worth (DFW) CSA, 
			column 3 uses data from the El Paso Las Cruces CSA, 
			and column 4 uses data from the Texoma area.
			Each column reports
			a separate regression estimated at the zip code year level 
			where the dependent variable is the
			log of the real house price index. 
			In columns 1–4, coefficients are reported for 
			the interaction of the Texas dummy with an indicator for
			whether the year of observation falls on or after 1998.
All specifications include zip code and year fixed effects,  
state time trends, and national oil prices interacted with MSA dummies.
Specifications using the CBCP sample also include 
contiguous border county  pair by year fixed effects.
			%Data are weighted
			%data is weighed by the inverse as described in the paper
			Standard errors are 
			%	clustered at the state level
			%	and 
			reported in parentheses.
			Data sources can be found in \hyperref[data_table]{Table \ref*{data_table}}.
			* p$<$.1; ** p$<$.05; *** p$<$.01.
			%Some specifications control for US oil prices and interest rates 
			%interacted with state dummies as described in the bottom row. 	
			%
			%Law change assoc w/ 100*(.0381)% = 3.81% rise in real house prices.
			%LDV: A 1% change in prices last year is assoc with a 0.88\%  change this year (Chinco, Mayer)   
			%\eta_{k} = .0337 , price change between 97 and year "k" was 3.37% higher in TX rel to control group, 
			%
			%		Significance levels 10\%, 5\%, and 1\% are denoted by *, **, and ***, respectively.
			\par}	
	\end{minipage}
	
\end{table}

\newpage
\subsection{Treatment Effect Heterogeneity}
\begin{table}[!htb]%[center]
		\centering
	%\label{Texarkana_static} %\ref{Texarkana_static}
	%	\small
	%	\caption{\bf{Texarkana Static}}
	\small
	\caption{Impact of HEL Legalization on Texas House Prices, Treatment Effect Heterogeneity}
	\label{US_DDD}
	%	\label{Texarkana_static} %\ref{Texarkana_static}
	%	\scalebox{1.00}{
	%{\scriptsize }
	%{\footnotesize }
	%\tiny
	\scalebox{1}
{\scriptsize 
\begin{tabular}{lcccc} \hline
 & (1) & (2) & (3) & (4) \\
Variables &  &  &  &  \\ \hline
 &  &  &  &  \\
$\text{Texas} \times \text{Post}$ & 0.063*** & -1.002*** & 0.080*** & -0.367*** \\
 & (0.011) & (0.246) & (0.015) & (0.122) \\
$\text{Texas} \times \text{Post} \times \text{Elasticity} $ & -0.009** &  &  &  \\
 & (0.004) &  &  &  \\
$\text{Texas} \times \text{Post} \times \text{LogRealIncomePre} $ &  & 0.109*** &  &  \\
 &  & (0.026) &  &  \\
$\text{Texas} \times \text{Post} \times \text{UnemploymentRatePre} $ &  &  & -0.007*** &  \\
 &  &  & (0.002) &  \\
$\text{Texas} \times \text{Post} \times \text{LogRealMedianHousePricesPre} $ &  &  &  & 0.036*** \\
 &  &  &  & (0.011) \\
 &  &  &  &  \\
\textit{N} & 111,748 & 142,077 & 143,676 & 132,431 \\
$R^2$ & 0.651 & 0.616 & 0.611 & 0.626 \\
Max $ H_{i} $ & 4.749 & 9.949 & 19.95 & 13.77 \\
Min $ H_{i} $ & 1.18 & 8.936 & 1.95 & 10.16 \\
TE Max $ H_{i} $ & .0201 & .0803 & -.0619 & .131 \\
TE Min $ H_{i} $ & .0525 & -.03 & .066 & .000534 \\
% std-err & state & state & state & state \\ 
 \hline
%\multicolumn{5}{c}{ Robust standard errors in parentheses} \\
%\multicolumn{5}{c}{ *** p$<$0.01, ** p$<$0.05, * p$<$0.1} \\
\end{tabular}
	 }
%{\scriptsize \input  {../../4_output_analysis/main/US_DDD.tex}	 }	
%	{\scriptsize \input{../../4_output_analysis/main/US_ECON.tex}}
\medskip

%*AZ: NOTE TO SELF, put max and min treatment at bottom of each column!

\begin{minipage}{0.9\textwidth}
	{\footnotesize 
%Treatment assoc w/ 6.63% rise in HP in TX rel to control,  -17.7%, etc  IF H=0
%1 unit higher elasticity, assoc w/ -1.01% lower TE
%1%higher HPI92 asso w/ $.0482\%$ higher TE	
%1%higher Pop92 asso w/ $.00914\%$ higher TE		
%1%higher Inc92 asso w/ $.119\%$ higher TE		
%1%higher UW92 assoc w/ $100*-.00892\% = -.892%$ higher TE		
		This table reports estimates of the effect of a law change 
		(which legalized HELs in Texas) on house prices. 
Each column reports
a separate regression	in which	
the treatment effect is allowed to vary based on 
%five 
four measures of 
heterogeneity: supply elasticity, 
prelaw 
%log population, 
log real income per capita,
%and 
%the 
unemployment, %rate
and
log real median house prices. 
Prelaw variables are set equal to their 
average before 1998. 
%value in 1992, the first 
%year of the sample. 
%the homeowner’s income as reported		
%		Each column reports
%		a separate regression estimated at the zip code year level 
%		where 
		The dependent variable is the
		log of the real house price index. 
%
%coefficients are reported 
The first row reports coefficients
for 
the interaction of the Texas dummy with an indicator for
whether the year of observation falls on or after 1998.
This represents the treatment effect if the measure of heterogeneity 
is equal to zero.
% H=0.
%
The %second to seventh 
remaining rows 
report the coefficient
on the triple interaction between 
the Texas dummy, 
an indicator for the treatment period,
and 
one of four measures of heterogeneity. % (and its square for median prices). 
All specifications include zip code and year fixed effects,  
state time trends, and national oil prices interacted with MSA dummies.
Specifications using the CBCP sample also include 
contiguous border county  pair by year fixed effects.
Standard errors are clustered at the state level
and are reported in parentheses.
When interpreting the parameters in this table,
take into account that some measures of heterogeneity are in logs while others are not. 
Data sources can be found in \hyperref[data_table]{Table \ref*{data_table}}.
* p$<$.1; ** p$<$.05; *** p$<$.01.
		\par}
\end{minipage}

\end{table}
%%%%%%%%%%%%%%%%%%%

%%%%%%%%%%%%%%%%%%%
\newpage
\subsection{Other Outcome Variables: Falsification, Channels,  Marginal Buyer}
%Related to House Prices
\begin{table}[!htb]%[center]
	\centering
	%\label{Texarkana_static} %\ref{Texarkana_static}
	%	\small
	%	\caption{\bf{Texarkana Static}}
	\small
	\caption{Impact of HEL Legalization on %Texas House Prices 
		Other Outcome Variables}
	\label{LHS_ECON}
	%{\scriptsize }
	%{\footnotesize }
	%\tiny
{\scriptsize 
\begin{tabular}{lccccccc} \hline
 & (1) & (2) & (3) & (4) & (5) & (6) & (7) \\
 &  &  & Log &  & Home & Home & Log \\
 & Log & Log & Real Income & Unemployment & Ownership & Owner & Single Family \\
Variables & Real Rent & Population & Per Capita & Rate & Rate & AHS & Permits \\ \hline
 &  &  &  &  &  &  &  \\
$\text{Texas} \times \text{Post}$ & 0.003 & 0.000 & -0.001 & 0.003** & 0.003 & 0.004 & 0.048 \\
 & (0.005) & (0.007) & (0.005) & (0.001) & (0.004) & (0.004) & (0.033) \\
 &  &  &  &  &  &  &  \\
\textit{N} & 16,276 & 40,690 & 40,404 & 21,606 & 947 & 171,966 & 36,614 \\
$R^2$ & 0.303 & 0.254 & 0.634 & 0.391 & 0.442 & 0.036 & 0.063 \\
% std-err & state & state & state & state & state & state & state \\ 
\hline
%\multicolumn{8}{c}{ Robust standard errors in parentheses} \\
%\multicolumn{8}{c}{ *** p$<$0.01, ** p$<$0.05, * p$<$0.1} \\
\end{tabular}
	 }	
%	{\scriptsize \input{../../4_output_analysis/main/LHS.tex}}

\medskip

\begin{minipage}{.9\textwidth}
	{\footnotesize 
		This table reports estimates of the effect of a law change 
		(which legalized HELs in Texas) on 
		six outcome variables: 
		Log Real Rent, 
		Log Population, 
		Log Real Income Per Capita, 
		the Unemployment Rate, 
		the Homeownership Rate,
		and
		Log Single-Family Building Permits.
		%house prices. 
		%
%		Each column reports
%		a separate regression estimated at the zip code year level. 
		%
		%In columns 1-6, 
		Coefficients are reported for 
		the interaction of the Texas dummy with an indicator for
		whether the year of observation falls on or after 1998.
		All specifications include year fixed effects
		along with location fixed effects at the most local level possible. 
%%		state time trends.
		%
		%		State
		%trends are estimated by interacting state dummies with a linear time trend.		 
		%Data are weighted
		%data is weighed by the inverse as described in the paper
		Standard errors are clustered at the state level
		and are reported in parentheses.
				Data sources can be found in \hyperref[data_table]{Table \ref*{data_table}}.
				* p$<$.1; ** p$<$.05; *** p$<$.01.
		\par}
\end{minipage}

\end{table}

\newpage
\restoregeometry
%\section{Appendix for online publication}
	%	\chapter{Appendix 1}
	%	
	%	\chapter{Appendix 2}

\section{For Online Publication: Online Appendix Figures}
%\newpage
%\newgeometry{left=1.5cm, right=1.5cm, bottom=0.0cm} 	
%\newpage
%\restoregeometry

\subsection{Treatment Effect Heterogeneity by Zip Code}
%\newpage
%\restoregeometry
\newpage
\section{For Online Publication: Online Appendix Tables}

%\begin{sidewaystable}
%	\centering
%	\caption{Your caption here}
%	\begin{tabular}{ll}
%		First First & First Second\\
%		Second First & Second Second
%	\end{tabular}
%\end{sidewaystable}
%

%\begin{sidewaystable}
%\begin{sidewaystable}%S[!htb]%[center]
%	\centering
%	%\label{Texarkana_static} %\ref{Texarkana_static}
%	%	\small
%	%	\caption{\bf{Texarkana Static}}
%	\small
%	\caption{US Sample}
%	\label{US_T1}
%	%	\label{Texarkana_static} %\ref{Texarkana_static}
%	%{\scriptsize } %{\footnotesize }%\tiny
%\scalebox{0.5}
%	{\scriptsize \input{../../4_output_analysis/0_clean_12_25/PAPER-allstates-Clusterstate-lnrhpi-MSATrend-from1992-to-2004-Min1992.tex}
%	}
%\end{sidewaystable}

\subsection{Datasets} \label{data_appx}
%%%%%%%%%%%%%%%%%%%%%%%%%%%%%%%%%%%%%%%%%%%%%%%%%%%%%%
%\newpage
%https://www.fhfa.gov/DataTools/Downloads/Pages/House-Price-Index-Datasets.aspx

\begin{table}[!htb]%[center]
	\centering
	%\label{Texarkana_static} %\ref{Texarkana_static}
	%	\small
	%	\caption{\bf{Texarkana Static}}
	\small
	\small
	\caption{Datasets}
	\label{data_table} %\ref{Texarkana_static}
	%	\scalebox{0.90}{
	\begin{tabular}
		{|l|l|l|c|c|}
		\hline
		$ \text{Variable}$ 
		& 
		$ \text{Level}$	
		& 
		$ \text{Source}$	
		%$ \text{frequency, period}$	
		& 
		$ \text{\# Locations}$	
		& 
		$ \text{TX}$	
		%		& 
		%		$ \text{TX border}$	%near TX border
		%
		\\ 
		\hline
		\hline
		House Price Index
		%		$ \text{House Price, 5 digit zips}$
		& {5-digit zip code} 
		& 
		%$ \text{FHFA}$ 
		\href{http://www.fhfa.gov/DataTools/Downloads/pages/house-price-index-datasets.aspx}
		{FHFA}
		%		$ \text{annual, 1975-2015}$	
		& 
		$ \text{17,936 
			%5 digit 
			zips}$	
		& 
		$ \text{918 
			%5 digit 
			zips}$	
		%		& 
		%		$ \text{1,640 
		%			%5 digit 
		%			zips}$	
		\\ [.15em]
		%All-Transactions Indexes (Estimated using Sales Prices and Appraisal Data)
		\hline
		House Price Index
		%		$ \text{House Price AT, state}$ 
		& state		
		& 
		\href{http://www.fhfa.gov/DataTools/Downloads/pages/house-price-index-datasets.aspx}
		{FHFA}
		%$ \text{FHFA}$ 
		%$ \text{quarterly, 1975-2016}$	
		%$ \text{quarterly, 1975Q1-2016Q1}$	
		& 
		$ \text{51: US \& DC}$	
		& 
		$ \text{1}$	
		%		& 
		%		$ \text{5 states}$	
		\\ [.15em]
		%All-Transactions Indexes (Estimated using Sales Prices and Appraisal Data)
		\hline
		LTV (1st purchase lien)
		%		$ \text{House Price AT, state}$ 
		& state		
		& 
		\href{https://www.fhfa.gov/DataTools/Downloads/Pages/Monthly-Interest-Rate-Data.aspx}
		{FHFA}
		%$ \text{FHFA}$ 
		%$ \text{quarterly, 1975-2016}$	
		%$ \text{quarterly, 1975Q1-2016Q1}$	
		& 
		$ \text{51: US \& DC}$	
		& 
		$ \text{1}$	
		%		& 
		%		$ \text{5 states}$	
		\\ [.15em]
		%All-Transactions Indexes (Estimated using Sales Prices and Appraisal Data)
		\hline
		Interest Rate
		%		$ \text{House Price AT, state}$ 
		& state		
		& 
		\href{https://www.fhfa.gov/DataTools/Downloads/Pages/Monthly-Interest-Rate-Data.aspx}
		{FHFA}
		%$ \text{FHFA}$ 
		%$ \text{quarterly, 1975-2016}$	
		%$ \text{quarterly, 1975Q1-2016Q1}$	
		& 
		$ \text{51: US \& DC}$	
		& 
		$ \text{1}$	
		%		& 
		%		$ \text{5 states}$	
		\\ [.15em]
		\hline
		Median House Prices
		%		$ \text{House Price AT, state}$ 
		& {5-digit zip code} 	
		& 
		\href{https://www.zillow.com/research/data/}
		{Zillow}
		%$ \text{FHFA}$ 
		%$ \text{quarterly, 1975-2016}$	
		%$ \text{quarterly, 1975Q1-2016Q1}$	
		& 
		$ 12.665$	zips
		& 
		$ 576$	zips
		\\ [.15em]
		\hline
		%*********************************************************************
		Median Rent (historic)  %QUARTERLY
		%		$ \text{House Price AT, state}$ 
		& {MSA} 	
		& 
		\href{https://www.zillow.com/research/data/}
		{Zillow}
		%$ \text{FHFA}$ 
		%$ \text{quarterly, 1975-2016}$	
		%$ \text{quarterly, 1975Q1-2016Q1}$	
		& 
		$ 313$	MSAs
		& 
		$ 21$	MSAs
		\\ [.15em]
		\hline
		Supply Elasticity
		%		$ \text{Saiz elasticity, MSA}$ 
		& MSA
		& 
		%\cite{Saiz2010}
		\href{https://scholar.google.com/citations?user=UXEZZS0AAAAJ&hl=en}
		{Saiz (2010)}		
		& 
		$ \text{269 MSAs}$	
		& 
		$ \text{21 MSAs}$	
		%		& 
		%		$ \text{5 states}$	
		\\ [.15em]
		\hline
		Employment
		%		$ \text{Employment, county}$ 
		& county
		& 
		\href{https://www.bls.gov/ces/}
		{BLS}
		& 
		2,581 fips 	
		& 
		153 fips
		%		& 
		%		$ \text{5 states}$	
		\\ [.15em]
		\hline
		Income
		%		$ \text{Income, county}$ 
		& county
		& 
		\href{https://www.bea.gov/regional/}
		{BEA}
		& 
		2,581 fips 	
		& 
		153 fips
		%		& 
		%		$ \text{5 states}$	
		\\ [.15em]
		\hline
		Population
		%		$ \text{Population, county}$ 
		& county
		& 
		\href{http://www.census.gov/data.html}
		{Census}
		& 
		2,581 fips 	
		& 
		153 fips
		%		& 
		%		$ \text{5 states}$	
		\\ [.15em]
		\hline
		Building permits
		%		$ \text{Building permits, county}$ 
		& county
		& 
		\href{http://www.census.gov/data.html}
		{Census}
		& 
		3,072 fips	
		& 
		229	fips
		%		& 
		%		$ \text{5 states}$	
		\\ [.15em]
		\hline
		Homeownership Rate
		%		$ \text{Homeownership, MSA}$ 
		& MSA
		& 
		\href{http://www.census.gov/data.html}
		{Census}
		& 
		75 MSAs	
		& 
		5 MSAs
		%		& 
		%		$ \text{5 states}$	
		\\ [.15em]
\hline
Homeownership 
%		$ \text{Homeownership, MSA}$ 
& Household Survey
& 
\href{https://www.census.gov/programs-surveys/ahs/data.html}
{AHS}
& 
 123 MSAs	
& 
 9 MSAs	
%		& 
%		$ \text{5 states}$			\\ [.15em]
		\\ [.15em]
		\hline
		Oil Prices
		%		$ \text{Oil Prices, US}$ 
		& US
		& 
		\href{http://www.eia.gov/petroleum/data.cfm}
		{EIA}
		& 
		-			
		& 
		-	
		%		& 
		%		$ \text{5 states}$	
		\\ [.15em]
		\hline
		10 year Treasury
		%		$10 \text{ year Treasury, US}$ 
		& US
		& 
		\href{https://www.treasury.gov/resource-center/data-chart-center/interest-rates/Pages/TextView.aspx?data=yield}
		{Treasury}
		& 
		-
		& 
		-
		%		& 
		%		$ \text{5 states}$	
		\\ [.15em]
		\hline
		Inflation Expectations
		%		$\text{Inflation Expectations, US}$ 
		& US
		& 
		\href{https://www.philadelphiafed.org/research-and-data/real-time-center/livingston-survey}
		{Livingston survey}
		& 
		-
		& 
		-
		%		& 
		%		$ \text{5 states}$	
		\\ [.15em]
		\hline
		CPI-U
		%		$\text{CPI-U, US}$ 
		& US
		& 
		\href{https://www.bls.gov/news.release/cpi.t01.htm}
		{BLS}
		& 
		-
		& 
		-
		%		& 
		%		$ \text{5 states}$	
		\\ [.15em]
		\hline
	\end{tabular}
	\medskip
	
	\begin{minipage}{0.9\textwidth}
		{\footnotesize 
			\textit{Note}. 
			This table lists sources for the different variables used in this paper. 
			All data are annual except for the 
			FHFA state house price index,
			%and 
			Zillow rent index,
			%which are
			%quarterly, 
			and the Zillow Home Value Index. %MONTHLY 
			The Metropolitan Statistical Areas (MSAs) are defined by the United States Office of Management and Budget. 
			The Federal Information Processing Standards (fips) codes uniquely identify counties.
			The AHS survey data is at the household level, but the most local 
			geographic level available is the MSA.
			%		For variable, the level, source and number of locations is given.
			%		figure plots point estimates 
			%		against housing supply elasticity 
			%		from triple difference regressions 
			%		presented in \hyperref[US_DDD]{Table \ref*{US_DDD}}. %column 7
			%		This figure 
			%		%uses data from 
			%		includes
			%		all 21 Texas MSAs. 
			%		Galveston, the most elastic MSA, has a 5.4\% effect, 
			%		whereas Sherman, 
			%		the most inelastic MSA, has a 1.86\% effect.
			\par}
	\end{minipage}
	
	%		}
\end{table}

\newpage
%AZ: hack in TEX CODE: replace "Border_State"  w/ "Border\_State"
\subsection{Standard Error Robustness}
\begin{table}[H]%[!htb]%[center]
	\centering
	%\small \caption{US Sample}
	\small \caption{Impact of HEL Legalization 
		on Texas House Prices, Standard Error Robustness}
	%	{\scriptsize \input  {../../4_output_analysis/main/SeRobust.tex}	 }	
	\label{SE_robust}
		\scalebox{1}
	{%\scriptsize 
		%\documentclass[]{article}\pagestyle{empty}\begin{document}
%\begin{center}
\begin{tabular}{lcccc}
\hline 
\noalign{\smallskip}
& (1) & (2) & (3) & (4) \\
METHOD & US & Border-State & Border & CBCP
\\
%\noalign{\smallskip}METHOD &   &   &   &  
%\\
\noalign{\smallskip}\hline \noalign{\smallskip}
TexasPost & 0.0350 & 0.0616 & 0.0476 & 0.0413\\
OLS & \begin{footnotesize}(0.0049)***\end{footnotesize} & \begin{footnotesize}(0.0045)***\end{footnotesize} & \begin{footnotesize}(0.0098)***\end{footnotesize} & \begin{footnotesize}(0.0116)***\end{footnotesize}\\
EHW & \begin{footnotesize}(0.0029)***\end{footnotesize} & \begin{footnotesize}(0.0037)***\end{footnotesize} & \begin{footnotesize}(0.0087)***\end{footnotesize} & \begin{footnotesize}(0.0127)***\end{footnotesize}\\
Zip5 & \begin{footnotesize}(0.0030)***\end{footnotesize} & \begin{footnotesize}(0.0045)***\end{footnotesize} & \begin{footnotesize}(0.0097)***\end{footnotesize} & \begin{footnotesize}(0.0115)***\end{footnotesize}\\
Zip3 & \begin{footnotesize}(0.0129)***\end{footnotesize} & \begin{footnotesize}(0.0164)***\end{footnotesize} & \begin{footnotesize}(0.0148)***\end{footnotesize} & \begin{footnotesize}(0.0120)***\end{footnotesize}\\
Fips & \begin{footnotesize}(0.0149)**\end{footnotesize} & \begin{footnotesize}(0.0179)***\end{footnotesize} & \begin{footnotesize}(0.0143)***\end{footnotesize} & \begin{footnotesize}(0.0120)***\end{footnotesize}\\
Msa & \begin{footnotesize}(0.0167)**\end{footnotesize} & \begin{footnotesize}(0.0210)***\end{footnotesize} & \begin{footnotesize}(0.0139)***\end{footnotesize} & \begin{footnotesize}(0.0116)***\end{footnotesize}\\
State & \begin{footnotesize}(0.0075)***\end{footnotesize} & \begin{footnotesize}(0.0221)**\end{footnotesize} & \begin{footnotesize}(0.0151)**\end{footnotesize} & \begin{footnotesize}(0.0039)***\end{footnotesize}\\
SHAC & \begin{footnotesize}(0.0099)***\end{footnotesize} & \begin{footnotesize}(0.0121)***\end{footnotesize} & \begin{footnotesize}(0.0098)***\end{footnotesize} & \begin{footnotesize}(0.0111)***\end{footnotesize}\\
Pair & \begin{footnotesize}\end{footnotesize} & \begin{footnotesize}\end{footnotesize} & \begin{footnotesize}\end{footnotesize} & \begin{footnotesize}(0.0128)***\end{footnotesize}\\
Pair Zip5 & \begin{footnotesize}\end{footnotesize} & \begin{footnotesize}\end{footnotesize} & \begin{footnotesize}\end{footnotesize} & \begin{footnotesize}(0.0126)***\end{footnotesize}\\
Pair State & \begin{footnotesize}\end{footnotesize} & \begin{footnotesize}\end{footnotesize} & \begin{footnotesize}\end{footnotesize} & \begin{footnotesize}(0.0083)***\end{footnotesize}\\
\noalign{\smallskip}$N$ & 143,637 & 11,310 & 1,365 & 2,041\\
$R^2$ & \begin{footnotesize}0.9772\end{footnotesize} & \begin{footnotesize}0.9826\end{footnotesize} & \begin{footnotesize}0.9817\end{footnotesize} & \begin{footnotesize}0.9851\end{footnotesize}\\
\noalign{\smallskip}\hline\end{tabular}%\\
%\end{center}
%\end{document}

		%{../../4_output_analysis/main/SeRobust.tex}	 
	}	
	%\\
	%\smallskip
	%{\scriptsize \input  {../../4_output_analysis/main/US_NoLagOilInt.tex}	 }
	%	{\scriptsize \input  {../../4_output_analysis/main/Main-Cluster-StateZip5OLS.tex}	 }
	\medskip
	\begin{minipage}{1\textwidth}
		{\footnotesize 
			\textit{Note}. 
			This table repeats the estimates in \hyperref[US_T1]{Table \ref*{US_T1}}: columns 1-4,
			using a variety of different methods to estimate standard errors.
			%			clustered at different levels of observation.
			%
			Each column corresponds to the sample 
			and each row (after the first) corresponds to the method for estimating standard errors.  
			All specifications include zip code and year fixed effects,  
			state time trends, and national oil prices interacted with MSA dummies.
			Specifications using the CBCP sample also include 
			contiguous border county  pair by year fixed effects.
%			Estimates of standard errors clustered at the state level are not 
%			not reported for the Border-State and Border sample as there are too few clusters. 
%			Similarly estimates are not reported for the Texarkana sample 
%			for clusters more local than the 5-digit zip code level.   
%			Estimates clustered at the FIPS and MSA levels used slightly different 
%			samples as a few zip codes could not be matched.  
			Data sources can be found in \hyperref[data_table]{Table \ref*{data_table}}.
			* p$<$.1; ** p$<$.05; *** p$<$.01.
			%		Significance levels 10\%, 5\%, and 1\% are denoted by *, **, and ***, respectively.
			\par}
	\end{minipage}
\end{table}

\newpage
%AZ: hack in TEX CODE: replace "Border_State"  w/ "Border\_State"
\subsection{House Price Index Robustness}
\begin{table}[H]%[!htb]%[center]
	\centering
	%\small \caption{US Sample}
	\small \caption{Impact of HEL Legalization 
		on Texas House Prices, House Price Index Robustness}
	%	{\scriptsize \input  {../../4_output_analysis/main/SeRobust.tex}	 }	
	\label{ZHVI_robust}
	\scalebox{.8}
	{%\scriptsize 
		
\begin{tabular}{lcccccccc} \hline
 & (1) & (2) & (3) & (4) & (5) & (6) & (7) & (8) 
 \\
 SAMPLES & US & US & Border-State & Border-State & Border & Border & CBCP & CBCP  
 \\ 
VARIABLES & (lnY) & (\$Y) & (lnY) & (\$Y) & (lnY) & (\$Y) & (lnY) & (\$Y)
%VARIABLES & (\%) & (\$) & (\%) & (\$) & (\%) &(\$) & (\%) & (\$) 
\\ 
\hline
 &  &  &  &  &  &  &  &  \\
$\text{Texas} \times \text{Post}$ & 0.046*** & 5,363.34*** & 0.031*** & 2,714.41*** & 0.042*** & 4,020.57*** & 0.050*** & 3,027.09*** \\
 & (0.008) & (971.54) & (0.003) & (136.47) & (0.002) & (100.10) & (0.007) & (380.86) \\
 &  &  &  &  &  &  &  &  \\
N & 1,277,495 & 1,277,495 & 126,806 & 126,806 & 14,662 & 14,662 & 14,948 & 14,948 \\
$R^2$ & 0.992 & 0.973 & 0.995 & 0.996 & 0.994 & 0.996 & 0.994 & 0.995 \\
% std-err & state & state & state & state & state & state & state and border pair & state and border pair \\ 
 \hline
%\multicolumn{9}{c}{ Robust standard errors in parentheses} \\
%\multicolumn{9}{c}{ *** p$<$0.01, ** p$<$0.05, * p$<$0.1} \\
\end{tabular}

		%{../../4_output_analysis/main/SeRobust.tex}	 
	}	
	%\\
	%\smallskip
	%{\scriptsize \input  {../../4_output_analysis/main/US_NoLagOilInt.tex}	 }
	%	{\scriptsize \input  {../../4_output_analysis/main/Main-Cluster-StateZip5OLS.tex}	 }
	\medskip
	\begin{minipage}{1\textwidth}
		{\footnotesize 
			\textit{Note}. 
			This table repeats the estimates in \hyperref[US_T1]{Table \ref*{US_T1}}: columns 1-4,
			using data from Zillow.
			Each sample estimates the treatment effect two ways: using the log house real price index and 
			using the real ZHVI (Zillow's measure of median home value in a zip-code).
			All specifications include zip code and year fixed effects,  
			state time trends, and national oil prices interacted with MSA dummies.
			Specifications using the CBCP sample also include 
			contiguous border county  pair by year fixed effects.
			Data sources can be found in \hyperref[data_table]{Table \ref*{data_table}}.
			* p$<$.1; ** p$<$.05; *** p$<$.01.
			%		Significance levels 10\%, 5\%, and 1\% are denoted by *, **, and ***, respectively.
			\par}
	\end{minipage}
\end{table}

%%%%%%%%%%%%%%%%%%%%%%%%%%%%%%%%%%%%%%%%%%%%%%%%%%%%%%
%%%%%%%%%%%%%%%%%%%%%%%%%%%%%%%%%%%%%%%%%%%%%%%%%%%%%%
\newpage
%%%%%%%%%%%%%%%%%%%%%%%%%%%%%%%%%%%%%%%%%%%%%%%%%%%%%%
\section{For Online Publication: Texas Legal Timeline }
\label{timeline}
This appendix provides a brief timeline of relevant laws 
regarding home 
equity borrowing in Texas.\footnote{\cite{AEJ2012, Forrester, Kumar, McKnight, Stolper2015, TLC}}

\begin{itemize}
	\item 
	Article XVI, Section 50 of Texas Constitution of 1876 protects homesteads 
	from foreclosure   for failure to pay all debts except  
	%\textbf{original} home 
	the purchase loan, 	property taxes,	or a mechanic's lien. 
	%(for work and materials) %debt incurred to finance home improvements. 
%
%Texas 1876 constitution, Article 16, section 50
%-purchase mortgage, property taxes, or mechanics lien (work & materials) 
%-https://tarltonapps.law.utexas.edu/constitutions/texas1876/a16
%“SEC. 50. The homestead of a family shall be, and is hereby protected from forced sale, for the payment of all debts except for the purchase money thereof, or a part of such purchase money, the taxes due thereon, or for work and materials used in constructing improvements thereon, and in this last case only when the work and material are contracted for in writing, with the consent of the wife given in the same manner as is required in making a sale and conveyance of the homestead; nor shall the owner, if a married man, sell the homestead without the consent of the wife, given in such manner as may be prescribed by law. No mortgage, trust, deed, or other lien on the household shall ever be valid, except for the purchase money therefor, or improvements made thereon, as herein before provided, whether such mortgage, or trust deed, or other lien, shall have been created by the husband alone, or together with his wife; and all pretended sales of the homestead involving any condition of defeasance shall be void.
%”
%
%
%	\item
%	If lender can't foreclose, this restricts home equity lending. 
%	Only one-shot purchase mortgage.
	\item 
	This section has only been amended twice before 1997; 
	the first amendment extended protections to single adults in 1973 
	and the second amendment of 1995 related to divorce proceedings.
	\item 
	December 1994, Texas Senate Interim Committee on Home Equity Lending 
	comes out in strong support of easing restrictions on lending 
	with limits on home equity lending to protect consumers, 
	and called for the amendment proposal to be put on the ballot 
	for voters to decide. The proposal did not gather the two-thirds 
	majority in the House of Representatives, but it did pass the Senate, 
	which was a first for such a proposal. 
	The committee’s report was incorporated into 
	House Joint Resolution 31 which passed the Texas House and Senate in May 1997.
	\item 
Voters approved Proposition 8 on 11-4-1997 with almost 
	60\% voting yes of the 1.17 million votes cast,
	begins \textbf{1-1-1998}, legalized Home Equity Loans without restrictions 
	on how the money could be used, however the total value of all liens on the 
	home \textbf{cannot exceed 80\% of the fair market value}. 
	Reverse mortgages and cash-out refinance loans were also legalized.
%	, 
%	but not Home Equity Lines of Credit. 
%	\\
%	$CLTV \leq .80 \times \text{Fair Market Value}$
%	\\
%	Prop 8 allowed lump-sum HELs, not lines of credit. 
	\item 
	Several problems with Proposition 8 in the first year. 
	\\
	HELs not allowed for those who live on more than one acre of land. 
	\\
	Reverse mortgage rules made the loans ineligible for purchase by Fannie Mae. 
	\item
%	Prop 8 shortcomings resolved, November 2, 1999.
%	\\
	Proposition 2 on the November 2, 1999 ballot corrected the shortcomings.
	\\
	Article XVI, Section 50 was amended to clear up
	the issues with reverse mortgages.
	\\
%	Proposition 6 amended 
	Article XVI, Section 51 was amended by increasing the acreage limit
	used to define urban households to 10 acres.
%	\item
%	Before 1997 law, no second liens, no HELs, REFI only up to current balance. 
%	Cash-outs were illegal. 
%	\\ \\
%	\item
%	Fall of 2002, the Texas Credit Union League, representing credit unions, started a campaign pushing for further home equity reform, including lines of credit. 
%	\item 
%	March 2003, Texas state comptroller issued report supporting home equity lines of credit. 
%	\item
%	Prop 16, legalized HELOCs, passed Sept 13, 2003, effective Sept 29, 2003
%	\\
%	Admin interpretations, 12-18-2003, and 2-20-2004
\end{itemize}

\newpage 
% % % % % % % % % % % % % % % % % % % % % % %
% % % % % % % % % % % % % % % % % % % % % % % % % % % % % %
\section{For Online Publication: Household's Problem}
\label{deriv1}
This appendix solves the household's problem presented in Section \ref{Model}.
\begin{align*}
\max \text{ } \E_{0} \sum_{t=0}^{\infty} \beta^{t} u(c_{t} , h_{t})
&
\\
\text{s.t.}
\\
c_{t}+p_{t}h_{t+1} + a_{t+1}
&
\leq 
y_{t} + p_{t} h_{t} (1-\delta_{t}) + (1+r_{t}) a_{t}
\tag{DBC $\lambda_{t}$}
\\
-a_{t+1} 
&\leq 
\kappa_{t} p_{t} h_{t}
\tag{CC $\mu_{t}$}
\end{align*}
Savings: $a_{t+1}$ at time $t$. 
If $a_{t+1}>0$, save, if $a_{t+1}<0$,
borrow.
\\
The multiplier on the dynamic budget constraint (DBC):
$\lambda(s^{t}) \beta^{t} \pi(s^{t}) > 0$
\\
The multiplier on the collateral constraint (CC):
$\mu(s^{t}) \beta^{t} \pi(s^{t}) \geq 0 $
\\
Complementary slackness:
$
\mu(s^{t}) \beta^{t} \pi(s^{t})
\left[ 
\kappa(s^{t})  p(s^{t}) h_{t}(s^{t-1} ) + a_{t+1}(s^{t} )
\right]
=
0
$
%$
%\mu(s^{t}) \beta^{t} \pi(s^{t})
%\left[ 
% p(s^{t}) h_{t}(s^{t-1} ) + a_{t+1}(s^{t} )
% \right]
%=
%0
%$
\\ \\
State variables in $s^{t}$: 
$y(s^{t})$, 
$h_{t}(s^{t-1})$,
$a_{t}(s^{t-1})$,
$r_{t}(s^{t-1})$,
$\delta(s^{t})$,
$\kappa(s^{t})$.
% % % % %
\\
Choice variables in $s^{t}$: 
$c(s^{t})$, 
$h_{t+1}(s^{t})$,
$a_{t+1}(s^{t})$.
%(Any two will pin down the third.)
%Equilibrium variables: $p_{t}, r_{t+1}, LTV_{t}$
\begin{align*} 
\Lag 
=
\sum_{t=0}^{\infty} 
\underset{s^{t}\in S^{t}}{\sum} 
%\left[ 
& 
\beta^{t} \pi(s^{t})
u( c(s^{t}) , h(s^{t}) )
\\
+&
\lambda(s^{t}) \beta^{t} \pi(s^{t})
\left[ 
 y(s^{t}) 
 + p(s^{t}) h_{t}(s^{t-1} ) (1-\delta(s^{t})) 
 + \left(1+r_{t}(s^{t-1} ) \right) a_{t}(s^{t-1} )
\right]  
%\right. 
%\\
%&
%\left.
\\-&
\lambda(s^{t}) \beta^{t} \pi(s^{t})
\left[   
 c(s^{t}) + p(s^{t}) h_{t+1}(s^{t}) + a_{t+1}(s^{t} )
 \right]
\\
+&
\mu(s^{t}) \beta^{t} \pi(s^{t})
\left[ 
\kappa(s^{t})  p(s^{t}) h_{t}(s^{t-1} ) + a_{t+1}(s^{t} )
 \right]
\end{align*}
% % % % %
In $s^{t}$ we have: 
$y(s^{t})$, 
$h_{t}(s^{t-1})$,
$a_{t}(s^{t-1})$,
$\delta(s^{t})$,
$\kappa(s^{t})$.
% % % % %
\\
In $s^{t}$ we choose: 
$c(s^{t})$, 
$h_{t+1}(s^{t})$,
$a_{t+1}(s^{t})$.
(Any two will pin down the third.)
% % % %
\begin{align*} 
\Lag_{c(s^{t})} 
&=
\beta^{t} \pi(s^{t})
u_{1}( s^{t} )
+
\lambda(s^{t}) \beta^{t} \pi(s^{t}) \left[-1\right]
\\&=0
\\
&\Leftrightarrow 
\\
\lambda(s^{t}) &= u_{1}(s^{t})
\end{align*} 
%\\ \\
% % % %
\begin{align*} 
\Lag_{h_{t+1}(s^{t})} 
&=
\sum_{s^{t+1}|s^{t}}
\beta^{t+1} \pi(s^{t+1}) u_{2}( s^{t+1} )
\\&
+
\lambda(s^{t}) \beta^{t} \pi(s^{t}) 
\left[
-p(s^{t})
\right]
\\&
+
\sum_{s^{t+1}|s^{t}}
\lambda(s^{t+1}) \beta^{t+1} \pi(s^{t+1}) 
\left[
p(s^{t+1})  (1-\delta(s^{t+1}))
\right]
\\&
+
\sum_{s^{t+1}|s^{t}}
\mu(s^{t+1}) \beta^{t+1} \pi(s^{t+1}) 
\left[
\kappa(s^{t+1})
p(s^{t+1})
\right]
\\&
=0
\end{align*} 
% % % %
\begin{align*} 
\Lag_{a_{t+1}(s^{t})} 
&=
\lambda(s^{t}) \beta^{t} \pi(s^{t}) 
\left[
-1
\right]
\\&
+
\mu(s^{t}) \beta^{t} \pi(s^{t}) 
\left[
1
\right]
\\&
+
\sum_{s^{t+1}|s^{t}}
\lambda(s^{t+1}) \beta^{t+1} \pi(s^{t+1}) 
%(1+r)
\left(1+r_{t+1}(s^{t} ) \right)
\\&
=0
\\&
\Leftrightarrow
\\
\lambda(s^{t}) \beta^{t} \pi(s^{t}) 
&=
\mu(s^{t}) \beta^{t} \pi(s^{t}) 
+
\sum_{s^{t+1}|s^{t}}
\lambda(s^{t+1}) \beta^{t+1} \pi(s^{t+1}) 
%(1+r)
\left(1+r_{t+1}(s^{t} ) \right)
\\&
\Leftrightarrow
\\
\lambda(s^{t})  
&=
\mu(s^{t}) 
+
\beta \left(1+r_{t+1}(s^{t} ) \right)
\E_{t}[   \lambda(s^{t+1})  ]
\\
\text{recall }
\lambda(s^{t})&=u_{1}(s^{t}) 
\text{ (consumption foc)}
\\&
\Leftrightarrow
\\
u_{1}(s^{t})  
&=
\mu(s^{t}) 
+
\beta \left(1+r_{t+1}(s^{t} ) \right)
\E_{t}[   u_{1}(s^{t+1})  ]
\end{align*} 
This is the famous liquidity-constrained Euler equation!
If the collateral constraint is not binding ($\mu=0$)
or doesn't exist, then this collapses to the usual
frictionless Euler equation.
\\ \\
Using $\lambda(s^{t})=u_{1}(s^{t})$, we combine the
consumption and housing FOCs.
\begin{align*}
u_{1}(s^{t}) \beta^{t} \pi(s^{t}) 
\left[
p(s^{t})
\right]
&=
\sum_{s^{t+1}|s^{t}}
\beta^{t+1} \pi(s^{t+1}) u_{2}( s^{t+1} )
\\&
+
\sum_{s^{t+1}|s^{t}}
u_{1}(s^{t+1}) \beta^{t+1} \pi(s^{t+1}) 
\left[
p(s^{t+1}) (1-\delta( s^{t+1} ) )
\right]
\\&
+
\sum_{s^{t+1}|s^{t}}
\mu(s^{t+1}) \beta^{t+1} \pi(s^{t+1}) 
\left[
 p(s^{t+1}) \kappa( s^{t+1} )
\right]
\end{align*} 
% % % % % % % % % % % % % % %
% % % % % % % % % % % % % % %
\begin{align*}
u_{1}(s^{t})   
\left[
p(s^{t})
\right]
&=
\E_{t}
\beta    u_{2}( s^{t+1} )
\\&
+
\E_{t}
u_{1}(s^{t+1}) \beta    
\left[
p(s^{t+1})  (1-\delta( s^{t+1} ) )
\right]
\\&
+
\E_{t}
\mu(s^{t+1}) \beta   
\left[
p(s^{t+1})  \kappa( s^{t+1} )
\right]
\end{align*} 
% % % % % % % % % % % % % % %
% % % % % % % % % % % % % % %
\begin{align*}
p(s^{t})
&=
\E_{t}
\beta   \frac{  u_{2}( s^{t+1} ) }{ u_{1}(s^{t})   }
\\&
+
\E_{t}
 \beta    
\frac{ u_{1}(s^{t+1})  }{ u_{1}(s^{t})   } 
\left[
p(s^{t+1})  (1-\delta( s^{t+1} ) )
\right]
\\&
+
\E_{t}
\beta
\frac{ \mu(s^{t+1}) }{ u_{1}(s^{t}) }    
\left[
p(s^{t+1})  \kappa( s^{t+1} )
\right]
\end{align*} 
% % % % % % % % % % % % % % %
% % % % % % % % % % % % % % %
\begin{align*}
p(s^{t})
&=
\E_{t}
\beta   
\frac{  u_{1}( s^{t+1} ) }{ u_{1}(s^{t})   }
\frac{  u_{2}( s^{t+1} ) }{ u_{1}(s^{t+1})   }
\\&
+
\E_{t}
 \beta    
\frac{ u_{1}(s^{t+1})  }{ u_{1}(s^{t})   } 
\left[
p(s^{t+1}) (1-\delta( s^{t+1} ) )
\right]
%\\&
+
\E_{t}
\beta
\frac{ u_{1} (s^{t+1}) }{ u_{1}(s^{t}) }    
\frac{ \mu(s^{t+1}) }{ u_{1}(s^{t+1}) }    
\left[
p(s^{t+1}) \kappa( s^{t+1} )
\right]
\end{align*} 
% % % % % % % % % % % % % % %
% % % % % % % % % % % % % % %
\begin{align*}
p(s^{t})
&=
\E_{t}
\left[ 
\beta   
\frac{  u_{1}( s^{t+1} ) }{ u_{1}(s^{t})   }
\left( 
\frac{  u_{2}( s^{t+1} ) }{ u_{1}(s^{t+1})   }
+   
\left[
p(s^{t+1}) (1-\delta( s^{t+1} ) )
\right]
%\\&
+
\frac{ \mu(s^{t+1}) }{ u_{1}(s^{t+1}) }    
\left[
p(s^{t+1}) \kappa( s^{t+1} )
\right]
\right) 
\right] 
\end{align*} 
% % % % % % % % % % % % % % %
% % % % % % % % % % % % % % %
\begin{align*}
\underbrace{p_{t}}_{\text{price}}
&=
\E_{t}
\left[ 
\underbrace{ 
\beta   
\frac{  u_{1}( t+1 ) }{ u_{1}( t )   }
}_{sdf}
\times 
\left( 
\underbrace{ 
\frac{  u_{2}(  t+1  ) }{ u_{1}( t+1 )   }
}_{\text{housing service flow}}
+
\underbrace{ 
\frac{ \mu( t+1 ) }{ u_{1}( t+1 ) }    
\kappa_{t+1} p_{t+1}
}_{\text{collateral service flow}}
+   
\underbrace{ (1-\delta_{t+1} ) p_{t+1} }_{\text{resale price}}
\right) 
\right] 
\end{align*} 
% % % % % % % % % % % % % % % %
\begin{align*}
p_{t} &= \E_{t}[ M_{t+1} \times 
\left(  s_{t+1} + CSF_{t+1} + (1-\delta_{t+1} ) p_{t+1} \right) ]
\end{align*}
We can use the equilibrium value of the multiplier:
\\
$\mu(t+1) = u_{1}(t+1) - \beta \left(1+r_{t+2}(s^{t+1} ) \right) \E_{t+1}[ u_{1}(t+2) ]$
%\\ \\
% % % % % % % % % % % % % % %
\begin{align*}
%\underbrace{ 
	\mu_{t} 
%}_{\text{price}}
&=
u_{c}( t  )
- \beta \left(1+r_{t+1}\right) \E_{t}[ u_{c}(t+1) ]
\end{align*} 
%\begin{align*}
%	\underbrace{ \mu_{t+1} }_{\text{price}}
%	&=
%	u_{c}( t+1 )
%	- \beta \left(1+r_{t+2}(s^{t+1} ) \right) \E_{t+1}[ u_{c}(t+2) ]
%\end{align*} 
% % % % % % % % % % % % % % % %
%
If there is no collateral constraint or the collateral 
constraint never binds ($\mu(s^{t}) = 0$), then housing 
has no collateral value ($CSF_{t}=0$).

\newpage
\subsection{Price Decomposition}
% % % % % % % % % % % % % % % %
%where 
%\begin{align*} 
%M_{t,j} 
%&\equiv M_{t+1}\times M_{t+2}\times \cdots \times M_{t+j}
%\\
%M_{t+1,j} 
%&\equiv 
%M_{t+2}\times M_{t+3}\times \cdots \times M_{t+j} \times M_{t+j+1}
%\\
%M_{t,j} 
%&=
%M_{t,1} \times M_{t+1,j-1}
%\end{align*} 
Define $M_{t,j}\equiv M_{t}\times M_{t+1}\cdots M_{j-1} \times M_{j}$
\\
Define the total dividend $d_{t} \equiv s_{t}+CSF_{t}$   
\begin{align*}
p_{t} 
&= 
\E_{t}\left[  
M_{t+1} \times 
\left(  d_{t+1} + (1-\delta ) p_{t+1} \right) 
\right]
\\
&= 
\E_{t}\left[  
M_{t+1} \times \left( d_{t+1}  \right) 
\right]
+ 
(1-\delta ) 
\E_{t}\left[  M_{t+1} \times p_{t+1}  \right]
\\
&= 
\E_{t}\left[  
M_{t+1} \times 
\left(  d_{t+1}  \right) 
\right]
+ 
(1-\delta ) 
\E_{t}\left[  M_{t+1} \times \E_{t+1}\left[M_{t+2} \times \left(  d_{t+2} + (1-\delta ) p_{t+2} \right)  \right]  \right]
\\
&= 
\E_{t}\left[  
M_{t+1} \times 
\left(  d_{t+1}  \right) 
\right]
+ 
(1-\delta ) 
\E_{t}\left[  M_{t+1} \times M_{t+2} \times \left(  d_{t+2}  + (1-\delta ) p_{t+2} \right)   \right]
\\
&= 
\E_{t}\left[  
M_{t+1} \times 
\left(  d_{t+1}   \right) 
\right]
+ 
(1-\delta ) 
\E_{t}\left[  M_{t+1,t+2}  \left(  d_{t+2}   \right)   \right]
+ 
(1-\delta )^{2} 
\E_{t}\left[  M_{t+1,t+2}   p_{t+2}    \right]
\\
&= 
\E_{t}\left[
\sum_{j=1}^{\infty} 
M_{t+1,t+j} \times 
\left(  d_{t+j}   \right) 
(1-\delta )^{j-1} 
\right]
\\
&= 
\E_{t}\left[
\sum_{j=1}^{\infty} 
M_{t+1,t+j} \times 
\left(  s_{t+j} + CSF_{t+j} \right) 
(1-\delta )^{j-1} 
\right]
\\
&= 
\E_{t}\left[
\sum_{j=1}^{\infty} 
M_{t+1,t+j} \times 
\left(  s_{t+j} (1-\delta )^{j-1} \right) 
\right]
+
\E_{t}\left[
\sum_{j=1}^{\infty}  
M_{t+1,t+j} \times 
\left(  CSF_{t+j} (1-\delta )^{j-1} \right) 
\right]
%
%\\
%&= 
%\sum_{j=1}^{\infty} 
%(1-\delta )^{j-1} 
%\E_{t}\left[ M_{t+1,t+j} \times 
%\left(  s_{t+j} \right) 
%\right]
%+
%\sum_{j=1}^{\infty} 
%(1-\delta )^{j-1} 
%\E_{t}\left[ M_{t+1,t+j} \times 
%\left(  CSF_{t+j} \right) 
%\right]
%
%\\
%&= 
%PDV_{t}\left( s \right)
%+
%PDV_{t}\left( CSF \right)
\end{align*}

Given a stochastic process $x\equiv \left\{ x_{t+j} \right\}_{j=1}^{j=\infty} $
we can define the PDV-operator
\begin{align*}
PDV_{t}\left( x_{t+j} \right)
&=
\E_{t}
\left[  
M_{t+1,t+j} 
%\times 
%\left( 
x_{t+j}  
%\right) 
\right]
\\
PDV_{t}\left( x \right)
&=
\E_{t}
\left[
\sum_{j=1}^{\infty}  
M_{t+1,t+j} \times 
\left(  x_{t+j}  \right) 
\right]
\\
&=
\sum_{j=1}^{\infty}  
PDV_{t}\left( x_{t+j} \right)
\\
&=
PDV_{t}\left( x_{t+1} \right)
+
PDV_{t} \left( 
\left\{ x_{t+j} \right\}_{j=2}^{j=\infty} 
\right) 
\end{align*}

\begin{align*}
p_{t}
&= 
\E_{t}\left[
\sum_{j=1}^{\infty} 
M_{t+1,t+j} \times 
\left(  s_{t+j} (1-\delta )^{j-1} \right) 
\right]
+
\E_{t}\left[
\sum_{j=1}^{\infty}  
M_{t+1,t+j} \times 
\left(  CSF_{t+j} (1-\delta )^{j-1} \right) 
\right]
\\
\underbrace{ 
	p_{t}
}_{\text{price}}
&= 
\underbrace{ 
	PDV_{t} \left( 
	\left\{ s_{t+j} \right\}_{j=1}^{j=\infty} 
	\right) 
}_{\text{pdv housing service flow} }
+
\underbrace{ 
	%PDV_{t}\left( CSF \right) 
	PDV_{t} \left( 
	\left\{ CSF_{t+j} \right\}_{j=1}^{j=\infty} 
	\right) 
}_{\text{pdv collateral service flow} }
%\\
%\frac{	p_{t} }{ rent_{t} }
%&= 
%\frac{ 
% PDV_{t}\left( s \right)  
%+
% PDV_{t}\left( CSF \right)  
%}
%{ rent_{t} }
\end{align*}
The process $s\equiv \left\{ s_{t+j} (1-\delta )^{j-1} \right\}_{j=1}^{j=\infty} $
will be written 
$s\equiv \left\{ s_{t+j}   \right\}_{j=1}^{j=\infty} $
to save space. 
%\\ \\
%Rent: $r_{h,t}=\E_{t}[ M_{t+1} \left( s_{t+1} \right) ]$
%\\
%price-rent ratio:
%$\frac{ p_{t} }{ r_{h,t} } = 
%\frac
%{\E_{t}[ M_{t+1} \left( s_{t+1} + CSF_{t+1} +
%	(1-\delta_{t+1}) p_{t+1} \right) ]}
%{\E_{t}[ M_{t+1} \left( s_{t+1} \right) ] } 
%$
%\\
%The price-rent ratio could have varied because the collateral
%value rose during the boom.

%\newpage
%Price impact: time of announcement vs enactment
%\\ 
\subsection{Price before the law}
We can re-write the price before the law assuming no expectations of the law:
\begin{align*}
p_{t}^{NL}
& =
PDV_{t}\left( s_{t+1}^{NL} \right)
+
PDV_{t} \left( 
\left\{ s_{t+j}^{NL} \right\}_{j=2}^{j=\infty} 
\right) 
%%
%\\
%& =
%\E_{t}\left[
%\sum_{j=1}^{\infty} 
%(1-\delta )^{j-1} 
%M_{t+1,t+j} 
%\times 
%s_{t+j} 
%\right]
%\\ &  =
%\E_{t}\left[ M_{t+1,t+1} \times s_{t+1} \right]
%+
%\E_{t}\left[
%\sum_{j=2}^{\infty} 
%(1-\delta )^{j-1} 
%M_{t+1,t+j} \times 
%s_{t+j} 
%\right]
%%
%%\\ \\
%\\ &  =
%\E_{t}\left[ M_{t+1,t+1} \times s_{t+1} \right]
%+
%\left(1-\delta \right) 
%\E_{t}\left[
%\sum_{j=2}^{\infty} 
%(1-\delta )^{j-2} 
%M_{t+1,t+j} \times 
%s_{t+j} 
%\right]
%%
%\\ &  =
%\E_{t}\left[ M_{t+1,t+1} \times s_{t+1} \right]
%+
%\left(1-\delta \right) 
%\E_{t}\left[
%\sum_{j=1}^{\infty} 
%(1-\delta )^{j-1} 
%M_{t+1,t+j+1} \times 
%s_{t+1+j} 
%\right]
%%
%\\ &  =
%\E_{t}\left[  M_{t+1,t+1} \times s_{t+1} \right]
%+
%\left(1-\delta \right) 
%\E_{t}
%\left[
% M_{t+1,t+1}
%\sum_{j=1}^{\infty} 
%(1-\delta )^{j-1} 
%M_{t+2,t+j+1} \times 
%s_{t+1+j} 
%\right]
%%
%\\ &  =
%\E_{t}\left[  M_{t+1,t+1}  \times s_{t+1} \right]
%+
%(1-\delta )
%\E_{t}\left[
%M_{t+1,t+1} \times p_{t+1}\right]
\\
p_{t+1}^{NL}
& =
PDV_{t+1} \left( 
\left\{ s_{t+j}^{NL} \right\}_{j=2}^{j=\infty} 
\right) 
%
%
%\E_{t+1}
%\left[
%\sum_{j=1}^{\infty} 
%(1-\delta )^{j-1} 
%M_{t+2,t+j+1} \times 
%s_{t+1+j} 
%\right]
\end{align*}

\subsection{Price change before the law} 
%Price change before the law
\begin{align*}
p_{t+1}^{NL} - p_{t}^{NL}
&=
PDV_{t+1} \left( 
\left\{ s_{t+j}^{NL} \right\}_{j=2}^{j=\infty} 
\right) 
-
\left(
PDV_{t}\left( s_{t+1}^{NL} \right)
+
PDV_{t} \left( 
\left\{ s_{t+j}^{NL} \right\}_{j=2}^{j=\infty} 
\right) 
\right)
%%%
%%%
%%%
\\
&=
\left( PDV_{t+1} - PDV_{t} \right) \left( 
\left\{ s_{t+j}^{NL} \right\}_{j=2}^{j=\infty} 
\right) 
-
PDV_{t} \left( 
s_{t+1}^{NL} 
\right) 
\\
&=
\text{news about future cash flow}
-
\text{current period cash flow}
%\\
%&=
%\E_{t+1}
%\left[
%\sum_{j=1}^{\infty} 
%(1-\delta )^{j-1} 
%M_{t+2,t+j+1} \times 
%s_{t+1+j} 
%\right]
%\\ &
%-
%\left(
%\E_{t}\left[  M_{t+1,t+1} \times s_{t+1} \right]
%+
%\left(1-\delta \right) 
%\E_{t}
%\left[
%M_{t+1,t+1}
%\sum_{j=1}^{\infty} 
%(1-\delta )^{j-1} 
%M_{t+2,t+j+1} \times 
%s_{t+1+j} 
%\right]
%\right)
%%-
%%\E_{t}\left[ M_{t,1} \times s_{t+1} \right]
%%-
%%\left(1-\delta \right) 
%%\E_{t}
%%\left[
%%M_{t,1}
%%\sum_{j=1}^{\infty} 
%%(1-\delta )^{j-1} 
%%M_{t+1,j} \times 
%%s_{t+1+j} 
%%\right]
%%
%\\
%%
%&=
%\left( \E_{t+1} - \left(1-\delta \right) \E_{t}  M_{t+1,t+1} \right) 
%\left[
%\sum_{j=1}^{\infty} 
%(1-\delta )^{j-1} 
%M_{t+2,t+j+1} \times 
%s_{t+1+j} 
%\right]
%-
%\E_{t}\left[ M_{t+1,t+1} \times s_{t+1} \right]
%%%
%%%
%%%
%\\
%&=
%PDV_{t+1} \left( 
%\left\{ s_{t+j} \right\}_{j=2}^{j=\infty} 
%\right) 
%-
%PDV_{t} \left( 
%\left\{ s_{t+j} \right\}_{j=1}^{j=\infty} 
%\right) 
%%%%
%%%%
%%%%
%\\
%&=
%PDV_{t+1} \left( 
%\left\{ s_{t+j} \right\}_{j=2}^{j=\infty} 
%\right) 
%-
%PDV_{t} \left( 
%\left\{ s_{t+j} \right\}_{j=2}^{j=\infty} 
%\right) 
%-
%PDV_{t} \left( 
%s_{t+1} 
%\right) 
\end{align*}

%\newpage
%%%
\subsection{Price Change at the time of the law}
%%%
The law is a surprise and occurs at $t+1$.
Homeowners can borrow at $t+1$, but the price will only reflect CSF starting at 
$t+2$.
\begin{align*}
p_{t+1}^{L} - p_{t}^{NL}
&=
PDV_{t+1} \left( 
\left\{ CSF_{t+j} \right\}_{j=2}^{j=\infty} 
\right) 
+
PDV_{t+1} \left( 
\left\{ s_{t+j}^{L} \right\}_{j=2}^{j=\infty} 
\right) 
\\&
-
\left(
PDV_{t}\left( s_{t+1}^{NL} \right)
+
PDV_{t} \left( 
\left\{ s_{t+j}^{NL} \right\}_{j=2}^{j=\infty} 
\right) 
\right)
%%%
%%%
%%%
%\\
%&=
%\E_{t+1}
%\left[
%\sum_{j=1}^{\infty} 
%(1-\delta )^{j-1} 
%M_{t+2,t+j+1} \times 
%s_{t+1+j} 
%\right]
%+
%\E_{t+1}
%\left[
%\sum_{j=1}^{\infty} 
%(1-\delta )^{j-1} 
%M_{t+2,t+j+1} \times 
%CSF_{t+1+j} 
%\right]
%\\ &
%-
%\left(
%\E_{t}\left[  M_{t+1,t+1} \times s_{t+1} \right]
%+
%\left(1-\delta \right) 
%\E_{t}
%\left[
%M_{t+1,t+1}
%\sum_{j=1}^{\infty} 
%(1-\delta )^{j-1} 
%M_{t+2,t+j+1} \times 
%s_{t+1+j} 
%\right]
%\right)
%%-
%%\E_{t}\left[ M_{t,1} \times s_{t+1} \right]
%%-
%%\left(1-\delta \right) 
%%\E_{t}
%%\left[
%%M_{t,1}
%%\sum_{j=1}^{\infty} 
%%(1-\delta )^{j-1} 
%%M_{t+1,j} \times 
%%s_{t+1+j} 
%%\right]
%%
%\\
%%
%&=
%\E_{t+1}
%\left[
%\sum_{j=1}^{\infty} 
%(1-\delta )^{j-1} 
%M_{t+2,t+j+1} \times 
%CSF_{t+1+j} 
%\right]
%\\ & +
%\left( \E_{t+1} - \left(1-\delta \right) \E_{t}  M_{t+1,t+1} \right) 
%\left[
%\sum_{j=1}^{\infty} 
%(1-\delta )^{j-1} 
%M_{t+2,t+j+1} \times 
%s_{t+1+j} 
%\right]
%-
%\E_{t}\left[ M_{t+1,t+1} \times s_{t+1} \right]
\\
&=
\text{PDV(CSF)}
+
\text{news about future cash flow}
-
\text{current period cash flow}
%
%%%
%%%
%%%
%\\
%&=
%PDV_{t+1}\left( CSF \right) 
%+ PDV_{t+1}\left( s \right) 
%- PDV_{t}\left( s \right) 
%- \E_{t}\left[ M_{t+1,t+1} \times s_{t+1} \right]
%%%
%%%
%%%
%\\
%&=
%PDV_{t+1} \left( 
%\left\{ CSF_{t+j} \right\}_{j=2}^{j=\infty} 
%\right) 
%+
%PDV_{t+1} \left( 
%\left\{ s_{t+j} \right\}_{j=2}^{j=\infty} 
%\right) 
%-
%PDV_{t} \left( 
%\left\{ s_{t+j} \right\}_{j=1}^{j=\infty} 
%\right) 
%%%%
%%%%
%%%%
%\\
%&=
%PDV_{t+1} \left( 
%\left\{ CSF_{t+j} \right\}_{j=2}^{j=\infty} 
%\right) 
%+
%PDV_{t+1} \left( 
%\left\{ s_{t+j} \right\}_{j=2}^{j=\infty} 
%\right) 
%-
%PDV_{t} \left( 
%\left\{ s_{t+j} \right\}_{j=2}^{j=\infty} 
%\right) 
%-
%PDV_{t} \left( 
%s_{t+1} 
%\right) 
%%%%
%%%%
%%%%
%\\
%&=
%PDV_{t+1} \left( 
%\left\{ CSF_{t+j} \right\}_{j=2}^{j=\infty} 
%\right) 
%+
%\left( PDV_{t+1} - PDV_{t} \right) \left( 
%\left\{ s_{t+j} \right\}_{j=2}^{j=\infty} 
%\right) 
%-
%PDV_{t} \left( 
%s_{t+1} 
%\right) 
\end{align*}

%\newpage
\subsection{Impact of law changes on house prices}\label{pricelaw}
%The law change affects asset prices in two ways: 
%(1) through the collateral option value $PDV(CSF)$,
%(2) through rents $PDV(s)$.
%%%%
%Suppose the law change occurs at $t+1$:
Compare the price change if the law change occurred at $t+1$ versus if it didn't:
\begin{align*}
\Delta^{\text{Law}}p_{t+1}
& \equiv 
p_{t+1}^{L} - p_{t}^{NL}
\\
&=
PDV_{t+1} \left( 
\left\{ CSF_{t+j} \right\}_{j=2}^{j=\infty} 
\right) 
+
PDV_{t+1} \left( 
\left\{ s_{t+j}^{L} \right\}_{j=2}^{j=\infty} 
\right) 
-
PDV_{t} \left( 
\left\{ s_{t+j}^{NL} \right\}_{j=1}^{j=\infty} 
\right)  
%\Delta^{\text{no law}}p_{t+1}
%+
%\end{align*}
%
%\begin{align*}
\\
\Delta^{\text{NoLaw}}p_{t+1}
& \equiv 
p_{t+1}^{NL} - p_{t}^{NL}
\\
&=
\left( 
PDV_{t+1} \left( 
\left\{ s_{t+j}^{NL} \right\}_{j=2}^{j=\infty} 
\right) 
-
PDV_{t} \left( 
\left\{ s_{t+j}^{NL} \right\}_{j=1}^{j=\infty} 
\right) 
\right) 
\\
\Delta^{\text{Law}}p_{t+1} - \Delta^{\text{NoLaw}}p_{t+1}
&=
p_{t+1}^{L} - p_{t+1}^{NL}
\\
&=
PDV_{t+1} \left( 
\left\{ CSF_{t+j} \right\}_{j=2}^{j=\infty} 
\right) 
+
PDV_{t+1} \left( 
\left\{ s_{t+j}^{L} - s_{t+j}^{NL} \right\}_{j=2}^{j=\infty} 
\right) 
\end{align*}
If we assume the law had no impact on rents 
$s_{t+j}^{L}=s_{t+j}^{NL}$, then the second term cancels out
\begin{align*}
\Delta^{\text{Law}}p_{t+1}
-
\Delta^{\text{NoLaw}}p_{t+1}
&= 
PDV_{t+1} \left( 
\left\{ CSF_{t+j} \right\}_{j=2}^{j=\infty} 
\right) 
\\
\frac{ 
	\Delta^{\text{Law}}p_{t+1}
}{ p_{t} }
-
\frac{ 
	\Delta^{\text{NoLaw}}p_{t+1}
}{ p_{t} }
&=
\frac{ 
	PDV_{t+1} \left( 
	\left\{ CSF_{t+j} \right\}_{j=2}^{j=\infty} 
	\right) 
}{ p_{t} }
\end{align*}

In the dynamic difference-in-differences regression the coefficient 
is 
\begin{align*}
\eta_{1998}
&=
\frac{ 
	\Delta^{\text{Law}}p_{t+1}
}{ p_{t} }
-
\frac{ 
	\Delta^{\text{NoLaw}}p_{t+1}
}{ p_{t} }
\\
&=
\frac{  PDV_{t+1}(CSF)  }{ p_{t} }
\\
&=
\frac{ 
	\E_{t+1}\left[
	\sum_{j=1}^{\infty} 
	(1-\delta )^{j-1} 
	M_{t+1,j} \times 
	CSF_{t+1+j} 
	\right]	
}{ p_{t} }
\end{align*}
Hence, if we assume that the law change had no impact on rents 
then $\eta_{1998}$ is not only the impact of the law change 
on house prices, but also the percent of a house price due 
to the collateral option value. 

If the law change 
increases 
%total demand for housing 
the supply of housing
and $h_{t+j}$ 
rises more than it would without the law, 
then rents should be lower 
$s_{t+j}^{L} \leq s_{t+j}^{NL}$
implying the coefficient underestimates the collateral option value
\begin{align*}
\eta_{1998}
&=
\frac{ 
	\Delta^{\text{Law}}p_{t+1}
}{ p_{t} }
-
\frac{ 
	\Delta^{\text{NoLaw}}p_{t+1}
}{ p_{t} }
\\
& \leq 
\frac{  PDV_{t+1}(CSF)  }{ p_{t} }
\\
&=
\frac{ 
	\E_{t+1}\left[
	\sum_{j=1}^{\infty} 
	(1-\delta )^{j-1} 
	M_{t+1,j} \times 
	CSF_{t+1+j} 
	\right]	
}{ p_{t} }
\end{align*}
The more elastic housing supply is in a given location, 
the bigger the rise in $h_{t+j}$,
the more we would be underestimating the collateral option value. 
However, regardless of the impact on rent (service flow),
as long as the parallel trends assumption holds, 
we are still able to identify the total impact of the law on house prices. 

%AZ:
%%%Another way the law change can affect rents is through consumption and income. 
%%%If the law change sufficiently stimulates the economy, income should rise
%%%and rents should rise. 
%%%In this case 
%%%$s_{t+j}^{L} \geq s_{t+j}^{NL}$.
%%%In this case we are over-estimating the collateral option value. 

In general 
\begin{align*}
\eta_{k}
&=
\E \left( y_{Texas,k} - y_{Control,k} \right) 
-
\E \left( y_{Texas,97} - y_{Control,97} \right) 
\\
\eta_{k}
&=
\frac{ \Delta^{\text{treatment}}p_{t+k} }{ p_{t} }
-
\frac{ \Delta^{\text{control}}p_{t+k} }{ p_{t} }
%\\ \\
%\beta_{k}
%&=
%\E \left( y_{Texas,post} - y_{Control,post} \right) 
%-
%\E \left( y_{Texas,pre} - y_{Control,pre} \right) 
\end{align*}

%%%%%%%%%%%%%%%%%%%%%%%%%%%%%%%%%%%%%%%%%%%%%%%%%%%%%%
\newpage 
\section{For Online Publication: Disentangling Purchase VS HEL Collateral Service Flow }
\label{Model_simple}
The objective of this note is to disentangle 
purchase debt capacity from HEL debt capacity. 
%PM: can be used for consumption \& Housing.
%HEL: can only be used for consumption in this model, as in Texas.
Consider a household who lives for three periods. 
The household: 
buys a home with a purchase mortgage $(b_{0}^{PM})$ at $t=0$, 
borrows via a HEL $(b_{1}^{HEL})$ at $t=1$, 
sells the home and repays the loans at $t=2$. 
\\ \\
The household maximizes:
%\\
$
U= u(c_0, h_0)
+ \beta u(c_1, h_1) 
%+ %HEL for certain? 
+ \beta^2 B(c_2) %u(c_2, h_2) %Bequest wealth instead?
$
\\
subject to:
\\ %\\
$c_0 + p_{0} h_{0} \leq y_{0} + b_{0}^{PM}$
\\
$ b_{0}^{PM} \leq \kappa^{PM} p_{0} h_{0} $
%and 
%$ b_{0}^{PM} \geq 0 $
\\ %\\
$c_1  \leq y_{1} + b_{1}^{HEL}$
\\ 
$ b_{1}^{HEL} \leq \kappa^{HEL} p_{1} h_{1}  - b_{1}^{PM} $
$\Leftrightarrow$
$ 
\frac{ b_{1}^{PM} + b_{1}^{HEL}  } { p_{1} h_{1}  } 
\leq \kappa^{HEL}
$
\\
Note: $\kappa^{HEL}$ is the combined LTV limit
for after purchase loans based on the appraised $p_{1}h_{1}$.
\\ %\\
$c_2 + b_2^{PM}  + b_2^{HEL}  \leq y_{2} + p_2 h_2 $
\\ \\
%Assume enough structure so that: 
We impose: 
%and ALL constraints bind
%to ignore extra conditions:
\\
$b_{t}^{PM}, b_{t}^{HEL} \geq 0$: %borrow not save! 
households don't want to save
\\
$b_{t+1}^{PM} = R b_{t}^{PM}$
and 
$b_{t+1}^{HEL} = R b_{t}^{HEL}$: 
households pay principal and interest at the end
\\
$h_0 =h_1 =h_2 
\equiv h$: housing is chosen once $(t=0)$ and doesn't depreciate
\\ 
The household's Lagrangian is:
\\
$
\Lag = U
\\ 
+\lambda_{0}\left(y_{0} + b_{0}^{PM} -c_0 - p_{0} h  \right)
+\mu_{0}\left(\kappa^{PM} p_{0} h - b_{0}^{PM}   \right)
\\
+\lambda_{1}\left(y_{1} + b_{1}^{HEL} -c_1  \right)
+\mu_{1}\left(\kappa^{HEL} p_{1} h- R b_{0}^{PM} - b_{1}^{HEL}   \right)
\\
+\lambda_{2}\left(y_{2} + p_2 h - c_2 - R^2 b_{0}^{PM} - R b_{1}^{HEL}    \right)
$
%\\
%Exogenous variables
%$ \left(
%y_{0}, y_{1}, y_{2}, p_{0}, p_{1}, p_{2}, R, \kappa^{PM}, \kappa^{HEL}
%\right)$
%\\
%6 Choice variables:
%$
%\left(c_0, c_1, c_2, h, b_{0}^{PM}, b_{1}^{HEL} \right)
%$
\\ \\
%6 
FOCs, Primal Feasibility, Dual Feasibility, and Complementary Slackness:
\\
$\Lag_{c_{0}}$:
$U_{c_{0}} - \lambda_{0} = 0$
\\
$\Lag_{c_{1}}$:
$U_{c_{1}} - \lambda_{1} = 0$
\\
$\Lag_{c_{2}}$:
$U_{c_{2}} - \lambda_{2} = 0$
\\ 
$\Lag_{\lambda_{0}}$:
$
\left(y_{0} + b_{0}^{PM} -c_0 - p_{0} h  \right) =0
$
\\ 
$\Lag_{\lambda_{1}}$:
$ \left(y_{1} + b_{1}^{HEL} -c_1  \right) =0 $
\\ 
$\Lag_{\lambda_{2}}$:
$
\left(y_{2} + p_2 h - c_2 - R^2 b_{0}^{PM} - R b_{1}^{HEL}    \right) =0
$
\\
$\Lag_{h}$:
$U_{h} -\lambda_{0} p_{0} + \mu_{0} \kappa^{PM} p_{0} 
+ \mu_{1} \kappa^{HEL} p_{1}
+\lambda_{2} p_{2}  = 0
$
\\
$\Lag_{b_{1}^{HEL}}$:
$\lambda_{1} - \mu_{1}  -\lambda_{2} R  
=
%\geq 
0$
\\
$\Lag_{b_{0}^{PM}}$:
$\lambda_{0} - \mu_{0} - \mu_{1} R -\lambda_{2} R^{2}  
=
%\geq 
 0$
\\
%Primal feasibility, dual feasibility, complementary slackness
$\Lag_{ \mu_{0} }$:
$
\left(\kappa^{PM} p_{0} h - b_{0}^{PM}   \right) \geq 0
$, 
$\mu_{0} \geq 0$, 
$
\mu_{0} \times \left(\kappa^{PM} p_{0} h - b_{0}^{PM}   \right) = 0
$
\\
%Primal feasibility, dual feasibility, complementary slackness
$\Lag_{ \mu_{1} }$:
$
\left(\kappa^{HEL} p_{1} h- R b_{0}^{PM} - b_{1}^{HEL}   \right) \geq 0
$, 
$\mu_{1} \geq 0$, 
$
\mu_{1} \times \left(\kappa^{HEL} p_{1} h- R b_{0}^{PM} - b_{1}^{HEL}   \right) = 0
$
\\ 
The housing FOC can be rearranged:
%$\Leftrightarrow$
%$
%\lambda_{0} p_{0}
%=
%U_{h} + \mu_{0} \kappa^{PM} p_{0} 
%+ \mu_{1} \kappa^{HEL} p_{1}
%+\lambda_{2} p_{2} 
%$
% % % % % % % % % % % % % % %
\begin{align*} 
%\label{eq:FOC}
\underbrace{p_{0}}_{\text{price}}
&=
%\E_{t}
%\left[ 
%\underbrace{ 
	\frac{ 1 }{ \lambda_{0}  }
%}_{\text{ discount factor}} %_{sdf}
\times 
\left( 
\underbrace{ 
	U_{h}
}_{
\substack{\text{housing} \\ \text{service flow (rent)}}
 %\underset{\text{service flow (rent)}} {\text{housing}}
}
%_{\text{service flow--rent}}
+
\underbrace{ 
	\mu_{0} \kappa^{PM} p_{0}
	%\frac{ \mu_{t+1} \kappa_{t+1}  p_{t+1} }{ u_{c}( t+1 ) }    
	%\kappa_{t+1}  p_{t+1}
}_{
\substack{\text{purchase collateral} \\ \text{service flow}}
%\text{collateral value}
}
+
\underbrace{ 
	\mu_{1} \kappa^{HEL} p_{1}
	%\frac{ \mu_{t+1} \kappa_{t+1}  p_{t+1} }{ u_{c}( t+1 ) }    
	%\kappa_{t+1}  p_{t+1}
}_{
	\substack{\text{HEL collateral} \\ \text{service flow}}
	%\text{collateral value}
}
+   
\underbrace{ \lambda_{2} p_{2}  }_{\text{resale price}}
\right) 
%\right] 
%\tag{1a}
%\\
%\label{eq:CC}
%\underbrace{ 
%\mu_{t+1} 
%}_{\text{price}}
%&=
%u_{c}( t+1  ) -  \E_{t+1}[ \beta \left(1+r_{t+2}\right) u_{c}(t+2) ]
%\tag{1b}
%u_{c}( t  ) -  \E_{t}[ \beta \left(1+r_{t+1}\right) u_{c}(t+1) ]
%
%\\ & =
%PDV_{t}\left( \text{s} \right) + PDV_{t}\left( \text{CV} \right)
\end{align*} 
Observe that both 
the purchase collateral service flow and 
the HEL collateral service flow enter the equation.
%
%AZ: CAN ADD
%1: unsecured interest for time=1 borrowing, w higher rate exog
%2: transaction costs for liquidating the home early 
%3: solve 2 versions of the model. A: no after purchase borrowing
%%B: yes
%\\
%%
%We can pin down all multipliers:
%\\
%$\lambda_{t}=U_{c_{t}}$ for $t=0,1,2$
%\\
%$\mu_{1} = \lambda_{1} - \lambda_{2} R 
%= U_{c_{1}} - U_{c_{2}} R$
%\\
%$\mu_{0} 
%=\lambda_{0}  - \mu_{1} R -\lambda_{2} R^{2}
%%\\ \\
%=\lambda_{0}  -R\left(\mu_{1} + \lambda_{2} R\right)  
%=\lambda_{0}  -R\left(\lambda_{1}\right)  
%= U_{c_{0}} - U_{c_{1}} R
%$

\newpage 
% % % % % % % % % % % % % % % % % % % % % % %
% % % % % % % % % % % % % % % % % % % % % % % % % % % % % %
\section{For Online Publication: Household's Problem With Owning and Renting}
\label{deriv2}
This appendix solves the household's problem presented in Section \ref{Model} except the household 
can own or rent a unit of housing at prices $p_{t}^{\text{own}}$, $p_{t}^{\text{rent}}$.
Both owner-occupied housing and rental housing provide shelter, however-owner occupied housing 
also provides collateral service flows. %as well as resale vale 
\begin{align*}
\max \text{ } \E_{0} \sum_{t=0}^{\infty} \beta^{t} u(c_{t} , h_t = h_{t}^{\text{own}} + h_{t}^{\text{rent}} )
&
\\
\text{s.t.}
\\
c_{t} 
+p_{t}^{\text{own}}h_{t+1}^{\text{own}}  +p_{t}^{\text{rent}}h_{t+1}^{\text{rent}}
+ a_{t+1}
&
\leq 
y_{t} + p_{t}^{\text{own}}h_{t}^{\text{own}} (1-\delta_{t}) + (1+r_{t}) a_{t}
\tag{DBC $\lambda_{t}$}
\\
-a_{t+1} 
&\leq 
\kappa_{t} p_{t}^{\text{own}}h_{t}^{\text{own}}
\tag{CC $\mu_{t}$}
\end{align*}
Savings: $a_{t+1}$ at time $t$. 
If $a_{t+1}>0$, save, if $a_{t+1}<0$,
borrow.
\\
The multiplier on the dynamic budget constraint (DBC):
$\lambda(s^{t}) \beta^{t} \pi(s^{t}) > 0$
\\
The multiplier on the collateral constraint (CC):
$\mu(s^{t}) \beta^{t} \pi(s^{t}) \geq 0 $
\\
Complementary slackness:
$
\mu(s^{t}) \beta^{t} \pi(s^{t})
\left[ 
\kappa(s^{t})  p^{\text{own}}(s^{t}) h_{t}^{\text{own}}(s^{t-1} ) + a_{t+1}(s^{t} )
\right]
=
0
$
%$
%\mu(s^{t}) \beta^{t} \pi(s^{t})
%\left[ 
% p(s^{t}) h_{t}(s^{t-1} ) + a_{t+1}(s^{t} )
% \right]
%=
%0
%$
\\ \\
State variables in $s^{t}$: 
$y(s^{t})$, 
$h_{t}^{\text{own}}(s^{t-1})$,
$h_{t}^{\text{rent}}(s^{t-1})$,
$a_{t}(s^{t-1})$,
$r_{t}(s^{t-1})$,
$\delta(s^{t})$,
$\kappa(s^{t})$.
% % % % %
\\
Choice variables in $s^{t}$: 
$c(s^{t})$, 
$h_{t+1}^{\text{own}}(s^{t})$,
$h_{t+1}^{\text{rent}}(s^{t})$,
$a_{t+1}(s^{t})$.
%(Any two will pin down the third.)
%Equilibrium variables: $p_{t}, r_{t+1}, LTV_{t}$
\begin{align*} 
\Lag 
=
\sum_{t=0}^{\infty} 
\underset{s^{t}\in S^{t}}{\sum} 
%\left[ 
& 
\beta^{t} \pi(s^{t})
u( c(s^{t}) , h(s^{t}) )
\\
+&
\lambda(s^{t}) \beta^{t} \pi(s^{t})
\left[ 
y(s^{t}) 
+ p^{\text{own}}(s^{t}) h_{t}^{\text{own}}(s^{t-1} ) (1-\delta(s^{t})) 
+ \left(1+r_{t}(s^{t-1} ) \right) a_{t}(s^{t-1} )
\right]  
%\right. 
%\\
%&
%\left.
\\-&
\lambda(s^{t}) \beta^{t} \pi(s^{t})
\left[   
c(s^{t}) 
+ p^{\text{own}}(s^{t}) h_{t+1}^{\text{own}}(s^{t}) 
+ p^{\text{rent}}(s^{t}) h_{t+1}^{\text{rent}}(s^{t}) 
+ a_{t+1}(s^{t} )
\right]
\\
+&
\mu(s^{t}) \beta^{t} \pi(s^{t})
\left[ 
\kappa(s^{t})  p^{\text{own}}(s^{t}) h_{t}^{\text{own}}(s^{t-1} ) 
+ a_{t+1}(s^{t} )
\right]
\end{align*}
% % % % %
%In $s^{t}$ we have: 
%$y(s^{t})$, 
%$h_{t}(s^{t-1})$,
%$a_{t}(s^{t-1})$,
%$\delta(s^{t})$,
%$\kappa(s^{t})$.
%% % % % %
%\\
%In $s^{t}$ we choose: 
%$c(s^{t})$, 
%$h_{t+1}(s^{t})$,
%$a_{t+1}(s^{t})$.
%(Any two will pin down the third.)
% % % %
\begin{align*} 
\Lag_{c(s^{t})} 
&=
\beta^{t} \pi(s^{t})
u_{1}( s^{t} )
+
\lambda(s^{t}) \beta^{t} \pi(s^{t}) \left[-1\right]
\\&=0
\\
&\Leftrightarrow 
\\
\lambda(s^{t}) &= u_{1}(s^{t})
\end{align*} 
%\\ \\
% % % %
\begin{align*} 
\Lag_{h_{t+1}^{\text{own}}(s^{t})} 
&=
\sum_{s^{t+1}|s^{t}}
\beta^{t+1} \pi(s^{t+1}) u_{2}( s^{t+1} )
\\&
+
\lambda(s^{t}) \beta^{t} \pi(s^{t}) 
\left[
-p^{\text{own}}(s^{t})
\right]
\\&
+
\sum_{s^{t+1}|s^{t}}
\lambda(s^{t+1}) \beta^{t+1} \pi(s^{t+1}) 
\left[
p^{\text{own}}(s^{t+1})  (1-\delta(s^{t+1}))
\right]
\\&
+
\sum_{s^{t+1}|s^{t}}
\mu(s^{t+1}) \beta^{t+1} \pi(s^{t+1}) 
\left[
\kappa(s^{t+1})
p^{\text{own}}(s^{t+1})
\right]
\\&
=0
\end{align*} 
% % % %
\begin{align*} 
\Lag_{h_{t+1}^{\text{rent}}(s^{t})} 
&=
\sum_{s^{t+1}|s^{t}}
\beta^{t+1} \pi(s^{t+1}) u_{2}( s^{t+1} )
\\&
+
\lambda(s^{t}) \beta^{t} \pi(s^{t}) 
\left[
-p^{\text{rent}}(s^{t})
\right]
%\\&
%+
%\sum_{s^{t+1}|s^{t}}
%\lambda(s^{t+1}) \beta^{t+1} \pi(s^{t+1}) 
%\left[
%p^{\text{own}}(s^{t+1})  (1-\delta(s^{t+1}))
%\right]
%\\&
%+
%\sum_{s^{t+1}|s^{t}}
%\mu(s^{t+1}) \beta^{t+1} \pi(s^{t+1}) 
%\left[
%\kappa(s^{t+1})
%p^{\text{own}}(s^{t+1})
%\right]
\\&
=0
\end{align*} 
% % % %
\begin{align*} 
\Lag_{a_{t+1}(s^{t})} 
&=
\lambda(s^{t}) \beta^{t} \pi(s^{t}) 
\left[
-1
\right]
\\&
+
\mu(s^{t}) \beta^{t} \pi(s^{t}) 
\left[
1
\right]
\\&
+
\sum_{s^{t+1}|s^{t}}
\lambda(s^{t+1}) \beta^{t+1} \pi(s^{t+1}) 
%(1+r)
\left(1+r_{t+1}(s^{t} ) \right)
\\&
=0
\\&
\Leftrightarrow
\\
\lambda(s^{t}) \beta^{t} \pi(s^{t}) 
&=
\mu(s^{t}) \beta^{t} \pi(s^{t}) 
+
\sum_{s^{t+1}|s^{t}}
\lambda(s^{t+1}) \beta^{t+1} \pi(s^{t+1}) 
%(1+r)
\left(1+r_{t+1}(s^{t} ) \right)
\\&
\Leftrightarrow
\\
\lambda(s^{t})  
&=
\mu(s^{t}) 
+
\beta \left(1+r_{t+1}(s^{t} ) \right)
\E_{t}[   \lambda(s^{t+1})  ]
\\
\text{recall }
\lambda(s^{t})&=u_{1}(s^{t}) 
\text{ (consumption foc)}
\\&
\Leftrightarrow
\\
u_{1}(s^{t})  
&=
\mu(s^{t}) 
+
\beta \left(1+r_{t+1}(s^{t} ) \right)
\E_{t}[   u_{1}(s^{t+1})  ]
\end{align*} 
%This is the famous liquidity-constrained Euler equation!
%If the collateral constraint is not binding ($\mu=0$)
%or doesn't exist, then this collapses to the usual
%frictionless Euler equation.
%%
%\\ \\
%
Using $\lambda(s^{t})=u_{1}(s^{t})$, we combine the
consumption and owner-occupied housing FOCs.
\begin{align*}
u_{1}(s^{t}) \beta^{t} \pi(s^{t}) 
\left[
p^{\text{own}}(s^{t})
\right]
&=
\sum_{s^{t+1}|s^{t}}
\beta^{t+1} \pi(s^{t+1}) u_{2}( s^{t+1} )
\\&
+
\sum_{s^{t+1}|s^{t}}
u_{1}(s^{t+1}) \beta^{t+1} \pi(s^{t+1}) 
\left[
p^{\text{own}}(s^{t+1}) (1-\delta( s^{t+1} ) )
\right]
\\&
+
\sum_{s^{t+1}|s^{t}}
\mu(s^{t+1}) \beta^{t+1} \pi(s^{t+1}) 
\left[
p^{\text{own}}(s^{t+1}) \kappa( s^{t+1} )
\right]
\end{align*} 
% % % % % % % % % % % % % % %
% % % % % % % % % % % % % % %
\begin{align*}
u_{1}(s^{t})   
\left[
p^{\text{own}}(s^{t})
\right]
&=
\E_{t}
\beta    u_{2}( s^{t+1} )
\\&
+
\E_{t}
u_{1}(s^{t+1}) \beta    
\left[
p^{\text{own}}(s^{t+1})  (1-\delta( s^{t+1} ) )
\right]
\\&
+
\E_{t}
\mu(s^{t+1}) \beta   
\left[
p^{\text{own}}(s^{t+1})  \kappa( s^{t+1} )
\right]
\end{align*} 
% % % % % % % % % % % % % % %
% % % % % % % % % % % % % % %
\begin{align*}
p^{\text{own}}(s^{t})
&=
\E_{t}
\beta   \frac{  u_{2}( s^{t+1} ) }{ u_{1}(s^{t})   }
\\&
+
\E_{t}
\beta    
\frac{ u_{1}(s^{t+1})  }{ u_{1}(s^{t})   } 
\left[
p^{\text{own}}(s^{t+1})  (1-\delta( s^{t+1} ) )
\right]
\\&
+
\E_{t}
\beta
\frac{ \mu(s^{t+1}) }{ u_{1}(s^{t}) }    
\left[
p^{\text{own}}(s^{t+1})  \kappa( s^{t+1} )
\right]
\end{align*} 
% % % % % % % % % % % % % % %
% % % % % % % % % % % % % % %
\begin{align*}
p^{\text{own}}(s^{t})
&=
\E_{t}
\beta   
\frac{  u_{1}( s^{t+1} ) }{ u_{1}(s^{t})   }
\frac{  u_{2}( s^{t+1} ) }{ u_{1}(s^{t+1})   }
\\&
+
\E_{t}
\beta    
\frac{ u_{1}(s^{t+1})  }{ u_{1}(s^{t})   } 
\left[
p^{\text{own}}(s^{t+1}) (1-\delta( s^{t+1} ) )
\right]
%\\&
+
\E_{t}
\beta
\frac{ u_{1} (s^{t+1}) }{ u_{1}(s^{t}) }    
\frac{ \mu(s^{t+1}) }{ u_{1}(s^{t+1}) }    
\left[
p^{\text{own}}(s^{t+1}) \kappa( s^{t+1} )
\right]
\end{align*} 
% % % % % % % % % % % % % % %
% % % % % % % % % % % % % % %
\begin{align*}
p^{\text{own}}(s^{t})
&=
\E_{t}
\left[ 
\beta   
\frac{  u_{1}( s^{t+1} ) }{ u_{1}(s^{t})   }
\left( 
\frac{  u_{2}( s^{t+1} ) }{ u_{1}(s^{t+1})   }
+   
\left[
p^{\text{own}}(s^{t+1}) (1-\delta( s^{t+1} ) )
\right]
%\\&
+
\frac{ \mu(s^{t+1}) }{ u_{1}(s^{t+1}) }    
\left[
p^{\text{own}}(s^{t+1}) \kappa( s^{t+1} )
\right]
\right) 
\right] 
\end{align*} 
%%%%%%%%%%%%%%%%%%%%%%%%%%%%%%%%
\begin{align*}
p_{t}^{\text{own}}
&=
\E_{t}
\left[ 
\underbrace{ 
	\beta   
	\frac{  u_{1}( t+1 ) }{ u_{1}( t )   }
}_{sdf}
\times 
\left( 
\underbrace{ 
	\frac{  u_{2}(  t+1  ) }{ u_{1}( t+1 )   }
}_{\text{housing service flow}}
+
\underbrace{ 
	\frac{ \mu( t+1 ) }{ u_{1}( t+1 ) }    
	\kappa_{t+1} p_{t+1}^{\text{own}}
}_{\text{collateral service flow}}
+   
\underbrace{ (1-\delta_{t+1} ) p_{t+1}^{\text{own}} }_{\text{resale price}}
\right) 
\right] 
\end{align*} 
% % % % % % % % % % % % % % %
Using $\lambda(s^{t})=u_{1}(s^{t})$, we combine the
consumption and rental housing FOCs.
\begin{align*} 
p^{\text{rent}}(s^{t})
&=
\sum_{s^{t+1}|s^{t}}
\frac{\beta^{t+1} \pi(s^{t+1}) }{ \beta^{t} \pi(s^{t}) }
\frac{ u_{2}( s^{t+1} ) }{ \lambda(s^{t}) } 
\\&=
\sum_{s^{t+1}|s^{t}}
\frac{\beta^{t+1} \pi(s^{t+1}) }{ \beta^{t} \pi(s^{t}) }
\frac{ u_{2}( s^{t+1} ) }{ u_{1}(s^{t}) } 
\\
p^{\text{rent}}_{t}
&=
\E_{t}
\left[ 
\underbrace{ 
	\beta   
	\frac{  u_{1}( t+1 ) }{ u_{1}( t )   }
}_{sdf}
\times 
\left( 
\underbrace{ 
	\frac{  u_{2}(  t+1  ) }{ u_{1}( t+1 )   }
}_{\text{housing service flow}}
\right) 
\right] 
\end{align*} 
% % % % % % % % % % % % % % %
\begin{align*}
p_{t}^{\text{own}}
&=
p^{\text{rent}}_{t}
+
\E_{t}
\left[ 
\underbrace{ 
	\beta   
	\frac{  u_{1}( t+1 ) }{ u_{1}( t )   }
}_{sdf}
\times 
\left( 
\underbrace{ 
	\frac{ \mu( t+1 ) }{ u_{1}( t+1 ) }    
	\kappa_{t+1} p_{t+1}^{\text{own}}
}_{\text{collateral service flow}}
+   
\underbrace{ (1-\delta_{t+1} ) p_{t+1}^{\text{own}} }_{\text{resale price}}
\right) 
\right] 
\end{align*} 
% % % % % % % % % % % % % % % %

\newpage 
% % % % % % % % % % % % % % % % % % % % % % %
% % % % % % % % % % % % % % % % % % % % % % % % % % % % % %
\section{For Online Publication: Household's Problem With Owning and Renting 2}
\label{deriv2b}
This appendix solves the household's problem presented in Section \ref{Model} except the household 
can own or rent a unit of housing at prices $p_{t}^{\text{own}}$, $p_{t}^{\text{rent}}$.
Both owner-occupied housing and rental housing provide shelter, 
%following 
$h$ units of rental housing provides $h$ units of housing services 
while 
$h$ units of owner-occupied housing provides $\omega h$ units of housing services, 
where $\omega \geq 1$ captures extra utility from homeownership
\citep{KMV}. 
In addition, a unit of owner-occupied housing 
also provides collateral service flows. %as well as resale vale 
\begin{align*}
\max \text{ } \E_{0} \sum_{t=0}^{\infty} \beta^{t} 
u(c_{t} , h_t = \omega h_{t}^{\text{own}} + h_{t}^{\text{rent}} )
&
\\
\text{s.t.}
\\
c_{t} 
+p_{t}^{\text{own}}h_{t+1}^{\text{own}}  +p_{t}^{\text{rent}}h_{t+1}^{\text{rent}}
+ a_{t+1}
&
\leq 
y_{t} + p_{t}^{\text{own}}h_{t}^{\text{own}} (1-\delta_{t}) + (1+r_{t}) a_{t}
\tag{DBC $\lambda_{t}$}
\\
-a_{t+1} 
&\leq 
\kappa_{t} p_{t}^{\text{own}}h_{t}^{\text{own}}
\tag{CC $\mu_{t}$}
\end{align*}
Savings: $a_{t+1}$ at time $t$. 
If $a_{t+1}>0$, save, if $a_{t+1}<0$,
borrow.
\\
The multiplier on the dynamic budget constraint (DBC):
$\lambda(s^{t}) \beta^{t} \pi(s^{t}) > 0$
\\
The multiplier on the collateral constraint (CC):
$\mu(s^{t}) \beta^{t} \pi(s^{t}) \geq 0 $
\\
Complementary slackness:
$
\mu(s^{t}) \beta^{t} \pi(s^{t})
\left[ 
\kappa(s^{t})  p^{\text{own}}(s^{t}) h_{t}^{\text{own}}(s^{t-1} ) + a_{t+1}(s^{t} )
\right]
=
0
$
%$
%\mu(s^{t}) \beta^{t} \pi(s^{t})
%\left[ 
% p(s^{t}) h_{t}(s^{t-1} ) + a_{t+1}(s^{t} )
% \right]
%=
%0
%$
\\ \\
State variables in $s^{t}$: 
$y(s^{t})$, 
$h_{t}^{\text{own}}(s^{t-1})$,
$h_{t}^{\text{rent}}(s^{t-1})$,
$a_{t}(s^{t-1})$,
$r_{t}(s^{t-1})$,
$\delta(s^{t})$,
$\kappa(s^{t})$.
% % % % %
\\
Choice variables in $s^{t}$: 
$c(s^{t})$, 
$h_{t+1}^{\text{own}}(s^{t})$,
$h_{t+1}^{\text{rent}}(s^{t})$,
$a_{t+1}(s^{t})$.
%(Any two will pin down the third.)
%Equilibrium variables: $p_{t}, r_{t+1}, LTV_{t}$
\begin{align*} 
\Lag 
=
\sum_{t=0}^{\infty} 
\underset{s^{t}\in S^{t}}{\sum} 
%\left[ 
& 
\beta^{t} \pi(s^{t})
u( c(s^{t}) , h(s^{t}) )
\\
+&
\lambda(s^{t}) \beta^{t} \pi(s^{t})
\left[ 
y(s^{t}) 
+ p^{\text{own}}(s^{t}) h_{t}^{\text{own}}(s^{t-1} ) (1-\delta(s^{t})) 
+ \left(1+r_{t}(s^{t-1} ) \right) a_{t}(s^{t-1} )
\right]  
%\right. 
%\\
%&
%\left.
\\-&
\lambda(s^{t}) \beta^{t} \pi(s^{t})
\left[   
c(s^{t}) 
+ p^{\text{own}}(s^{t}) h_{t+1}^{\text{own}}(s^{t}) 
+ p^{\text{rent}}(s^{t}) h_{t+1}^{\text{rent}}(s^{t}) 
+ a_{t+1}(s^{t} )
\right]
\\
+&
\mu(s^{t}) \beta^{t} \pi(s^{t})
\left[ 
\kappa(s^{t})  p^{\text{own}}(s^{t}) h_{t}^{\text{own}}(s^{t-1} ) 
+ a_{t+1}(s^{t} )
\right]
\end{align*}
% % % % %
%In $s^{t}$ we have: 
%$y(s^{t})$, 
%$h_{t}(s^{t-1})$,
%$a_{t}(s^{t-1})$,
%$\delta(s^{t})$,
%$\kappa(s^{t})$.
%% % % % %
%\\
%In $s^{t}$ we choose: 
%$c(s^{t})$, 
%$h_{t+1}(s^{t})$,
%$a_{t+1}(s^{t})$.
%(Any two will pin down the third.)
% % % %
\begin{align*} 
\Lag_{c(s^{t})} 
&=
\beta^{t} \pi(s^{t})
u_{1}( s^{t} )
+
\lambda(s^{t}) \beta^{t} \pi(s^{t}) \left[-1\right]
\\&=0
\\
&\Leftrightarrow 
\\
\lambda(s^{t}) &= u_{1}(s^{t})
\end{align*} 
%\\ \\
% % % %
\begin{align*} 
\Lag_{h_{t+1}^{\text{own}}(s^{t})} 
&=
\sum_{s^{t+1}|s^{t}}
\beta^{t+1} \pi(s^{t+1})  \omega u_{2}( s^{t+1} )
\\&
+
\lambda(s^{t}) \beta^{t} \pi(s^{t}) 
\left[
-p^{\text{own}}(s^{t})
\right]
\\&
+
\sum_{s^{t+1}|s^{t}}
\lambda(s^{t+1}) \beta^{t+1} \pi(s^{t+1}) 
\left[
p^{\text{own}}(s^{t+1})  (1-\delta(s^{t+1}))
\right]
\\&
+
\sum_{s^{t+1}|s^{t}}
\mu(s^{t+1}) \beta^{t+1} \pi(s^{t+1}) 
\left[
\kappa(s^{t+1})
p^{\text{own}}(s^{t+1})
\right]
\\&
=0
\end{align*} 
% % % %
\begin{align*} 
\Lag_{h_{t+1}^{\text{rent}}(s^{t})} 
&=
\sum_{s^{t+1}|s^{t}}
\beta^{t+1} \pi(s^{t+1}) u_{2}( s^{t+1} )
\\&
+
\lambda(s^{t}) \beta^{t} \pi(s^{t}) 
\left[
-p^{\text{rent}}(s^{t})
\right]
%\\&
%+
%\sum_{s^{t+1}|s^{t}}
%\lambda(s^{t+1}) \beta^{t+1} \pi(s^{t+1}) 
%\left[
%p^{\text{own}}(s^{t+1})  (1-\delta(s^{t+1}))
%\right]
%\\&
%+
%\sum_{s^{t+1}|s^{t}}
%\mu(s^{t+1}) \beta^{t+1} \pi(s^{t+1}) 
%\left[
%\kappa(s^{t+1})
%p^{\text{own}}(s^{t+1})
%\right]
\\&
=0
\end{align*} 
% % % %
\begin{align*} 
\Lag_{a_{t+1}(s^{t})} 
&=
\lambda(s^{t}) \beta^{t} \pi(s^{t}) 
\left[
-1
\right]
\\&
+
\mu(s^{t}) \beta^{t} \pi(s^{t}) 
\left[
1
\right]
\\&
+
\sum_{s^{t+1}|s^{t}}
\lambda(s^{t+1}) \beta^{t+1} \pi(s^{t+1}) 
%(1+r)
\left(1+r_{t+1}(s^{t} ) \right)
\\&
=0
\\&
\Leftrightarrow
\\
\lambda(s^{t}) \beta^{t} \pi(s^{t}) 
&=
\mu(s^{t}) \beta^{t} \pi(s^{t}) 
+
\sum_{s^{t+1}|s^{t}}
\lambda(s^{t+1}) \beta^{t+1} \pi(s^{t+1}) 
%(1+r)
\left(1+r_{t+1}(s^{t} ) \right)
\\&
\Leftrightarrow
\\
\lambda(s^{t})  
&=
\mu(s^{t}) 
+
\beta \left(1+r_{t+1}(s^{t} ) \right)
\E_{t}[   \lambda(s^{t+1})  ]
\\
\text{recall }
\lambda(s^{t})&=u_{1}(s^{t}) 
\text{ (consumption foc)}
\\&
\Leftrightarrow
\\
u_{1}(s^{t})  
&=
\mu(s^{t}) 
+
\beta \left(1+r_{t+1}(s^{t} ) \right)
\E_{t}[   u_{1}(s^{t+1})  ]
\end{align*} 
%This is the famous liquidity-constrained Euler equation!
%If the collateral constraint is not binding ($\mu=0$)
%or doesn't exist, then this collapses to the usual
%frictionless Euler equation.
%%
%\\ \\
%
Using $\lambda(s^{t})=u_{1}(s^{t})$, we combine the
consumption and owner-occupied housing FOCs.
\begin{align*}
u_{1}(s^{t}) \beta^{t} \pi(s^{t}) 
\left[
p^{\text{own}}(s^{t})
\right]
&=
\sum_{s^{t+1}|s^{t}}
\beta^{t+1} \pi(s^{t+1}) \omega u_{2}( s^{t+1} )
\\&
+
\sum_{s^{t+1}|s^{t}}
u_{1}(s^{t+1}) \beta^{t+1} \pi(s^{t+1}) 
\left[
p^{\text{own}}(s^{t+1}) (1-\delta( s^{t+1} ) )
\right]
\\&
+
\sum_{s^{t+1}|s^{t}}
\mu(s^{t+1}) \beta^{t+1} \pi(s^{t+1}) 
\left[
p^{\text{own}}(s^{t+1}) \kappa( s^{t+1} )
\right]
\end{align*} 
% % % % % % % % % % % % % % %
% % % % % % % % % % % % % % %
\begin{align*}
u_{1}(s^{t})   
\left[
p^{\text{own}}(s^{t})
\right]
&=
\E_{t}
\beta \omega   u_{2}( s^{t+1} )
\\&
+
\E_{t}
u_{1}(s^{t+1}) \beta    
\left[
p^{\text{own}}(s^{t+1})  (1-\delta( s^{t+1} ) )
\right]
\\&
+
\E_{t}
\mu(s^{t+1}) \beta   
\left[
p^{\text{own}}(s^{t+1})  \kappa( s^{t+1} )
\right]
\end{align*} 
% % % % % % % % % % % % % % %
% % % % % % % % % % % % % % %
\begin{align*}
p^{\text{own}}(s^{t})
&=
\E_{t}
\beta \omega  \frac{  u_{2}( s^{t+1} ) }{ u_{1}(s^{t})   }
\\&
+
\E_{t}
\beta    
\frac{ u_{1}(s^{t+1})  }{ u_{1}(s^{t})   } 
\left[
p^{\text{own}}(s^{t+1})  (1-\delta( s^{t+1} ) )
\right]
\\&
+
\E_{t}
\beta
\frac{ \mu(s^{t+1}) }{ u_{1}(s^{t}) }    
\left[
p^{\text{own}}(s^{t+1})  \kappa( s^{t+1} )
\right]
\end{align*} 
% % % % % % % % % % % % % % %
% % % % % % % % % % % % % % %
\begin{align*}
p^{\text{own}}(s^{t})
&=
\E_{t}
\beta   \omega
\frac{  u_{1}( s^{t+1} ) }{ u_{1}(s^{t})   }
\frac{  u_{2}( s^{t+1} ) }{ u_{1}(s^{t+1})   }
\\&
+
\E_{t}
\beta    
\frac{ u_{1}(s^{t+1})  }{ u_{1}(s^{t})   } 
\left[
p^{\text{own}}(s^{t+1}) (1-\delta( s^{t+1} ) )
\right]
%\\&
+
\E_{t}
\beta
\frac{ u_{1} (s^{t+1}) }{ u_{1}(s^{t}) }    
\frac{ \mu(s^{t+1}) }{ u_{1}(s^{t+1}) }    
\left[
p^{\text{own}}(s^{t+1}) \kappa( s^{t+1} )
\right]
\end{align*} 
% % % % % % % % % % % % % % %
% % % % % % % % % % % % % % %
\begin{align*}
p^{\text{own}}(s^{t})
&=
\E_{t}
\left[ 
\beta   
\frac{  u_{1}( s^{t+1} ) }{ u_{1}(s^{t})   }
\left( 
\omega  \frac{  u_{2}( s^{t+1} ) }{ u_{1}(s^{t+1})   }
+   
\left[
p^{\text{own}}(s^{t+1}) (1-\delta( s^{t+1} ) )
\right]
%\\&
+
\frac{ \mu(s^{t+1}) }{ u_{1}(s^{t+1}) }    
\left[
p^{\text{own}}(s^{t+1}) \kappa( s^{t+1} )
\right]
\right) 
\right] 
\end{align*} 
%%%%%%%%%%%%%%%%%%%%%%%%%%%%%%%%
\begin{align*}
p_{t}^{\text{own}}
&=
\E_{t}
\left[ 
\underbrace{ 
	\beta   
	\frac{  u_{1}( t+1 ) }{ u_{1}( t )   }
}_{sdf}
\times 
\left( 
\underbrace{ 
\omega	\frac{  u_{2}(  t+1  ) }{ u_{1}( t+1 )   }
}_{\text{housing service flow}}
+
\underbrace{ 
	\frac{ \mu( t+1 ) }{ u_{1}( t+1 ) }    
	\kappa_{t+1} p_{t+1}^{\text{own}}
}_{\text{collateral service flow}}
+   
\underbrace{ (1-\delta_{t+1} ) p_{t+1}^{\text{own}} }_{\text{resale price}}
\right) 
\right] 
\end{align*} 
% % % % % % % % % % % % % % %
Using $\lambda(s^{t})=u_{1}(s^{t})$, we combine the
consumption and rental housing FOCs.
\begin{align*} 
p^{\text{rent}}(s^{t})
&=
\sum_{s^{t+1}|s^{t}}
\frac{\beta^{t+1} \pi(s^{t+1}) }{ \beta^{t} \pi(s^{t}) }
\frac{ u_{2}( s^{t+1} ) }{ \lambda(s^{t}) } 
\\&=
\sum_{s^{t+1}|s^{t}}
\frac{\beta^{t+1} \pi(s^{t+1}) }{ \beta^{t} \pi(s^{t}) }
\frac{ u_{2}( s^{t+1} ) }{ u_{1}(s^{t}) } 
\\
p^{\text{rent}}_{t}
&=
\E_{t}
\left[ 
\underbrace{ 
	\beta   
	\frac{  u_{1}( t+1 ) }{ u_{1}( t )   }
}_{sdf}
\times 
\left( 
\underbrace{ 
	\frac{  u_{2}(  t+1  ) }{ u_{1}( t+1 )   }
}_{\text{housing service flow}}
\right) 
\right] 
\end{align*} 
% % % % % % % % % % % % % % %
\begin{align*}
p_{t}^{\text{own}}
&=
\omega p^{\text{rent}}_{t}
+
\E_{t}
\left[ 
\underbrace{ 
	\beta   
	\frac{  u_{1}( t+1 ) }{ u_{1}( t )   }
}_{sdf}
\times 
\left( 
\underbrace{ 
	\frac{ \mu( t+1 ) }{ u_{1}( t+1 ) }    
	\kappa_{t+1} p_{t+1}^{\text{own}}
}_{\text{collateral service flow}}
+   
\underbrace{ (1-\delta_{t+1} ) p_{t+1}^{\text{own}} }_{\text{resale price}}
\right) 
\right] 
\end{align*} 
% % % % % % % % % % % % % % % %
If $\omega >1$ then a unit of owner-occupied housing generates 
more utility than a unit of rental housing. 
The paper uses rents as a proxy for housing service flows to inspect the mechanism through which 
the HEL legalization affected house prices. 
Even if $\omega >1$, rent remains a valid proxy as long as the additional utility from shelter
was not affected by the law change.

%%%%%%%%%%%%%%%%%%%%%%%%%%%%%%%%%%%%%%%%%%%%%%%%%%%%%%
\newpage 
\section{For Online Publication: Comparison of Collateral Value Models in the Literature}
\label{Comparison}
The model in Section \ref{Model} assumes the collateral constraint is 
$-a_{t+1} \leq \kappa_{t} p_{t} h_{t}$. This section shows  the results are preserved 
under different specifications of the collateral constraint which appear in the literature.

The collateral constraint can be written in 
%one of 
the following 
%four
ways:
\begin{align*}
-a_{t+1} 
& \leq 
\kappa_{t} p_{t} h_{t}
%%%
\\
-a_{t+1} 
&  \leq 
\kappa_{t} \E_{t}[ p_{t+1} ] h_{t }
\\
-a_{t+1} 
&  \leq 
\kappa_{t} p_{t} h_{t+1}
\\
-a_{t+1} 
&  \leq 
\kappa_{t} \E_{t}[ p_{t+1} ] h_{t+1}
\end{align*}
Depending on whether lenders let you borrow against the
(1) present value of the asset you currently own, 
(2) expected future value of the asset you currently own,  
(3) present value of the asset you bought for tomorrow, 
(4) expected future value of the asset you bought for tomorrow. 
%Several of these issues disappear in continuous time.   
\\ \\
%\textbf{Zevelev 2014}
%\\ 
(1) Households borrow against the present value 
of housing they have today.
%\\
%$-a_{t+1} 
%\leq 
%\kappa_{t} p_{t} h_{t}
%$
\begin{align*}
c_{t}+p_{t}h_{t+1} + a_{t+1}
&
\leq 
y_{t} + p_{t} h_{t} (1-\delta_{t}) + (1+r_{t}) a_{t}
\tag{DBC $\lambda_{t}$}
\\
-a_{t+1} 
&\leq 
\kappa_{t} p_{t} h_{t}
\tag{CC $\mu_{t}$}
\end{align*}
% % % % % % % % % % % % % % %
% % % % % % % % % % % % % % %
\begin{align*}
\underbrace{p_{t}}_{\text{price}}
&=
\E_{t}
\left[ 
\underbrace{ 
	\beta   
	\frac{  u_{1}( t+1 ) }{ u_{1}( t )   }
}_{sdf}
\times 
\left( 
\underbrace{ 
	\frac{  u_{2}(  t+1  ) }{ u_{1}( t+1 )   }
}_{\text{service flow}}
+
\underbrace{ 
	\frac{ \mu( t+1 ) }{ u_{1}( t+1 ) }    
	\kappa_{t+1}  p_{t+1}
}_{\text{collateral value}}
+   
\underbrace{ (1-\delta_{t+1} ) p_{t+1} }_{\text{resale price}}
\right) 
\right] 
\end{align*} 
%\textbf{Zevelev 2014}
%\\ 
(2) Households borrow against the expected future
value of new housing. 
%\\
%$-a_{t+1} 
%\leq 
%\kappa_{t} p_{t} h_{t}
%$
\begin{align*}
c_{t}+p_{t}h_{t+1} + a_{t+1}
&
\leq 
y_{t} + p_{t} h_{t} (1-\delta_{t}) + (1+r_{t}) a_{t}
\tag{DBC $\lambda_{t}$}
\\
-a_{t+1} 
&\leq 
\kappa_{t} \E_{t}[ p_{t+1} ] h_{t+1}
\tag{CC $\mu_{t}$}
\end{align*}
% % % % % % % % % % % % % % %
% % % % % % % % % % % % % % %
% % % % % % % % % % % % % % %
% % % % % % % % % % % % % % %
\begin{align*}
\underbrace{p_{t}}_{\text{price}}
&=
\E_{t}
\left[ 
\underbrace{ 
	\beta   
	\frac{  u_{1}( t+1 ) }{ u_{1}( t )   }
}_{sdf}
\times 
\left( 
\underbrace{ 
	\frac{  u_{2}(  t+1  ) }{ u_{1}( t+1 )   }
}_{\text{service flow}}
+
\underbrace{ 
	\frac{ \mu_{t} \kappa_{t} }{ \beta u_{1}( t+1 ) }    
	p_{t+1}
}_{\text{collateral value}}
+   
\underbrace{ (1-\delta_{t+1} ) p_{t+1} }_{\text{resale price}}
\right) 
\right] 
\end{align*} 
\\ \\
(3)
Households borrow against the 
present value of  land they buy at $t$ for $t+1$ \citep{BianchiBozMendoza2012}.
\begin{align*}
q_{t} k_{t+1} + c_{t} + \frac{b_{t+1}}{R_{t}} 
&
\leq 
q_{t} k_{t} + b_{t} + \epsilon_{t} Y(k_{t})
\tag{DBC $\lambda_{t}$}
\\
-\frac{b_{t+1} }{R_{t} } 
&\leq 
\kappa_{t} q_{t}  k_{t+1}
\tag{CC $\mu_{t}$}
\end{align*}
\begin{align*} 
\underbrace{q_{t}}_{\text{price}} 
&=
\underbrace{ 
	\frac{ q_{t} \mu_{t} \kappa_{t} }{  u'(t) }
}_{\text{collateral value}}
+
\E_{t}^{s}
\left[
\underbrace{ 
	\beta \frac{ u'(t+1) }{ u'(t) }
}_{\text{sdf}} 
\times 
\left(
\underbrace{ 
	\epsilon_{t+1} Y_{k}(k_{t+1})
}_{\text{service flow}}
+ 
\underbrace{ q_{t+1} }_{\text{resale price}}
\right) 
\right]
\end{align*} 
This equation is not in their paper, this
is a rearrangement
of their FOC to illustrate collateral value.
%
%
%\textbf{Iacoviello 2005}
\\
(4) Households borrow against the 
expected future value of new housing collateral 
purchased at time $t$ \citep{Iacoviello2005}.
\begin{align*}
%\max \text{ } \E_{0} \sum_{t=0}^{\infty} 
%\gamma^{t} \log(c_{t})
%&
%\\
%\text{s.t.}
%\\
c_{t}+ q_{t} h_{t} + \frac{R_{t-1} b_{t-1} }{\pi_{t}} + w_{t}' L_{t}
&
\leq 
\frac{Y_{t}}{ X_{t} }  + q_{t} h_{t-1}   + b_{t}
\tag{DBC }
\\
b_{t}
&\leq 
m \E_{t}\left[ \frac{ q_{t+1} h_{t} \pi_{t+1} }{R_{t} } \right] 
\tag{CC $\lambda_{t}$}
%\\
%%Y_{t}
%&=
%A (h_{t-1})^{\nu} (L_{t})^{1-\nu}
\end{align*}
\begin{align*}
\underbrace{ q_{t} }_{\text{price}}
&=
\E_{t}
\left[
\underbrace{ \gamma  \frac{  u'(c_{t+1}) }{ u'(c_{t}) } }_{\text{sdf}}
\times
\left(
\underbrace{ \nu \frac{ Y_{t+1} }{ X_{t+1} h_{t}} }_{\text{service}}
+
\underbrace{q_{t+1}}_{\text{resale price}}
\right)  
+ 
\underbrace{ \lambda_{t} m \pi_{t+1} q_{t+1} }_{\text{collateral value}}
\right]
\end{align*}
%$
%p_{t} = E_{t}[ M_{t+1}( s_{t+1} + p_{t+1} ) + CV_{t+1} ]
%$
%\\
%The collateral value is not discounted in the sdf. 
%
%
%\textbf{Kiyotaki Moore 1997}
%\\
Farmers borrow against the land they buy at $t$ at next period's
price \citep{KiyotakiMoore1997}.
\begin{align*}
%\max 
%\text{ } 
%\E_{0} \sum_{t=0}^{\infty} \left(\beta^{F}\right)^{t} x_{t}
%&
%\\
%\text{s.t.}
%\\
x_{t}+ q_{t} k_{t} + R b_{t-1}
&
\leq 
(a+c) k_{t-1} + q_{t} k_{t-1}   + b_{t}
\tag{DBC $\lambda_{t}$}
\\
R b_{t}
&\leq 
q_{t+1} k_{t}
\tag{CC $\mu_{t}$}
\end{align*}
%
%
%\textbf{Fostel, Geanakoplos, AER 2008}
%Leverage Cycles and the Anxious Economy
%\\ 
%Equation (9)
%\\
The collateral value of asset $j$ in state $s$
to agent $i$ is  the marginal benefit from
being able to take out loans backed by asset $j$ \citep{FostelGeanakoplos2008}.
\begin{align*}
CV_{s,j}^{i}
&=
\left[
\frac{1}{1+r_{s}}
-
\frac{1}{1+\omega_{s}^{i} }
\frac{1}{1+r_{s}}
\right]
\phi_{s,j}^{i}
=
\frac{1}{1+r_{s}}
\frac{\omega_{s}^{i}}{1+\omega_{s}^{i} }
\phi_{s,j}^{i}
\end{align*} 
Where $\omega_{s}^{i} $ is the liquidity wedge and $\phi_{s,j}^{i}$
is the collateral capacity.

%%%%%%%%%%%%%%%%%%%%%%%%%%%%%%%%%%%%%%%%%%%%%%%%%%%%%%
%%%%%%%%%%%%%%%%%%%%%%%%%%%%%%%%%%%%%%%%%%%%%%%%%%%%%%
%\newpage
%%%%%%%%%%%%%%%%%%%%%%%%%%%%%%%%%%%%%%%%%%%%%%%%%%%%%%
%
%
%\noindent
\cite{HeWrightZhu2015} have a model where housing loans can only be used as collateral for non-housing 
consumption in a separate ``KM" market. 
%\textbf{He, Wright, Zhu  RED 2015}
%Housing and Liquidity
%\\
%What 
%%we call 
%this paper calls
%``collateral value", they call ``Liquidity Value".
%\\ 
%Equation (5)
% % % % % % % % % % % % % % %
% % % % % % % % % % % % % % %
\begin{align*}
\underbrace{\psi_{t}}_{\text{price}}
&=
%\E_{t}
%\left[ 
\underbrace{ 
	\beta   
	%\frac{  u_{1}( t+1 ) }{ u_{1}( t )   } 
}_{sdf}
\times 
\left( 
\underbrace{ 
	U_{2}\left( x_{t+1} , h_{t+1} \right)
}_{\text{service flow}}
+
\underbrace{ 
	\alpha D_{1} \psi_{t+1} \lambda\left( y_{t+1} \right)
}_{\text{collateral value}}
+   
\underbrace{ 
	%(1-\delta_{t+1} )
	\psi_{t+1} }_{\text{resale price}}
\right) 
%\right] 
\end{align*} 
Their work is closely related to the current paper. 
Housing is used as collateral for future non-housing consumption, not to buy housing. 
When the liquidity value is positive, they find that house prices can display fascinating 
dynamics. 
Figure 2 in their paper shows that steady state house prices are hump shaped in the LTV 
ratio. 
In Section 6 they allow houses to depreciate and construction. 
%\\ 
%Equation (13)
% % % % % % % % % % % % % % %
% % % % % % % % % % % % % % %
\begin{align*}
\underbrace{\psi_{t}}_{\text{price}}
&=
%\E_{t}
%\left[ 
\underbrace{ 
	\beta   
	%\frac{  u_{1}( t+1 ) }{ u_{1}( t )   } 
}_{sdf}
\times 
\left( 
\underbrace{ 
	\Omega\left(  h_{t+1} \right)
}_{\text{service flow}}
+
\underbrace{ 
	\alpha D_{1} (1-\delta  ) \psi_{t+1} \Lambda\left[ (1-\delta  ) \psi_{t+1} h_{t+1} \right]
}_{\text{collateral value}}
+   
\underbrace{ 
	(1-\delta  )
	\psi_{t+1} }_{\text{resale price}}
\right) 
%\right] 
\end{align*}

\newpage
\section*{For Online Publication: Appendix References }
%  \bibliographystyle{apalike}
%\nocite{*}
\bibliographystyle{jf}
%\bibliography{AZ_Reference}

\end{appendices}
%%%%%%%%%%%%%%%%%%%%%%%%%%%%%%%%%%%%%%%%%%%%%%%%%%%%%%
\end{document}